\documentclass{ws-rv975x65}
\usepackage{ws-rv-thm}     % comment this line when `amsthm / theorem / ntheorem` package is used
\usepackage{ws-rv-van}     % numbered citation/references (default)
\usepackage{bbm}
\usepackage{graphics}
\usepackage{amsbsy}
\usepackage{bm}
\usepackage{color}

\usepackage{amsmath}
\usepackage{amsfonts}

\usepackage{amssymb,mathtools}
\usepackage[colorlinks=true, citecolor=blue,linkcolor=blue]{hyperref}

\def\empile#1\over#2{\mathrel{\mathop{\kern 0pt#1}\limits_{#2}}}
\def\bs{\boldsymbol}
\def\wt#1{\widetilde{#1}}

\def\TODO#1{}

\def\p{{\boldsymbol p}}

\def\k{{\boldsymbol k}}

\def\x{{\boldsymbol x}}
\def\y{{\boldsymbol y}}

\def\r{{\boldsymbol r}}
\def\z{{\boldsymbol z}}
\def\u{{\boldsymbol u}}
\def\v{{\boldsymbol v}}

\def\s{{\boldsymbol s}}

\newcommand{\slL}{\raise.15ex\hbox{$/$}\kern-.53em\hbox{$L$}}
\newcommand{\slP}{\raise.15ex\hbox{$/$}\kern-.53em\hbox{$P$}}
\newcommand{\slD}{\raise.15ex\hbox{$/$}\kern-.67em\hbox{$D$}}
\newcommand{\slp}{\raise.1ex\hbox{$/$}\kern-.63em\hbox{$p$}}
\newcommand{\slq}{\raise.1ex\hbox{$/$}\kern-.53em\hbox{$q$}}
\newcommand{\slv}{\raise.1ex\hbox{$/$}\kern-.63em\hbox{$v$}}
\newcommand{\slR}{\raise.15ex\hbox{$/$}\kern-.53em\hbox{$R$}}
\newcommand{\slQ}{\raise.15ex\hbox{$/$}\kern-.53em\hbox{$Q$}}
\newcommand{\slK}{\raise.15ex\hbox{$/$}\kern-.53em\hbox{$K$}}
\newcommand{\slk}{\raise.15ex\hbox{$/$}\kern-.53em\hbox{$k$}}
\newcommand{\slSigma}{\raise.15ex\hbox{$/$}\kern-.53em\hbox{$\Sigma$}}
\newcommand{\slcalP}{\raise.15ex\hbox{$/$}\kern-.63em\hbox{$\cal P$}}
\newcommand{\slcalA}{\raise.15ex\hbox{$/$}\kern-.63em\hbox{$\cal A$}}
\newcommand{\slA}{\raise.15ex\hbox{$/$}\kern-.73em\hbox{$A$}}
\newcommand{\slbfA}{\raise.15ex\hbox{$/$}\kern-.73em\hbox{${\imb A}$}}
\newcommand{\slpartial}{\raise.15ex\hbox{$/$}\kern-.53em\hbox{$\partial$}}
\newcommand{\sla}{\raise.15ex\hbox{$/$}\kern-.53em\hbox{$a$}}
\newcommand{\slb}{\raise.15ex\hbox{$/$}\kern-.53em\hbox{$b$}}
\newcommand{\slc}{\raise.15ex\hbox{$/$}\kern-.53em\hbox{$c$}}
\newcommand{\slC}{\raise.15ex\hbox{$/$}\kern-.63em\hbox{$C$}}
\newcommand{\sln}{\raise.15ex\hbox{$/$}\kern-.575em\hbox{$n$}}

\def\colora{}
\def\colorb{}
\def\colorc{}
\def\colord{}

\makeindex

\begin{document}

%\chapter[The initial stages of heavy ion collisions]{Initial state and thermalization\\in the Color Glass Condensate framework}

\begin{center}
  {\large\bf Initial state and thermalization\\in the Color Glass Condensate framework}
\end{center}

%\label{chap:init}

\date{December 2014}

\author[F. Gelis]{Fran\c{}ois Gelis\footnote{email: francois.gelis@cea.fr}}
%\index[aindx]{Author, F.} % or \aindx{Author, F.}
%\index[aindx]{Author, S.} % or \aindx{Author, S.}

\address{Institut de Physique Th\'eorique,\\
CEA / DSM / Saclay,\\ 91191 Gif sur Yvette cedex, France}

\begin{abstract}
  In this review, I present the description of the early stages of
  heavy ion collisions at high energy in the Color Glass Condensate
  framework, from the pre-collision high energy nuclear wavefunction to
  the point where hydrodynamics may start becoming applicable.
\end{abstract}
%\markright{Initial state and Thermalization} % default is Chapter Title.
\body

\section{Introduction}
Heavy ion collisions pose a challenge for Quantum Chromodynamics
(QCD) because a comprehensive description of these collisions
involves a mix of hard short distance phenomena, and non perturbative
long distance soft physics. Even the aspects of these collisions where
a hard scale may justify a weak coupling approach are not perturbative
in the naive sense of a strict loop expansion. Indeed, resummations
are often required even though the coupling is small, usually because
the bosonic constituents of the system have a large occupation number
that may compensate the smallness of the coupling. 

One of the areas where a weak coupling QCD approach is expected to be
most effective is the description of the early stages of a heavy ion
collision. The term ``early stages'' usually encompasses the description
of the relevant degrees of freedom in the wavefunctions of the two
projectiles prior to their collision, the interactions that happen
during the very brief duration of the collision itself, and the
subsequent evolution of the produced gluons and quarks shortly after
the collision. Roughly speaking, the temperature (or the fourth root
of the energy density if the system is not yet thermalized) of the
system can serve as a measure of the applicability of weak coupling
techniques, since this scale sets the value of the running coupling
constant.

A lot of progress has been made in the last 20 years in understanding
how to apply QCD to these collisions. The starting point was the
realization that a hard scale naturally emerges from the non-linear
interactions among the gluons when their density is
large~\cite{GriboLR1,MuellQ1,BlaizM2}, as is the case in the
wavefunction of high energy hadrons or nuclei. Thus, the bulk of
particle production in high energy heavy ion collisions is amenable to
a weak coupling treatment. The formalism for doing this --known as the
Color Glass Condensate (CGC)-- was progressively established and refined
during this period, and has by now reached a mature state allowing
quantitative and systematic calculations.

An outstanding problem, that has not yet reached a satisfactory state
of understanding, is the transition from the Color Glass Condensate
description to a more macroscopic description such as
hydrodynamics. The main question is to explain, within the CGC
framework, why an hydrodynamical description is possible in the first
place.  In other words, why does the system produced in a heavy ion
collision flow as well as it seems to do? A satisfactory matching
between the CGC and hydrodynamics implies that there should be a
certain range of time in which the two descriptions predict the same
evolution. At the moment, we are not there yet, even though
considerable progress has been made in the past years.

After a brief account of why heavy ion collisions are interesting from
the point of view of QCD (section \ref{sec:hic}), the rest of this
review follows the time-line of a collision. We recall the main
physical ideas behind the parton model in the section
\ref{sec:parton}, and we describe {\sl gluon saturation} in the
section \ref{sec:sat}. The section \ref{sec:cgc} is devoted to the
Color Glass Condensate, the QCD-based effective theory for the
saturated regime, and in the section \ref{sec:LO} we show how to apply
it in order to make leading order calculations in heavy ion
collisions.  Next-to-Leading order contributions are considered in the
section \ref{sec:NLO}, as well as the scale evolution of the gluon
distribution in the projectiles via the JIMWLK equation. In the
section \ref{sec:matching}, we first introduce the main issues and
puzzles posed by attempts to match CGC calculations and hydrodynamics.
In the section \ref{sec:weibel}, we discuss the instabilities that
exist in the solutions of the classical Yang-Mills equations, and
their disastrous consequences for fixed loop-order CGC predictions
beyond leading order.  A resummation that cures these pathologies is
presented in the section \ref{sec:resum}, leading to a scheme known as
the {\sl Classical Statistical Approximation}. We also present
alternative derivations of this approximation scheme in order to make
connections with other approaches. We present in the section
\ref{sec:appl} the results obtained by using this approximation in the
context of heavy ion collisions, using different types of initial
conditions.  A discussion of some known shortcomings of this
approximation is presented in the section \ref{sec:limit}.

\section{Heavy ion collisions}
\label{sec:hic}
\subsection{Reminder on QCD}
\begin{figure}[htbp]
\resizebox*{!}{4.5cm}{\includegraphics{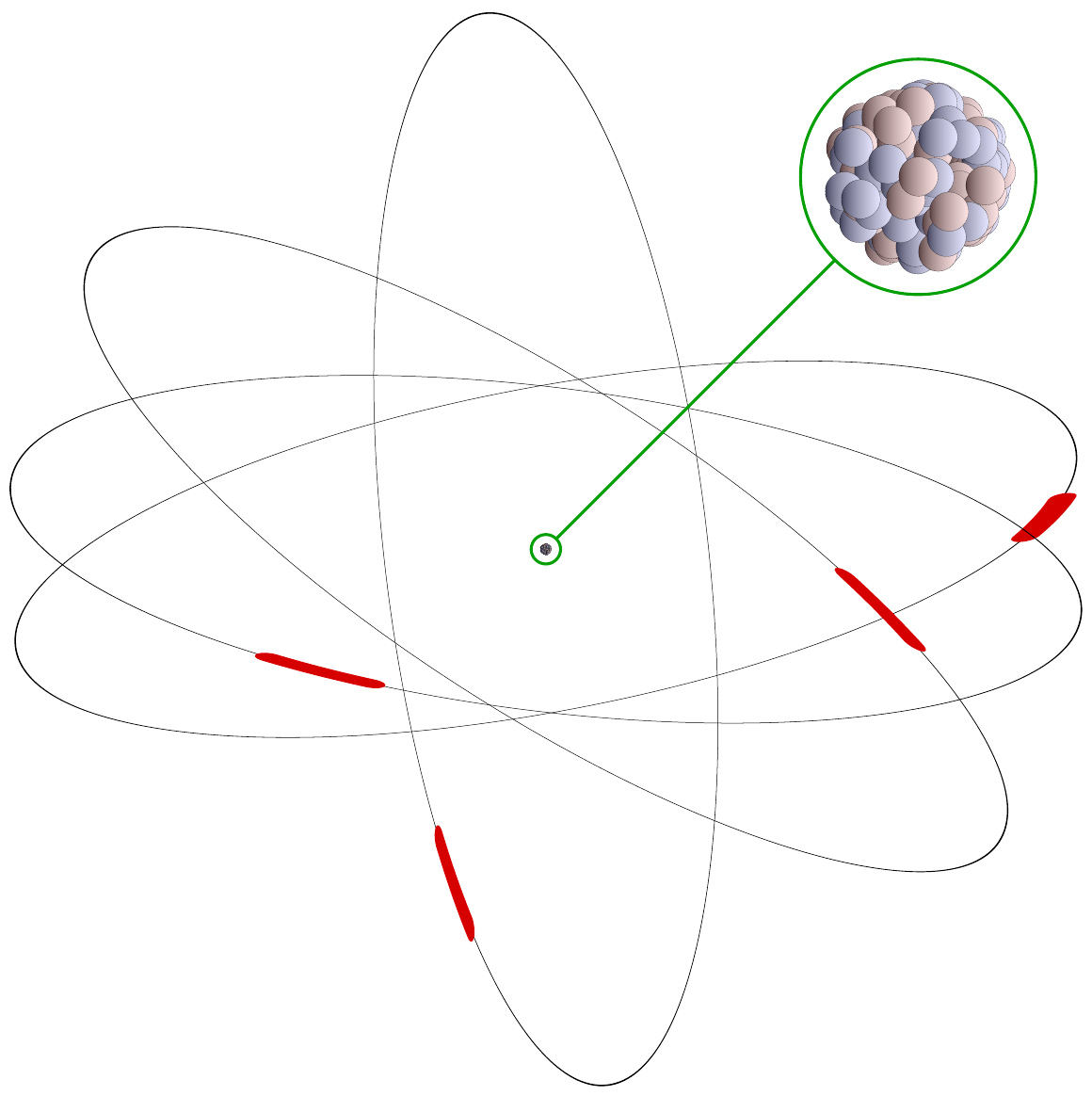}}
\hfill
\resizebox*{!}{4.5cm}{\includegraphics{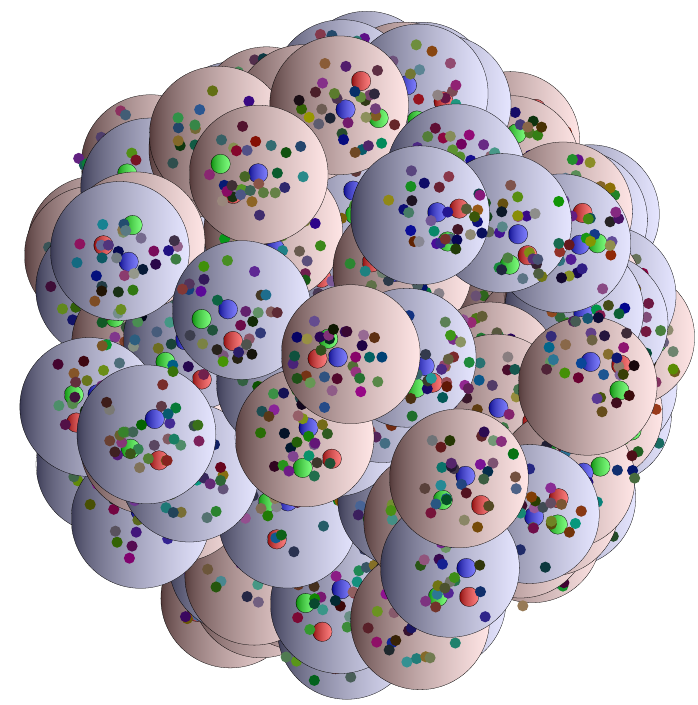}}
\caption{\label{fig:atom}Atoms and nuclei.}
\end{figure}
Although they occupy only a tiny fraction of the volume of atoms, the
atomic nuclei make up for most of the mass of ordinary matter.  The
protons and neutrons that are contained in nuclei each contain three
valence quarks, that give them their quantum numbers. However, these
valence quarks account only for a small part of the nucleon mass. Most of
it comes from binding energy, i.e. from the cloud of gluons and
virtual quark-antiquark pairs that surrounds the valence quarks. This
predominance of binding energy in the mass of hadrons reflects a
crucial property of the force which is responsible of the cohesion of
hadrons and nuclei: this force becomes strong on distance scales
comparable to the proton size, around $10^{-15}$~m. On the other hand,
the measurements of structure functions in deep inelastic scattering
experiments, first performed at SLAC in the 1960's, can be understood
if one assumes that this force becomes weak on distance scales that
are much smaller than the proton size.

The combination of these two properties paved the way to the
development of Quantum Chromodynamics (QCD), as the microscopic theory
that governs the interactions between quarks and gluons. On the
surface, QCD is a gauge theory that resembles very much Quantum
Electrodynamics. The matter degrees of freedom are spin 1/2 quarks,
that interact by the exchange of vector particles, the gluons.
\setbox1=\hbox to 1.5cm{\resizebox*{1.5cm}{!}{\includegraphics{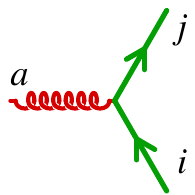}}}
\setbox2=\hbox to 1.5cm{\resizebox*{1.5cm}{!}{\includegraphics{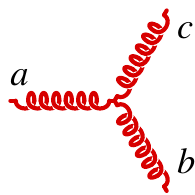}}}
\setbox3=\hbox to 1.5cm{\resizebox*{1.5cm}{!}{\includegraphics{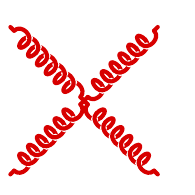}}}
\begin{equation}
\raise -5.2mm\box1\quad \sim\;\; {\colord g}\, (t^a)_{ij}\qquad
\raise -5.2mm\box2\quad \sim\;\; {\colord g}\, (T^a)_{bc}\qquad
\raise -5.2mm\box3
\end{equation}
The quark-gluon coupling in QCD is very similar to the electron-photon
coupling in QED, except that it has more ``structure'' since it
involves a matrix of the fundamental representation of SU(3),
$t^a_{ij}$. In this object, the index $a$ (running from 1 to 8, the
dimension of the SU(3) algebra) is the {\sl color
  charge} of the gluon, and the indices $i$ and $j$ (running from 1 to
3, the dimension of the matrices in the fundamental representation of
SU(3)) are the color charges of the incoming and outgoing quarks.  The
fact that the gluons themselves carry a color charge is the essential
difference between QCD and QED, since it leads to novel interaction
vertices that involve only gluons. These new interactions are a
requirement of gauge symmetry, and can be derived from the following
gauge invariant Lagrangian
\begin{equation}
{\cal L}=-\frac{1}{4}{\colorb F^2} + \sum_f {\colora\overline\psi_f}(i{\colorb\slD}-{\colora m_f}){\colora\psi_f}\; .
\end{equation}
At the classical level, the only free parameters in QCD are the quark
masses $m_f$ and a coupling constant $g$. In the quantized theory, the
coupling is usually traded for a scale\footnote{Note that, in the
  absence of quarks (or with only massless quarks), QCD is scale
  invariant at the classical level. Loop corrections induce a breaking
  of this scale invariance, which is the reason for the appearance of
  $\Lambda_{_{\rm QCD}}$ in the quantized theory.} $\Lambda_{_{\rm
    QCD}}$ that emerges from the renormalization of the coupling.
This new scale arises in the running of the coupling constant
$\alpha_s\equiv g^2/(4\pi)$. At one loop, this is given by~\cite{GrossW1,GrossW2,GrossW3,Polit1,Polit2}
\begin{equation}
      {\colord\alpha_s}({\colora E})=\frac{2\pi N_c}{({\colord 11N_c}-{\colorb 2N_f})
        \log(E/\Lambda_{_{QCD}})}\; ,
\end{equation}
where $E$ is the energy scale, $N_c$ the number of colors and $N_f$
the number of quark flavors.
\begin{figure}[htbp]
\begin{center}
\resizebox*{!}{6cm}{\includegraphics{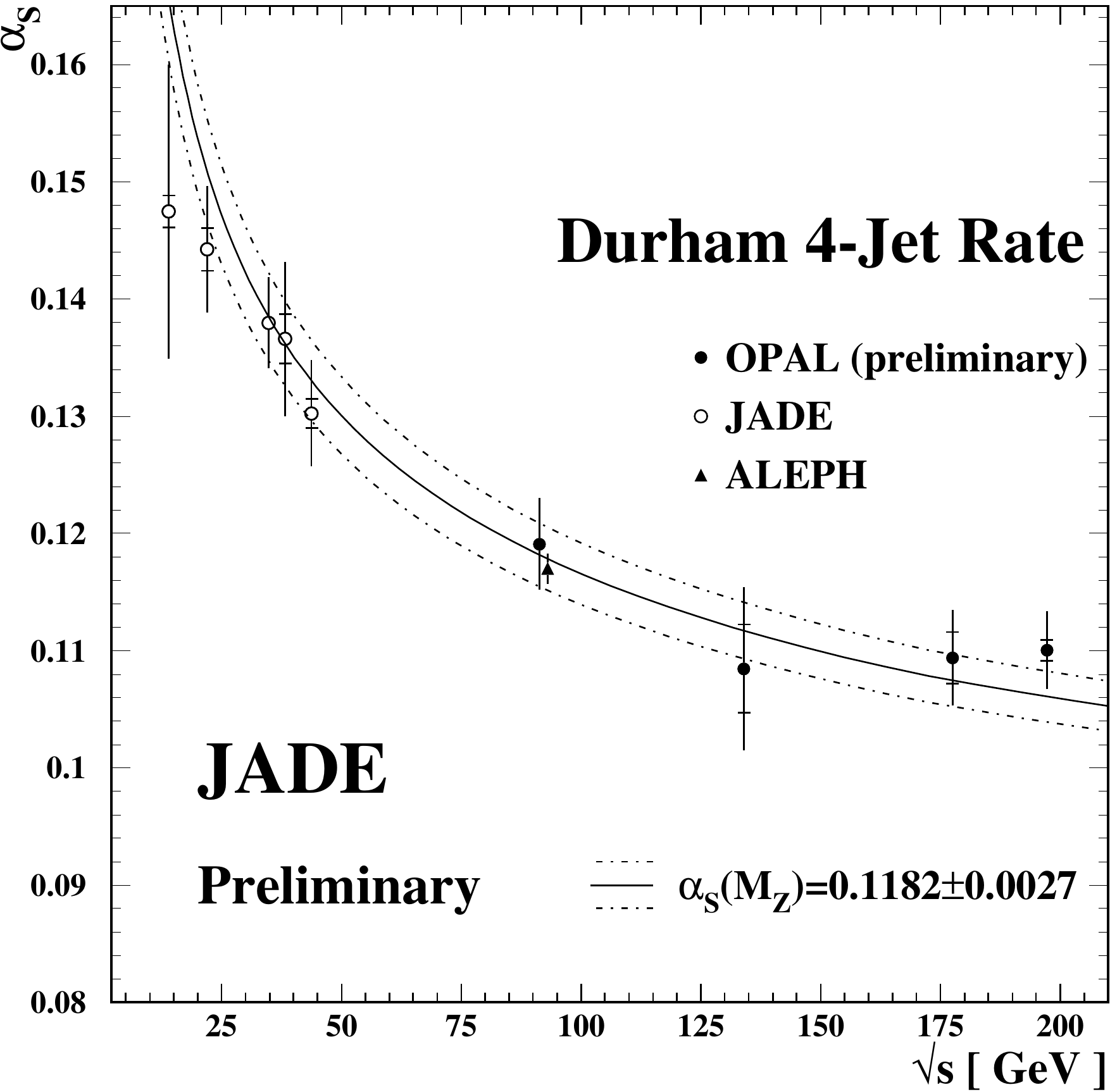}}
\end{center}
\caption{\label{fig:run}Running coupling in QCD.}
\end{figure}
The main difference compared to QED, due to the self-interactions of
the gluons, is the fact that the coupling becomes smaller at short
distances as shown in the figure \ref{fig:run}, a property known as
{\sl asymptotic freedom}.

\begin{figure}[htbp]
\begin{center}
\resizebox*{!}{6.5cm}{\includegraphics{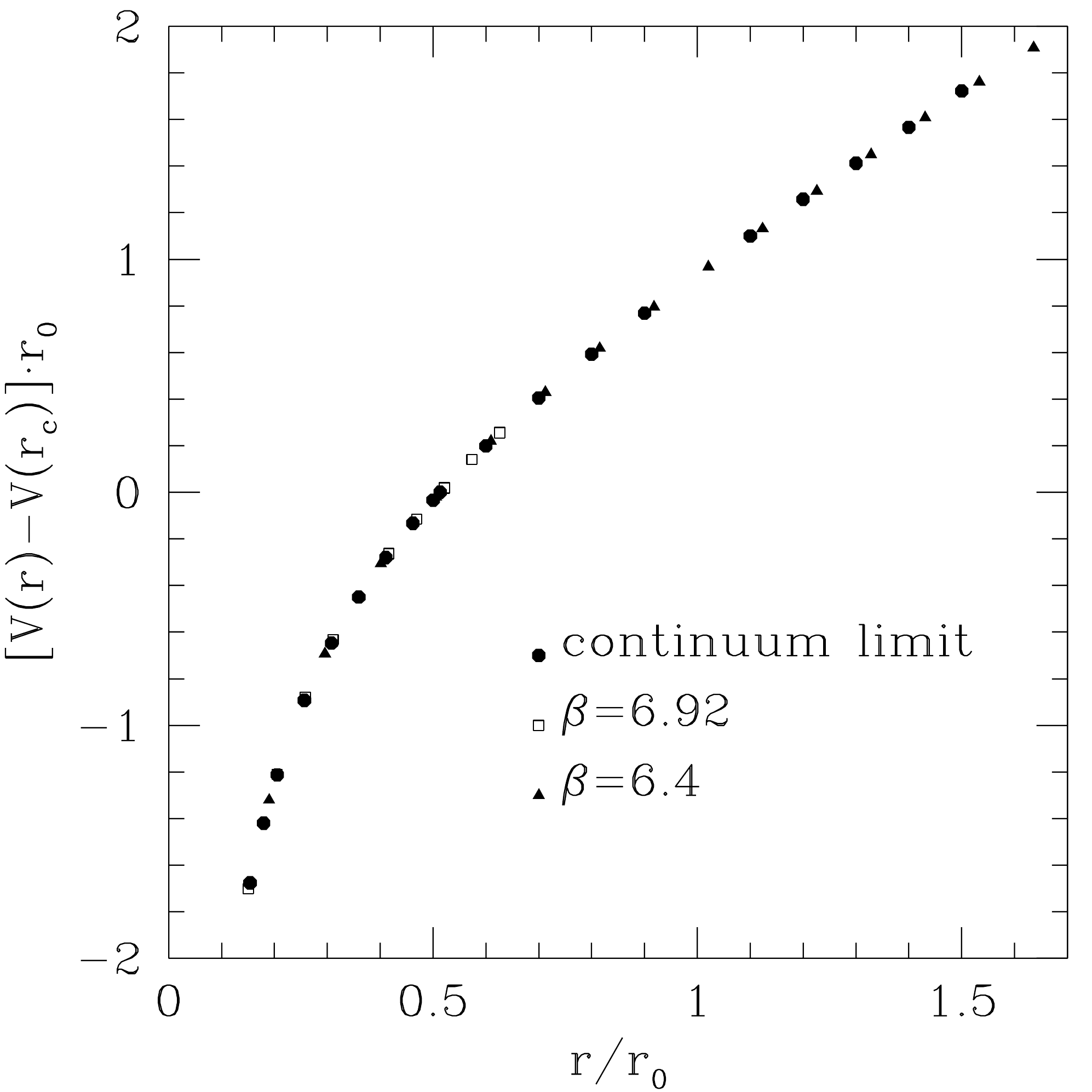}}
\end{center}
\caption{\label{fig:pot}Coulomb potential of a heavy quark and
  antiquark pair, from lattice QCD.}
\end{figure}

\begin{figure}[htbp]
\begin{center}
\resizebox*{!}{5cm}{\includegraphics{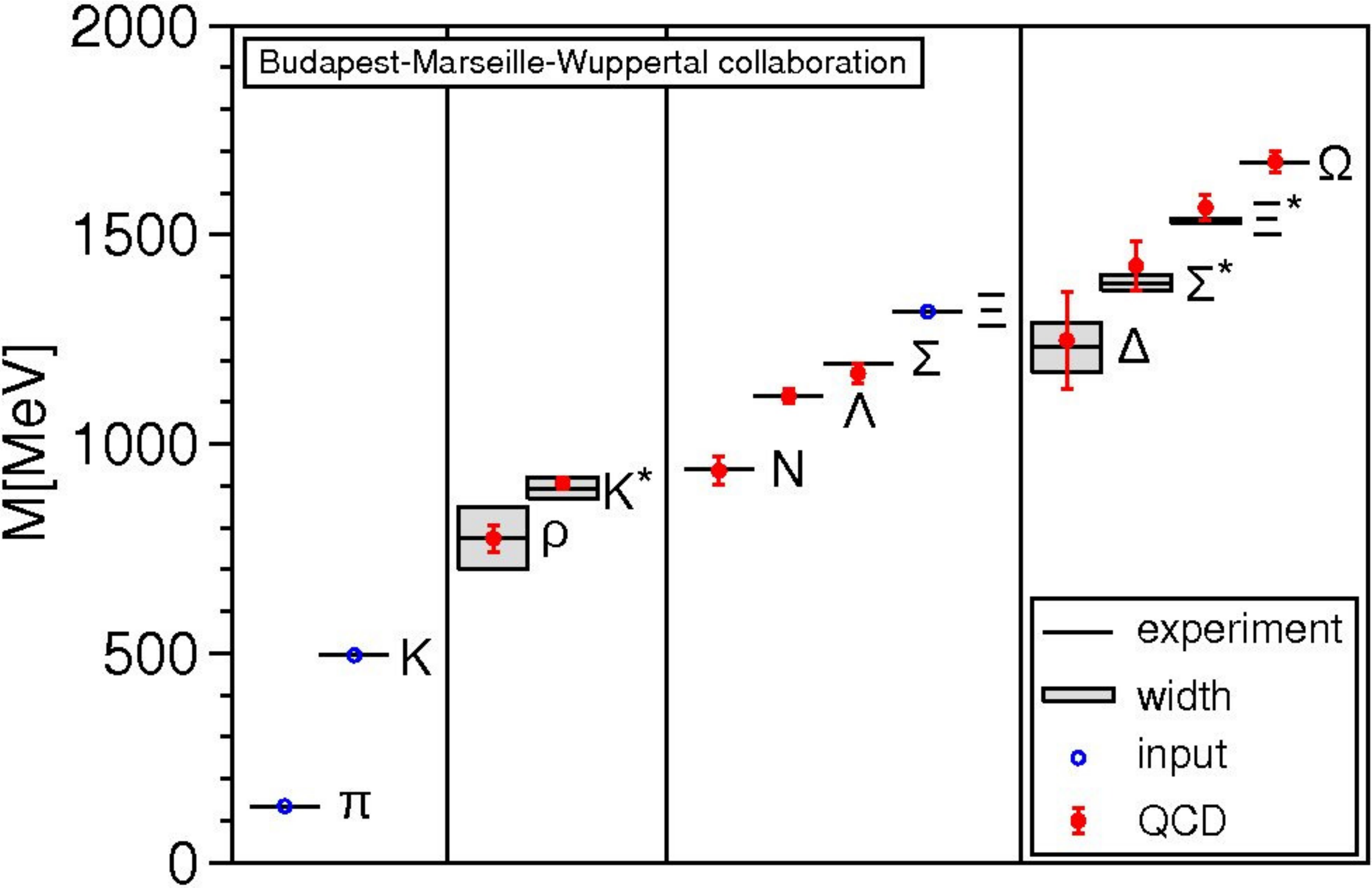}}
\end{center}
\caption{\label{fig:spectrum}Hadron spectrum from lattice QCD.}
\end{figure}

A related property of QCD is the long distance behavior of the
interaction potential between a quark and an antiquark. This can be
calculated numerically in lattice simulations for heavy quarks (that
are therefore static). This potential, shown in the figure
\ref{fig:pot}, behaves as a standard $1/r$ Coulomb potential at short
distance, but increases linearly at large distance, in sharp contrast
with electromagnetic interactions. 
This leads to {\sl color confinement}, that is the fact that free
color charges cannot exist in Nature. Quarks only appear in color
singlet bound states called hadrons, made of 3 quarks (baryons) or
quark-antiquark pairs (mesons). The spectrum of these bound states can
in principle be determined from first principles from the QCD
Lagrangian, and it depends only on the quark masses and on the QCD
scale $\Lambda_{_{\rm QCD}}$. However, this dependence is
non-perturbative and lattice simulations are the only way to perform
these calculations. Presently, lattice calculations can reproduce the
spectrum of light hadrons with an accuracy of the order of $5$\%, as
illustrated in the figure \ref{fig:spectrum}.

The QCD running coupling shown in the figure \ref{fig:run} can also be
viewed with a different perspective: it suggests that if one squeezes
many hadrons in a small volume, then the average inter-quark distance
will be small and their interactions will be weak. In such a
situation, the quarks would not be confined into individual hadrons,
and would instead form a plasma made of deconfined quarks and
gluons. This idea is substantiated by lattice calculations of the QCD
partition function as a function of temperature, that indicate a rapid
increase of the number of effective degrees of freedom at a
temperature around $160$~MeV\footnote{This is for QCD with 3 light
  quark flavors.  The transition temperature is higher for pure glue
  QCD.}. This suggests that the relevant degrees of freedom are no
longer the color singlet light hadrons (pions, kaons,...) and have
been replaced by quarks and gluons (that are more numerous because of
the uncovered color degree of freedom).

\subsection{Heavy ion collisions}
Experimentally, the conditions of such a transition can be realized by
colliding heavy nuclei at high energy. Such experiments are presently
being performed by the RHIC (gold nuclei collided at 200~GeV) and by
the LHC (lead nuclei collided at 5.5~TeV). Just after the impact of
the two nuclei, the energy density reaches values that are more than
ten times the normal nuclear matter density, well above the energy
density at the deconfinement transition inferred from lattice calculations.
\begin{figure}[htbp]
\begin{center}
\resizebox*{!}{5cm}{\includegraphics{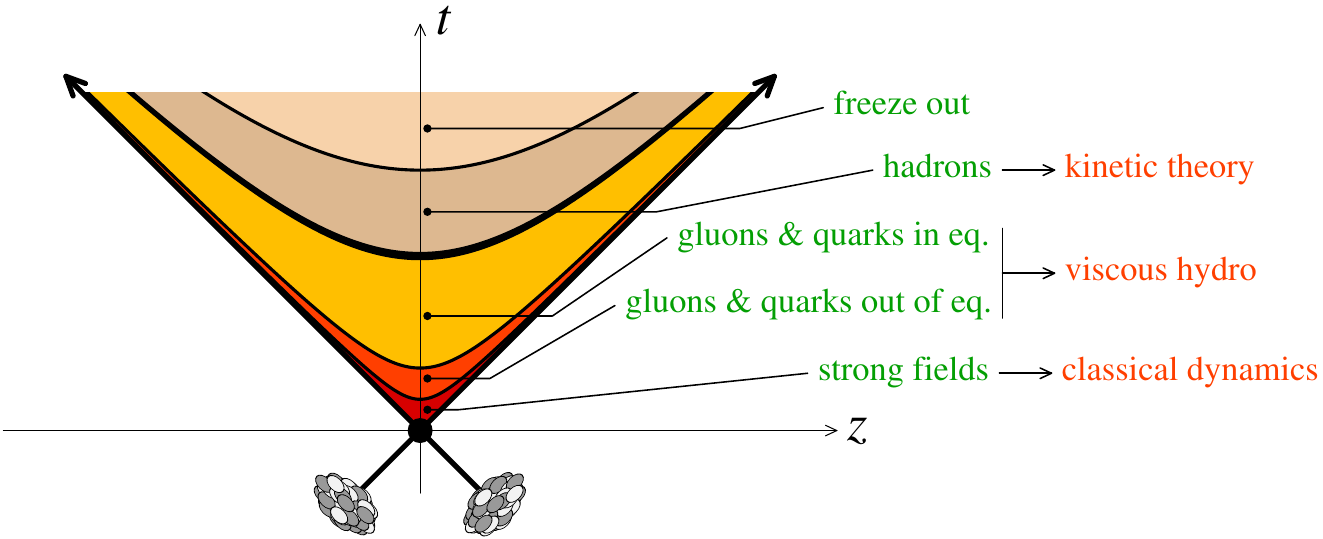}}
\end{center}
\caption{\label{fig:stages}Successive stages of a heavy ion collision.}
\end{figure}
Such a collision, whose total duration is of the order of $10$~fm/c,
can be divided into several stages, as shown in the figure
\ref{fig:stages}. In this figure, we have also indicated what kind of
tool one may employ for each of these stages. It turns out that
macroscopic descriptions such as relativistic hydrodynamics are quite
successful at describing the bulk evolution of the system. Somewhat
surprisingly, the matter produced in these collisions seems to behave
almost like a perfect fluid with close to no viscosity. A very small
viscosity suggests that this matter is not the siege of strong
dissipative processes that would rearrange its microscopic degrees of
freedom.  

In these lectures, we will be primarily interested in the beginning of
the collision, up to the point where a hydrodynamical description may
become plausible.  We will adopt a weak coupling
perspective\footnote{In the discussion on gluon saturation, we will
  see that the emergence of the {\sl saturation scale}, that increases
  with the collision energy, justifies this assumption.}, and we will
try to follow a heavy ion collision in a description which is as close
as possible to QCD. Indeed, in collisions at very high energy, the
initial energy density is so large that the early stages of such a
collision should be amenable to a weak coupling description, thanks to
the asymptotic freedom of QCD.  Note that a small viscosity, such that
could explain the success of hydrodynamics, is more naturally obtained
in the strong coupling limit because the viscosity is inversely
proportional to the scattering cross-section of the quarks and
gluons. However, it is also be possible to get strong interactions at
weak coupling, provided that the occupation number is inversely
proportional to the coupling $g^2$. In this case, the coupling
disappears from the scattering rate, and the system has many of the
features of a strongly coupled system.

\section{Parton model}
\label{sec:parton}
\subsection{Kinematics}
As discussed earlier, free quarks and gluons do not exist in normal
nuclear matter. Instead they are confined into color singlet bound
states, whose spectrum depends non-perturbatively on the parameters of
the QCD Lagrangian.  The same is true of the energy
levels of a nucleus: they could in principle be derived from the
underlying QCD dynamics, but this is even more complicated than in the
case of light hadrons and at the moment far out of reach of lattice
computations.  

\begin{figure}[htbp]
\begin{center}
\resizebox*{!}{2.5cm}{\includegraphics{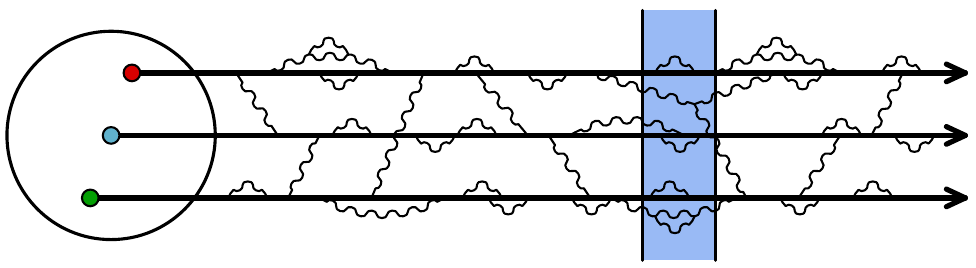}}
\end{center}
\caption{\label{fig:slow}Dynamics of the constituents inside a slow nucleon.}
\end{figure}

Does this mean that we should give up any hope of using QCD to
describe collisions between such objects?  Fortunately, the answer is
no, {\sl for collisions at sufficiently high energy}.  The kinematics
of these collisions is the key to overcome this difficulty. Let us
consider first a nucleon at low energy (i.e. when the nucleon is
almost at rest in the observer's frame), shown in the figure
\ref{fig:slow}.
In this cartoon, the thick lines represent the three valence quarks,
and the horizontal axis represents time. Only gluon constituents are
shown, not the sea quarks. In such a frame, the valence quarks orbit
with a period comparable to the proton size (they are
ultrarelativistic). These quarks exchange gluons that provide the
binding force, which also happens on scales of the order of the proton
size.  Moreover, the quarks and gluons can briefly fluctuate: for
instance, a quark can temporarily become a quark+gluon state. The
lifetime of these virtual states can be anything smaller than the
proton size\footnote{But since QCD is a renormalizable theory, the
  physics of the strong interactions at hadronic energy scales does
  not depend on what happens on much higher energy scales. Therefore,
  these short lived fluctuations have essentially no relevance in
  hadronic physics.}. When studying reactions involving hadrons, one
should compare the typical timescale of the collision (shown as a blue
strip in the figure) with the timescales of the internal dynamics of
the nucleon. In collisions involving low energy hadrons, the hadron
has a complicated internal dynamics on timescales comparable to the
duration of the collision, which makes these collisions untractable in
perturbative QCD.

\begin{figure}[htbp]
\begin{center}
\resizebox*{!}{2.5cm}{\includegraphics{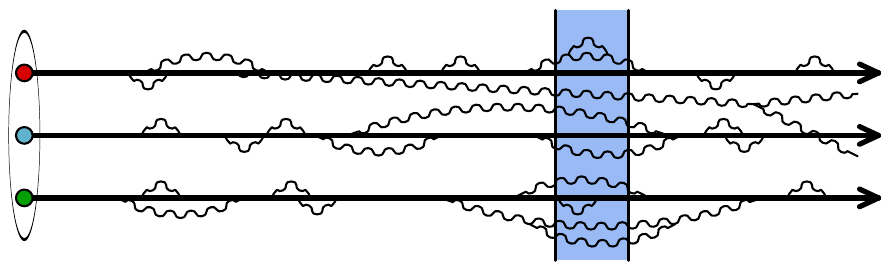}}
\end{center}
\caption{\label{fig:fast}Dynamics of the constituents inside a highly boosted nucleon.}
\end{figure}

Contrast this with what happens in a collision at very high
energy. Although scattering amplitudes are boost invariant and may be
discussed in any frame, it is convenient to imagine that we do not
change the momentum of one of the hadrons, and that all the energy
increase is achieved by boosting the second hadron.  This is
illustrated in the figure \ref{fig:fast}. The blue strip, unchanged
compared to the low energy case, may be viewed as the size of the
first hadron, that we did not boost. All the changes are in the
internal dynamics of the second hadron, whose timescales are now
stretched by Lorentz time dilation.
The gluon exchanges between the valence quarks are now happening over
timescales that are much longer than the duration of the collision,
which means that the constituents of the nucleon can be viewed as free
during the collision. The same happens to the fluctuations of the
constituents. The lifetime of these virtual states is increased much
beyond the collision timescales, making these off-shell constituents
undistinguishable from on-shell particles\footnote{The concept of
  on-shell or off-shell particles depends on the duration of the
  measurement. The only way to know that a particle is exactly
  on-shell is to perform an infinitely long measurement. Indeed, an
  off-shell particle may be viewed as a particle of momentum $\p$
  whose energy differs from the on-shell energy $E_\p$ (given by the
  dispersion relation). A measurement that lasts $\Delta t$ can only
  resolve energy differences of order $1/\Delta t$ or larger.}. Since
there are fluctuations at arbitrary small timescales in a nucleon at
rest, increasing the energy will uncover more and more of these
fluctuations. These simple kinematical considerations are the essence
of the {\sl parton model}~\cite{Feynm4}, that approximates a high
energy nucleon or nucleus as a collection of quasi-free constituents
(called partons), whose density grows with energy.

\subsection{Factorization}
From this discussion, it seems that a QCD description of high energy
collisions between hadrons may be feasible, provided we can provide
``snapshots'' of their partonic content at the time of the
collision. What information is necessary in this snapshot is not
completely obvious at this point, and may vary depending on the
observable one intends to calculate, but for instance one may think of
the following:
\begin{itemize}
\item flavor and color of each parton,
\item transverse position and longitudinal momentum.
\end{itemize}
Of course, these properties of a hadron cannot be known event by
event, which means that at best a probabilistic description may be
achieved, that would allow to compute expectation values for event
averaged observables. However, the possibility of describing hadronic
collisions with only a probabilistic partonic description of the
incoming hadrons is highly non-trivial, because it is an approximation
that amounts to discarding certain quantum interferences. Without
doing any approximation, the transition probability from a pair of
hadrons $h_1 h_2$ to some final state $X$ is obtained by summing all
the relevant reaction channels before squaring the amplitude,
\begin{equation*}
    {{\mbox{transition probability}}\atop{\mbox{from hadrons to X}}}
    \quad\equiv\quad
    \Big|\sum {{\mbox{Amplitudes}}\atop{h_1h_2\to X}} \Big|^2\; .
\end{equation*}
In contrast, the parton model as described above approximates the
transition probability as follows,
\begin{equation*}
        {{\mbox{transition probability}}\atop{\mbox{from hadrons to X}}}
    \approx
    \sum_{{\scriptsize\mbox{partons}}\atop{\{q,g\}}}
    {{\mbox{probability to find}}\atop{\{q,g\}\mbox{ in }\{h_1,h_2\}}}
    \;\otimes\;\Big|\sum {{\mbox{Amplitudes}}\atop{\{q,g\}\to X}}\Big|^2
\end{equation*}
which is clearly not equivalent to the previous formula. This
approximation is called {\sl initial state factorization}. Roughly
speaking, the physical motivation for such a factorization is that the
neglected terms are interferences between a hard process that occurs
on the timescale of the collision and a process internal to one of the
projectiles, happening on much longer timescales. The vast separation
in their timescales is what makes the corresponding interference
small.  At a more formal level, this factorization can be established
in QCD, with various degrees of sophistication\footnote{The weakest of
  these {\sl factorization theorems} are based on {\sl leading log
    factorization}, where the two formulas are shown to be equivalent
  for an infinite series of terms of the form $(\alpha_s\log(Q))^n$
  (where $Q$ is some hard scale), but not for terms of the form
  $\alpha_s(\alpha_s\log(Q))^n$. {\sl Next-to-leading log
    factorization} extends the proof of this equivalence to include
  terms in $\alpha_s(\alpha_s\log(Q))^n$, and so on. {\sl All-orders
    factorization theorems}~\cite{ColliSS1,ColliSS2,ColliSS3} prove
  that the two formulas are equivalent up to terms that decrease as
  inverse powers of the hard scale.}  depending on the observable.

\subsection{Single parton distributions}
The most developed framework for this type of factorization is the
DGLAP formalism, in which one describes the incoming hadrons by {\sl
  single parton distributions}. These distributions depend on the
hadron and on the parton under consideration, on the fraction $x$ of
longitudinal momentum carried by the parton, and on a momentum scale
$Q$ that can be viewed as the inverse of the transverse spatial resolution
with which the hadron is probed.
\begin{figure}[htbp]
\begin{center}
\resizebox*{!}{7cm}{\includegraphics{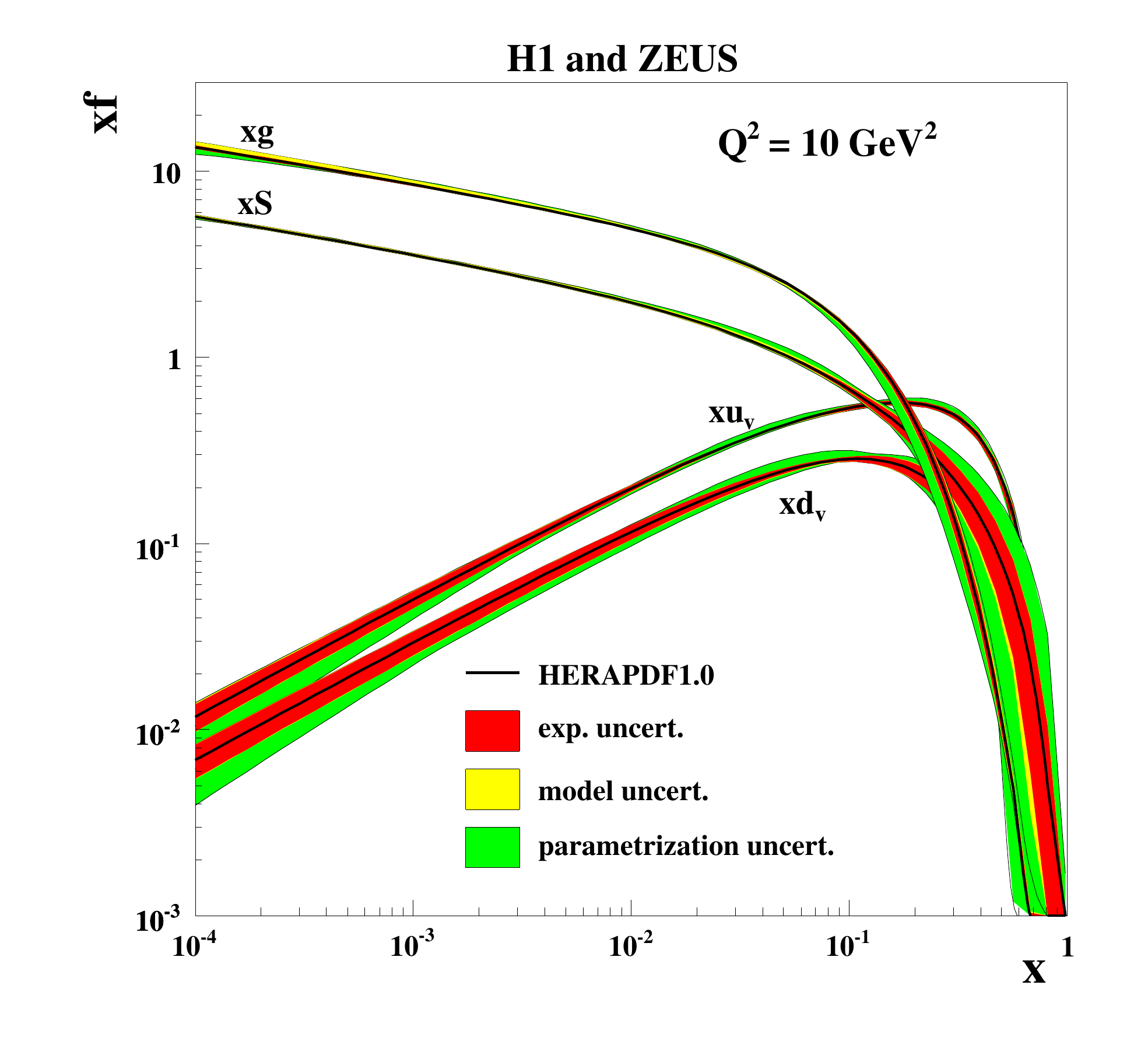}}
\end{center}
\caption{\label{fig:pdf}Parton distributions of a proton, at the resolution scale $Q^2=10~\mbox{GeV}{}^2$. From \cite{Aarona2}.}
\end{figure}
In the figure \ref{fig:pdf}, the single parton distributions of a
proton, extracted from deep inelastic scattering data, are shown at a
fixed resolution scale $Q$. Although these distributions are non
perturbative and cannot be computed easily from the QCD
Lagrangian\footnote{Lattice QCD can be used to evaluate the first few
  Mellin moments of parton distributions, since they are given by
  expectation values of local operators that can be evaluated in the
  Euclidean theory.}, QCD predicts how they change if one increases
the resolution scale $Q$, via the {\sl DGLAP
  equation}~\cite{AltarP1,GriboL1,GriboL2,Doksh1}. From this figure,
one sees that the valence quark distribution is predominant at large
values of the momentum fraction $x\gtrsim 0.1$, and is totally
negligible at small $x$. At any value $x\lesssim 0.1$, the gluons are
the dominant species of partons, and their density increases like a
power of $1/x$ when $x\to 0$. The sea quarks follow the trend set by
the gluons, but with a suppression factor of order $\alpha_s$ since
they are produced by the process $g\to q\overline{q}$.

\section{Gluon saturation}
\label{sec:sat}
\subsection{Dense regime of QCD}
\begin{figure}[htbp]
\begin{center}
\resizebox*{!}{3.2cm}{\includegraphics{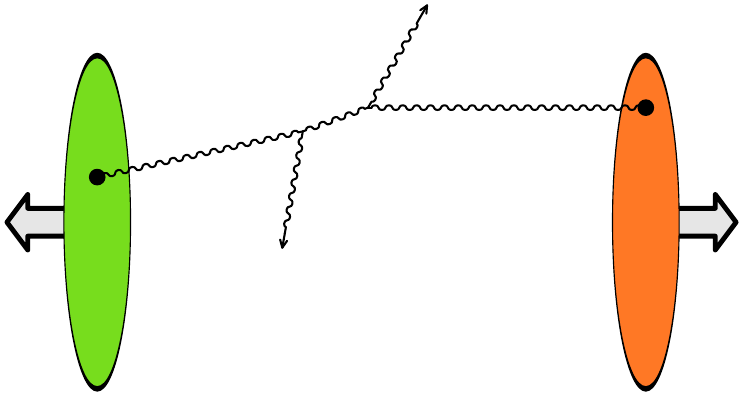}}
\hfill
\resizebox*{!}{3.2cm}{\includegraphics{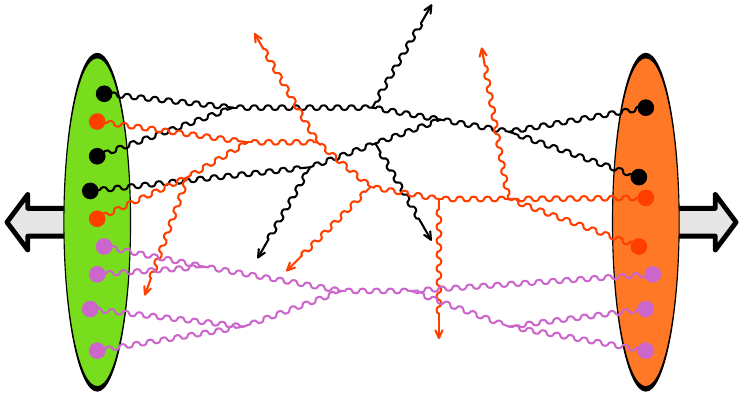}}
\end{center}
\caption{\label{fig:dense-dilute}Typical partonic processes in a
  collision between dilute (left) and dense (right) projectiles.}
\end{figure}

Since the DGLAP factorization framework is based solely on the single
parton distributions, it is expected to become insufficient at large
parton densities. The problem that will arise in this regime is
illustrated in the figure \ref{fig:dense-dilute}, that shows side to
side a typical scattering process in the dilute (left) and dense
(right) regimes.
 In the dilute situation, the incoming hadrons are ``mostly empty'',
and hard scatterings are rare processes. Moreover, reactions involving
more than one parton in each projectile are extremely rare (their rate
scales as the square of the probability to find a parton).  But when
the parton density is large, processes initiated by multiple partons
become more likely to happen. Moreover, the partons that lie nearby in
phase-space may be correlated, and therefore multiparton distributions
cannot be inferred simply from single parton distributions. A
framework that would enable one to calculate these processes should
provide information about multiparton distributions in hadrons and
nuclei, and thus should go beyond the DGLAP framework. Moreover, when
the parton density becomes of the order of the inverse coupling $1/g^2$, a
strongly interacting regime --called {\sl gluon
  saturation}~\cite{GriboLR1,MuellQ1,BlaizM2}-- is reached, where an
infinite series of Feynman graphs contribute at each order in $g^2$.

\begin{figure}[htbp]
\begin{center}
\resizebox*{!}{6.5cm}{\includegraphics{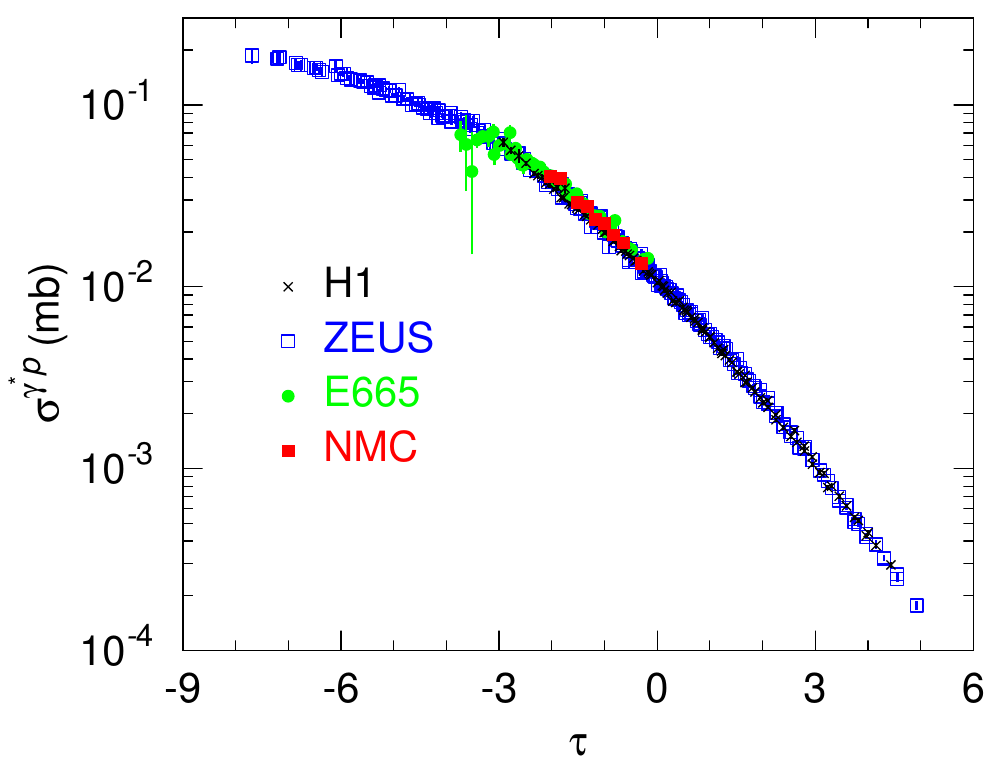}}
\end{center}
\caption{\label{fig:scaling}Geometrical scaling in the DIS cross-section at small $x$. The horizontal axis represents the variable $\tau\equiv x^{0.32}Q^2$.}
\end{figure}

A hint of the fact that the small $x$ saturation regime is
qualitatively different from the dilute regime appears when plotting
the deep inelastic scattering cross section slightly differently. This
cross-section depends on two independent Lorentz invariant quantities,
$x$ and the 4-momentum squared $Q^2$ of the photon exchanged in the
scattering. However, when plotted against the combination
$x^{0.32}Q^2$, this data appears to line up on a unique curve (see the
figure \ref{fig:scaling}).
This scaling signals the emergence of an $x$ dependent momentum scale,
that behaves roughly as $Q_s^2(x)\sim x^{-0.32}$. This scale, known as
the {\sl saturation momentum}, appears as a consequence of the
nonlinear interactions among the gluons, that become important at high
density.

\subsection{Saturation condition}
\begin{figure}[htbp]
\begin{center}
\resizebox*{!}{2cm}{\includegraphics{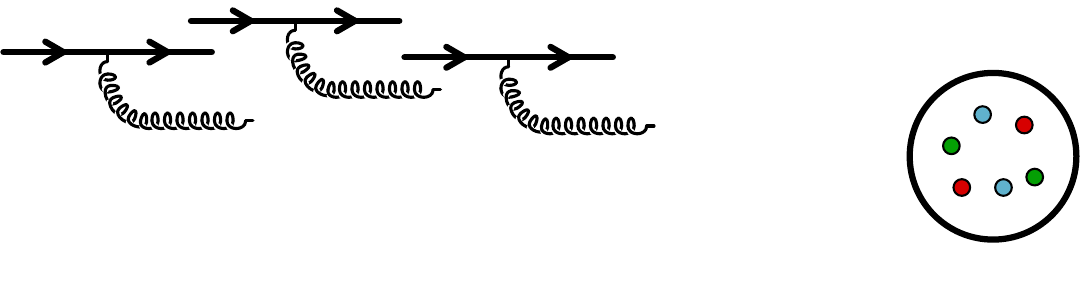}}
\hfill
\resizebox*{!}{2cm}{\includegraphics{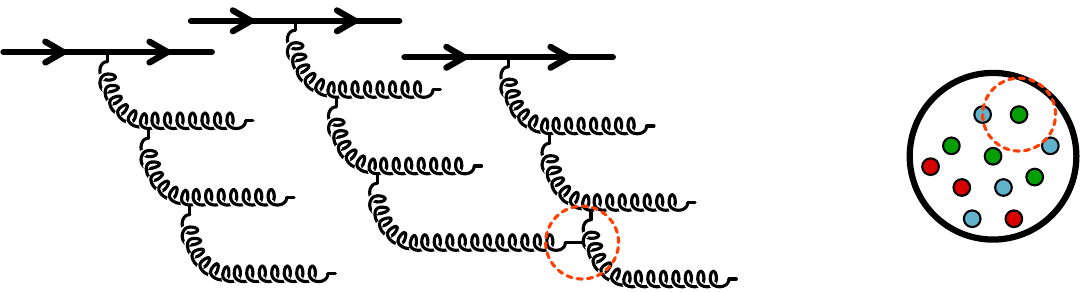}}
\end{center}
\caption{\label{fig:brem}Gluon cascades in the small $x$ evolution of
  a hadron. Top: dilute regime. Bottom: onset of the recombination corrections.}
\end{figure}

To understand the onset of gluon saturation, it is instructive to go
back to the dilute regime at large $x$. In this situation, a hadron
appears as a loose collection of a few partons. When the hadron is
progressively boosted, these partons radiate more gluons by
bremsstrahlung\footnote{As discussed before, these gluons are not
  truly on-shell, but can be viewed as real gluons if the lifetime of
  the quantum fluctuation that gave them birth is longer than the
  observation time.}, as illustrated in the top panel of the figure
\ref{fig:brem}. As long as the density remains low enough, these
cascades of gluons develop independently and the
evolution\footnote{Although this terminology is commonly used, it is
  somewhat of a misnomer, since the hadron does not truly
  ``evolve''. It is the observer's view of the hadron content that
  changes as the observer's frame is increasingly boosted with respect
  to the hadron.} of the hadron structure is governed by the linear
{\sl BFKL equation}~\cite{BalitL1,KuraeLF1}.
Since these additional gluons are contained within the geometrical
volume of the hadron, their density increases rapidly. At some point,
their wavefunctions start to overlap and their interactions are no
longer negligible.  Gluons from two different cascades can recombine,
which tames the growth of the gluon density. Moreover, this
recombination process makes the $x$ evolution of the gluon
distribution non-linear. 

\begin{figure}[htbp]
\begin{center}
\resizebox*{!}{5.1cm}{\includegraphics{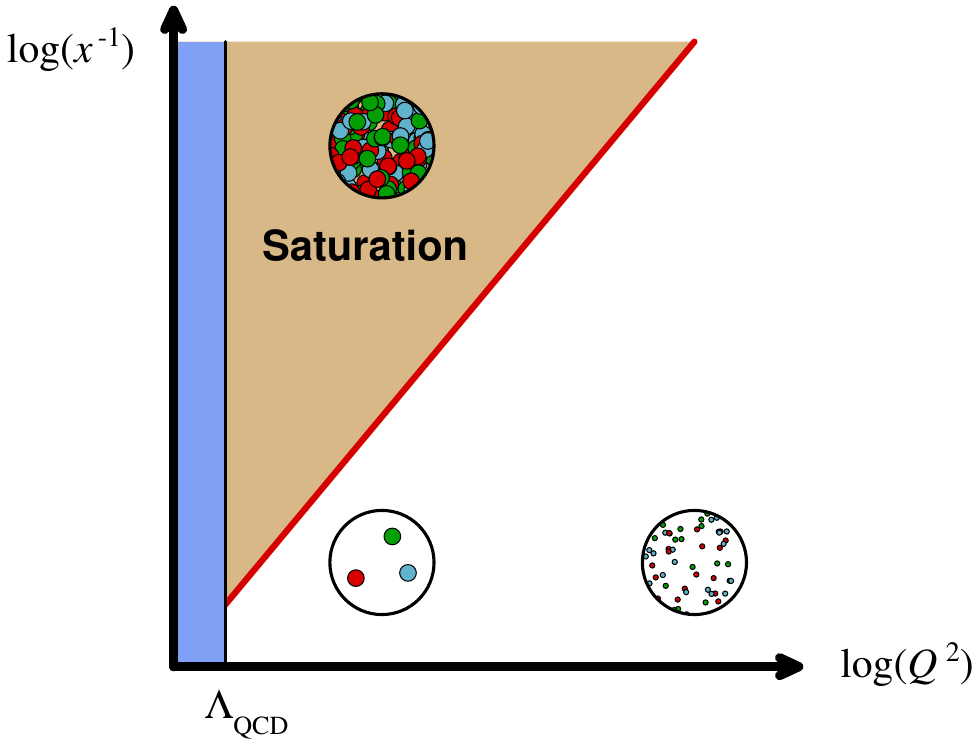}}\hfill
\resizebox*{!}{5.1cm}{\includegraphics{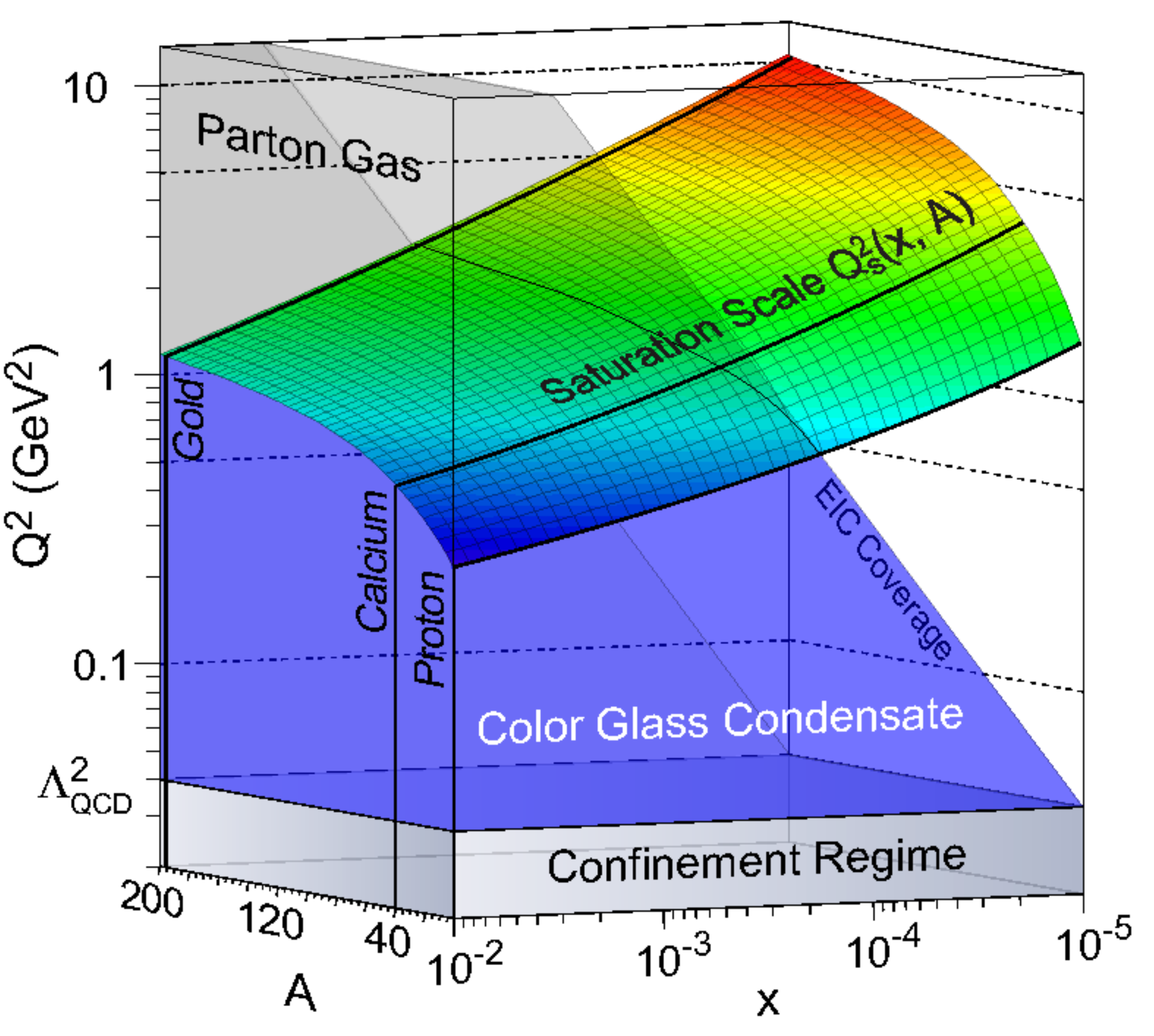}}
\end{center}
\caption{\label{fig:satdom}Saturation domain in the $x$ and $Q$
  plane. The 3-dimensional plot on the right adds information about
  the $A$ dependence. From \cite{DeshpEM1}.}
\end{figure}

Before going into a more quantitative description of gluon saturation,
it is easy to derive a simple criterion for the onset of
saturation. Gluon recombination becomes likely when the product of the
number of gluons per unit area with the cross-section for recombining
two gluons into one becomes larger than one,
\begin{equation}
        \underbrace{\alpha_s Q^{-2}}_{\sigma_{gg\to g}} 
        \;\times\; 
        \underbrace{A^{-2/3} xG(x,Q^2)}_{\mbox{\scriptsize surface density}} 
        \;\ge\; 1\; .
      \end{equation}
This condition can be rearranged in order to obtain an inequality on $Q$~:
 \begin{equation}
        Q^2 \le 
        \underbrace{{\colorb Q_s^2}\equiv 
        \frac{{\colord\alpha_s}{\colorb xG(x,Q_s^2)}}{{\colora A^{2/3}}}}_{\mbox{\scriptsize saturation momentum}}
        \;\sim\; {\colora A^{1/3}}{\colorb x^{-0.3}}\; .
\label{eq:Qs}
\end{equation}
This argument justifies the emergence of the saturation momentum,
which characterizes the physics of gluon saturation. Saturation is
important when the typical momentum scales in a process are smaller
than $Q_s$. The region where this condition is satisfied is shown in
the figure \ref{fig:satdom}. From the more quantitative plot on the
right panel of this figure, a typical value to keep in mind is that
$Q_s^2$ is in the range $2$--$4~\mbox{GeV}{}^2$ for nuclei at the
energy of the LHC.
In the saturation domain, non-linear gluon interactions are important,
which arguably makes the calculation of such processes more
complicated. However, this also has an unexpected positive side:
$Q_s$ now supersedes all the softer momentum scales in determining the
typical momentum of the relevant partons, and thus also controls the
running of the coupling. Since $Q_s$ increases when $x$ decreases
(i.e. when going at higher energy), this opens an avenue for an ab
initio weak coupling treatment of multiparton interactions (sometimes
called the ``underlying event'' in other contexts) in high energy
hadronic scatterings. From eq.~(\ref{eq:Qs}), one sees that the
saturation momentum also increases with the mass number of
nuclei. This implies that, at a given energy, saturation effects are
stronger in nucleus-nucleus collisions, for large nuclei\footnote{For
  gold or lead nuclei, the factor $A^{1/3}$ that appears in $Q_s^2$ is
  approximately 6.}.  This is important for heavy ion collisions at
the RHIC and the LHC, because in these collisions the bulk of particle
production is controlled by saturation physics.

\section{Color Glass Condensate}
\label{sec:cgc}
\subsection{Degrees of freedom}
The Color Glass Condensate\footnote{For more detailed reviews of the
  Color Glass Condensate, one can consult
  Refs.~\cite{IancuV1,Lappi6,Weige2,GelisIJV1,Gelis15}.} (CGC) is a QCD based
effective theory whose aim is to describe quantitatively the gluon
saturation regime.
\begin{figure}[htbp]
\begin{center}
\resizebox*{12cm}{!}{\includegraphics{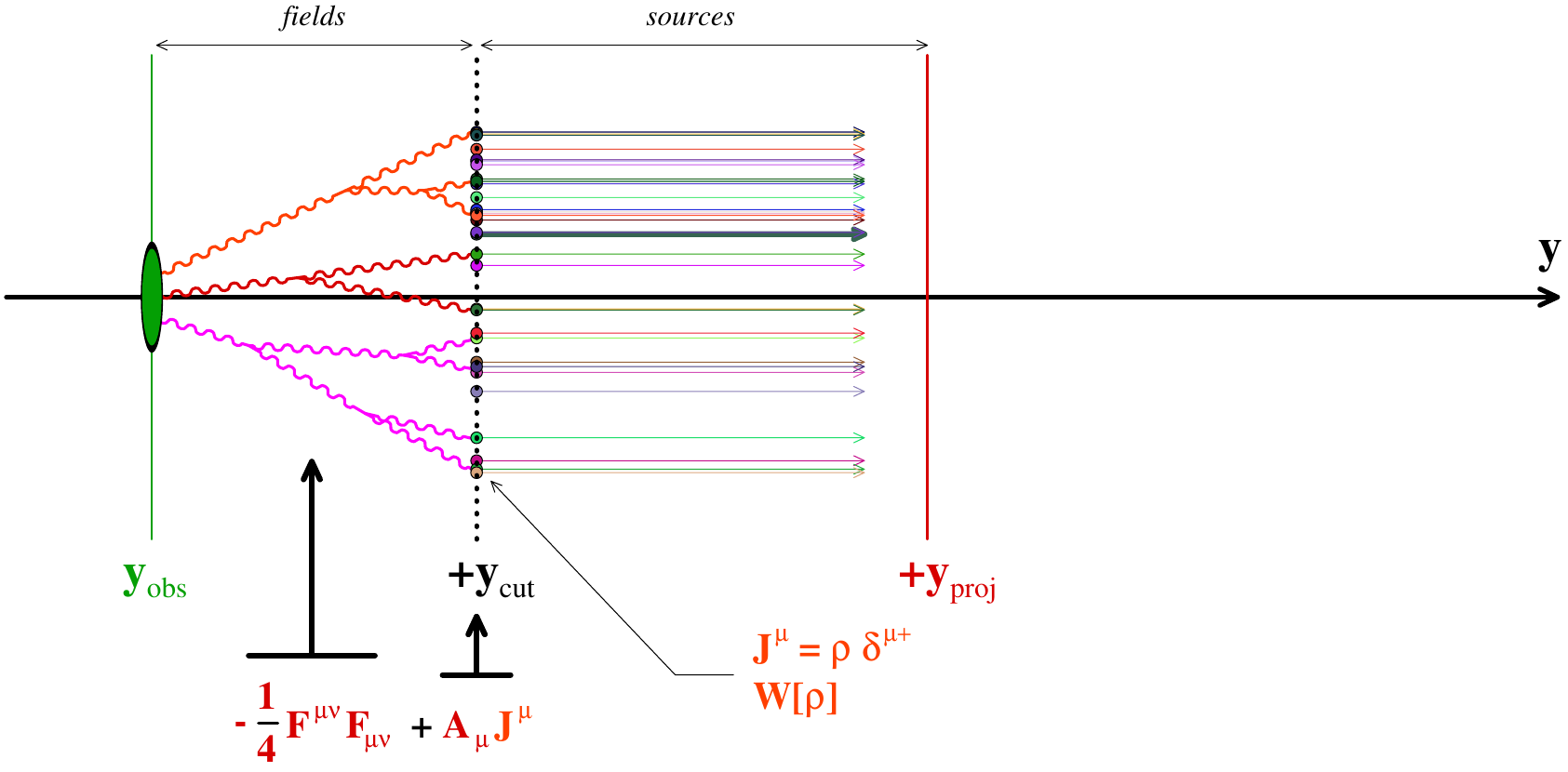}}
\end{center}
\caption{\label{fig:DoF}Degrees of freedom in the CGC effective theory.}
\end{figure}
The CGC exploits the high energy kinematics in order to simplify the
description of the non-perturbative valence partons. The main idea was
already encountered in the qualitative discussion of the parton model:
a high energy hadronic collision is so brief that the fact that the
partons are strongly bound by confinement is totally irrelevant. In
fact, over such short timescales, the internal motion of the partons
inside the hadron appears completely frozen. Thus, one may view the
partons as static in the transverse plane, with a large longitudinal
momentum~\cite{McLerV1,McLerV2,McLerV3}. For an observer sitting in
the center of mass frame of the collision, the only information that
matters about these partons is the color current $J^\mu_a$ that they
carry along the beam direction.  The dominant component of this
4-vector is the longitudinal one. In light-cone coordinates, for a
hadron moving in the $+z$ direction, it reads
\begin{equation}
J^\mu_a(x)=\rho_a(x^-,\x_\perp)\,\delta^{\mu+}\; ,
\end{equation}
where the function $\rho_a(x)$ is the density of color charges of the
partons. It does not depend on the light-cone ``time'' $x^+$ because
of time dilation.  Moreover, due to the concomitant Lorentz
contraction, its $x^-$ dependence is very peaked around $x^-=0$.

It is important to realize that this drastic simplification cannot be
used for all partons: it is applicable only to those partons whose
longitudinal momentum (in the observer's frame) is large enough.  The
partons that have a rapidity close to the observer's rapidity have
comparable transverse and longitudinal momenta, and thus cannot be
approximated by a longitudinal current. Moreover, for these partons,
the Lorentz boost factor that slows down their time evolution is not
large and one cannot neglect their dynamics. Therefore, these slower
partons must be treated as full fledged quantum fields.  

The situation is summarized in the figure \ref{fig:DoF}: a cutoff
$y_{\rm cut}$ must be introduced somewhere between the rapidity of the
observer and the rapidity of the hadron under consideration. The
partons close to the observer (mostly gluons, at least at leading
order in $\alpha_s$) are described as gauge fields, while those that
are close to the projectile are approximated as a static longitudinal
color current.

\subsection{CGC effective theory}
Thanks to the rapidity separation between the slow and the fast
degrees of freedom in the CGC, their coupling can be treated as
eikonal, i.e. via a term of the form $J^\mu A_\mu$. Therefore, the CGC
can be summarized by the following effective action
\begin{equation}
{\cal S}_{_{\rm CGC}}=\int d^4x\; \Big(-\frac{1}{4}F_{\mu\nu}F^{\mu\nu}+J^\mu A_\mu\Big)\; .
\label{eq:S-CGC}
\end{equation}
(For a collision of two hadrons, the current $J^\mu$ is the sum of two
terms, one for each hadron.) The function $\rho_a(x^-,\x_\perp)$ that
appears in the current $J^\mu$ reflects the particular arrangement of
the fast partons at the time of the collision. It is not a quantity
that can be predicted event by event, and one can only have a
statistical knowledge of this object. Therefore, the CGC also
introduces a probability distribution $W[\rho]$. As we shall see
later, all observable quantities must be averaged over all the
possible configurations of $\rho$, according to the distribution
$W[\rho]$,
\begin{equation}
\left<{\colorb {\cal O}}\right>
=
\int\left[D\rho\right]\;
{\colorb W[\rho]}\;
{\cal O}[\rho]\; .
\end{equation}
In words, one should first calculate the observable for an arbitrary
configuration of the color charge density $\rho$ (in a collision of
two hadrons, there is a $\rho_1$ and a $\rho_2$), and then perform a
weighted average over all the possible $\rho$'s. The justification of
this procedure will be given in the following two sections.

\section{CGC at Leading Order}
\label{sec:LO}
\subsection{Power counting}
\begin{figure}[htbp]
\begin{center}
\resizebox*{7cm}{!}{\includegraphics{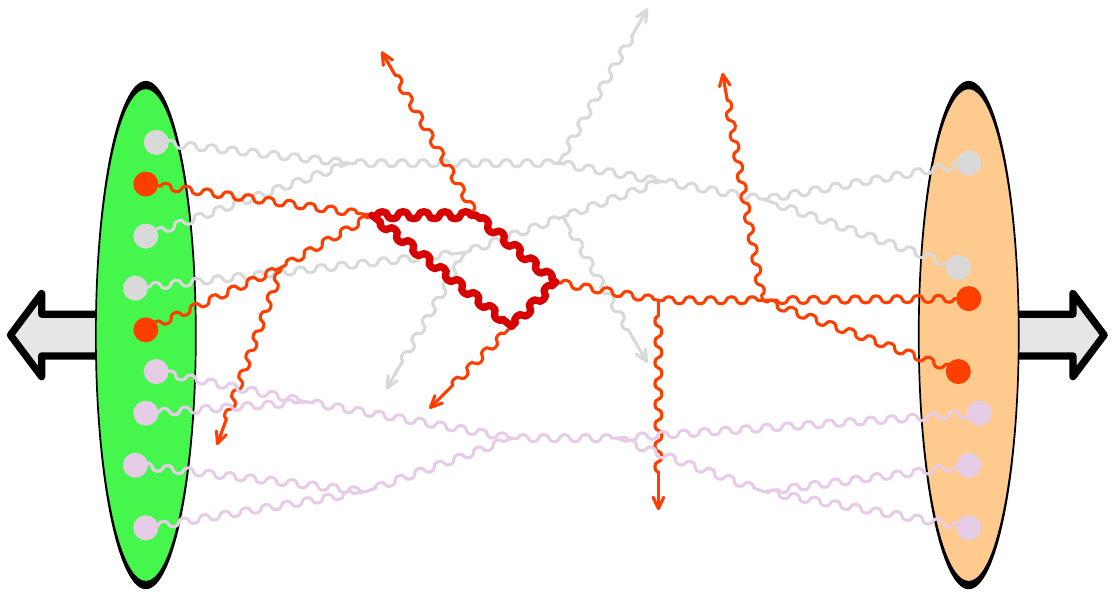}}
\end{center}
\caption{\label{fig:PC}Generic connected graph in the CGC effective
  theory. The dots represent the sources $\rho$. The lines terminated
  by arrows are the gluons produced in the final state.}
\end{figure}

So far, we have not assumed anything about the magnitude of the color
charge density $\rho_a$ that describes the fast partons in the CGC
effective theory. For the CGC to be applicable to the saturated
regime, we must allow $\rho_a$ to be as large as the inverse coupling
$1/g$. Indeed, the recombinations due to non-linear gluon interactions
can stabilize the gluon occupation number at a value of order $1/g^2$,
where the gluon splittings and the recombinations balance each
other. Since the occupation number is quadratic in the gauge field,
such a value corresponds to $\rho\sim g^{-1}$. As we shall see, such a
large value of the source $\rho$ simplifies the dynamics by making it
classical at leading order, but complicates things by making an
infinite set of graphs contribute at each order in $g^2$. In order to
see this, let us first examine the power counting in the CGC effective
theory. Consider a generic connected graph\footnote{Typical graphs are
  not connected: they are made of many disconnected subgraphs. But
  it is sufficient to discuss the properties of one of these
  subgraphs.}, as shown in the figure \ref{fig:PC}.
For such a graph, one finds that the order of magnitude depends only
on the number of produced gluons and the number of loops, via the
following formula~\cite{GelisV2,GelisV3}
\begin{equation}
{\colord\frac{1}{g^2}}\;\; 
{\colord g}^{\colorb\rm \#\ produced\ gluons}\;\; 
{\colord g}^{{\colord 2}({\colorc\rm \#\ loops})}\; .
\end{equation}
The main consequence of assuming that $\rho\sim g^{-1}$ is that this
power counting does not depend on the number of sources that are
included into the graph. The reason for this is that each additional
source (of order $g^{-1}$) is attached to the rest of the graph by a vertex
(of order $g$), and therefore does not contribute to the overall
magnitude of the graph. This also means that, at each order in $g^2$, one
must sum an infinite set of graphs.

For instance, the inclusive gluon spectrum has the following expansion
in powers of $g^2$
\begin{equation}
\frac{d{\colora N_1}}{d^3\vec\p}
=
\frac{1}{{\colorb g^2}}\;\Big[
c_0+c_1\,{\colorb g^2}+c_2\,{\colorb g^4}+\cdots
\Big]\; ,
\end{equation}
where each of the coefficients $c_0,c_1,\cdots$ is itself an infinite
series of terms of the form $(g\rho)^n$,
\begin{equation}
c_i\equiv\sum_{n=0}^\infty c_{i,n}\,({\colorb g}{\colorc\rho_{_{1,2}}})^n\; .
\end{equation}

At this point, we should make an important remark regarding exclusive
versus inclusive observables. From the above power counting, we see
that the average number of produced gluons in a high energy
nucleus-nucleus collision is of order $1/g^2$.  If we assume for
simplicity that the multiplicity distribution is
Poissonian\footnote{This is not exactly true in the CGC, but this fact
  does not change the essence of this argument.}, the probability to
have a given final state (e.g. a final state with a prescribed number
of gluons) is exponentially suppressed by a factor $\exp(-\#/g^2)$.
This factor may be viewed as a Sudakov factor that arises from
excluding all the other final states. It turns out that these
exclusive quantities are very difficult to calculate. In particular,
the (disconnected) graphs that involve spectator partons contribute to
exclusive observables, because these spectator partons may end up
producing the unwanted final states.

In contrast, many simplifications occur in the calculation of
inclusive quantities, that involve an average over all possible final
states
\begin{equation}
\big<{\colorb{\cal O}}\big>\equiv
\sum_{{\mbox{\scriptsize all final}}\atop{\mbox{\scriptsize states}\ {\bs f}}}
{\cal P}(AA\to{\bs f})\;{\colorb{\cal O}[{\bs f}]}\; .
\label{eq:incl}
\end{equation}
In particular, the high energy factorization results that will be
discussed in the next section can only be established for these
inclusive quantities, and their proof fails if one tries to generalize
it to exclusive quantities.

\subsection{Calculation of inclusive observables}
\begin{figure}[htbp]
\begin{center}
\resizebox*{7cm}{!}{\includegraphics{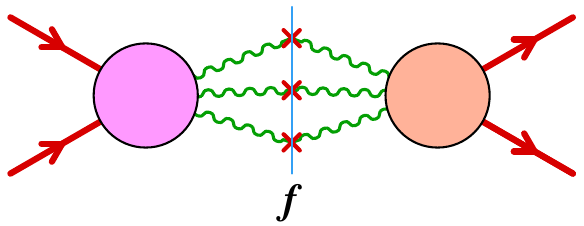}}
\end{center}
\caption{\label{fig:SK}Illustration of the Schwinger-Keldysh formalism.}
\end{figure}

The definition of eq.~(\ref{eq:incl}) suggests an elementary method for
calculating inclusive quantities: compute the exclusive probabilities
to end up in a given final state $f$, and sum over all possible
$f$'s. However, it turns out that one can calculate them in a much
more effective way without having to perform explicitly the sum over
the final states. For this, one should use the {\sl Schwinger-Keldysh
  formalism}~\cite{Schwi1,Keldy1}, in which this sum is already ``built
in''. Any contribution to eq.~(\ref{eq:incl}) is the product of an
amplitude, a complex conjugate amplitude going to the same final
state, and the observable evaluated on this final state, as
illustrated in the figure \ref{fig:SK}.
The diagrammatic rules for the amplitude (right of the dotted line) are
the usual time-ordered Feynman rules. The propagator is the usual
Feynman propagator, which for a (massless) scalar particle reads
\begin{equation}
G_{++}^0(p)=\frac{i}{p^2+i\epsilon}\; .
\end{equation}
For the complex conjugate
amplitude (left of the dotted line), one needs the complex conjugate
of the vertices and propagators.  The propagator is therefore
\begin{equation}
G_{--}^0(p)=\frac{-i}{p^2-i\epsilon}\; .
\end{equation}
Across the dotted line, one must use special propagators that
represent the on-shell particles of the final state $f$,
\begin{equation}
G_{+-}^0(p)=2\pi\theta(-p^0)\delta(p^2)\; .
\end{equation}
The Schwinger-Keldysh formalism amounts to the following:
\begin{itemize}
\item Draw all the graphs $AA\to AA$ that have a given order in $g^2$
  (the power counting is the same as before, with each loop adding one
  power of $g^2$).
\item Sum over all the possibilities of assigning the labels $+$ and
  $-$ to the internal vertices.
\item Only connected graphs contribute, because when summed over the $+$ and
  $-$ labels, the subgraphs that are not attached to the
  observable vanish.
\end{itemize}
These rules will automatically provide the sum over final states that
was included in the formula (\ref{eq:incl}). Note that when used in
this context, the Schwinger-Keldysh formalism is equivalent to
Cutkosky's cutting rules~\cite{Cutko1,t'HooV1}, that were developed as
a tool to compute the imaginary part of transition amplitudes. The
superficial description of the Schwinger-Keldysh formalism that we
have given here can be made more rigorous by writing the generating
functional for its Green's functions. It can be obtained as follows
from the generating functional $Z[j]$ of time-ordered perturbation
theory~:
\begin{equation}
Z[j_+,j_-]
=
\exp\left[\int \!\!d^4xd^4y\;G_{+-}^0(x,y)\,\square_x\square_y\,
\frac{\delta^2}{\delta j_+(x)\delta j_-(y)}\right]
Z[j_+]\,Z^*[j_-]\; .
\end{equation}
This formula makes more obvious the fact that the Schwinger-Keldysh
formalism is made of two copies of the ordinary Feynman perturbation
theory (one of them complex conjugated), ``stitched'' together by the
on-shell propagators $G_{+-}^0$.

At this point, we have not really simplified the calculation of
eq.~(\ref{eq:incl}). It has just been rephrased in a more systematic
language. The simplifications come from noticing the following
identities,
\begin{eqnarray}
&&
G_{++}+G_{--}=G_{+-}+G_{-+}
\nonumber\\
&&
G_{++}-G_{+-}=G_{-+}-G_{--}=G_{_R}\qquad\mbox{(retarded propagator)}\; .
\end{eqnarray}
When using the Schwinger-Keldysh formalism to calculate inclusive
observables at leading order, the sum over the $+$ and $-$ indices
always generates the combinations of propagators that appear in the
second of these equations, and these propagators therefore become
retarded propagators.

\subsection{Classical equation of motion}
This simplification is particularly dramatic for inclusive observables
at leading order. The starting point is a double sum over all possible
tree diagrams (they all have the same order in $g^2$ when the source
is $\rho\sim g^{-1}$) and over all the indices $+$ and $-$ of the
Schwinger-Keldysh formalism. Thanks to the previous remark, the second
sum merely replaces all the propagators by retarded propagators. One
is thus left with a sum over all the tree diagrams built with retarded
propagators, whose first few terms would be
\setbox1\hbox to 10cm{\resizebox*{10cm}{!}{\includegraphics{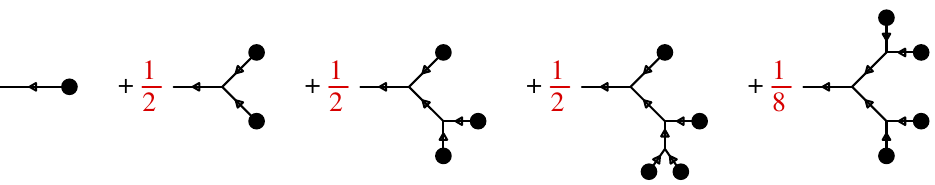}}}
\begin{equation*}
\raise -4mm\box1
\end{equation*}
(here for a $\phi^3$ scalar field theory.) It is then easy to see that
this sum is the solution of the classical field equations of motion
that vanishes when $x^0\to -\infty$ (this retarded boundary condition
follows from the fact that we are summing trees that are made of
retarded propagators\footnote{One can see here why it is important to
  consider inclusive observables for this simplification to happen. It
  is the sum over the final states that leads to the sum over the
  indices $+$ and $-$. Without this sum, one would be left with
  time-ordered propagators, which would make the boundary conditions
  of the classical solution untractable.}).  Although in interacting
theories the classical equation of motion is a non-linear wave
equation, this is a considerable simplification because we have now a
problem that can be solved numerically.

The same simplifications work in the case of the CGC: at leading
order, it is sufficient to solve the classical Yang-Mills equations
with null retarded boundary conditions
\begin{equation}
\big[D_\mu,F^{\mu\nu}\big]=\rho_1 \;\delta^{\nu+} +\rho_2 \;\delta^{\nu-}\quad,\quad\lim_{x^0\to -\infty}A^\mu(x)=0\; .
\label{eq:YM}
\end{equation}
Assuming that we have solved this equation, all the inclusive
observables at leading order can be expressed in terms of its solution
${\cal A}^\mu$. For instance, the single inclusive gluon spectrum is
given by
\begin{equation}
\left.
\frac{d{{\colora{N_1}}}}{dY d^2\vec\p_\perp}\right|_{_{\rm LO}}=
\frac{1}{16\pi^3}\int_{x,y}\; {\colorc e^{ip\cdot (x-y)}}\;
\square_x\square_y\;
\sum_\lambda \epsilon^\mu_\lambda \epsilon^\nu_\lambda\;\;
{\colord {\cal A}_\mu(x)}{\colord {\cal A}_\nu(y)}\; ,
\end{equation}
and the inclusive multigluon spectra simply read
\begin{equation}
\left.
\frac{d{{\colora{N_n}}}}{d^3\p_1\cdots d^3\p_n}\right|_{_{\rm LO}}=
\left.
\frac{d{{\colora{N_1}}}}{d^3\p_1}\right|_{_{\rm LO}}
\times\cdots\times
\left.
\frac{d{{\colora{N_1}}}}{d^3\p_n}\right|_{_{\rm LO}}\; .
\label{eq:Nn}
\end{equation}
Similarly, the components of the energy-momentum tensor have simple
expressions in terms of the classical chromo-electric and
chromo-magnetic fields ${\bs E}^i$ and ${\bs B}^i$,
\begin{eqnarray}
&&
T^{00}_{_{\rm LO}}
=
\frac{1}{2}\big[{{\bs E}^2+{\bs B}^2}\big]
\qquad
T^{0i}_{_{\rm LO}}=\big[{\bs E}\times {\bs B}\big]^i
\\
&&
T^{ij}_{_{\rm LO}}
=
\frac{\delta^{ij}}{2}\big[{{\bs E}^2+{\bs B}^2}\big]
-\big[{\bs E}^i{\bs E}^j+{\bs B}^i{\bs B}^j\big]\; .
\label{eq:Tclass}
\end{eqnarray}

\subsection{Numerical implementation}
\begin{figure}[htbp]
\begin{center}
\resizebox*{5.5cm}{!}{\includegraphics{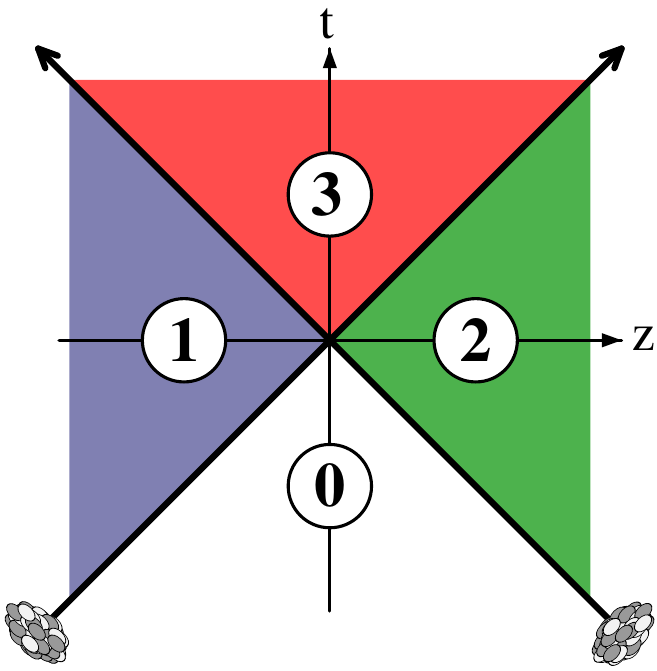}}
\hfill
\resizebox*{5cm}{!}{\includegraphics{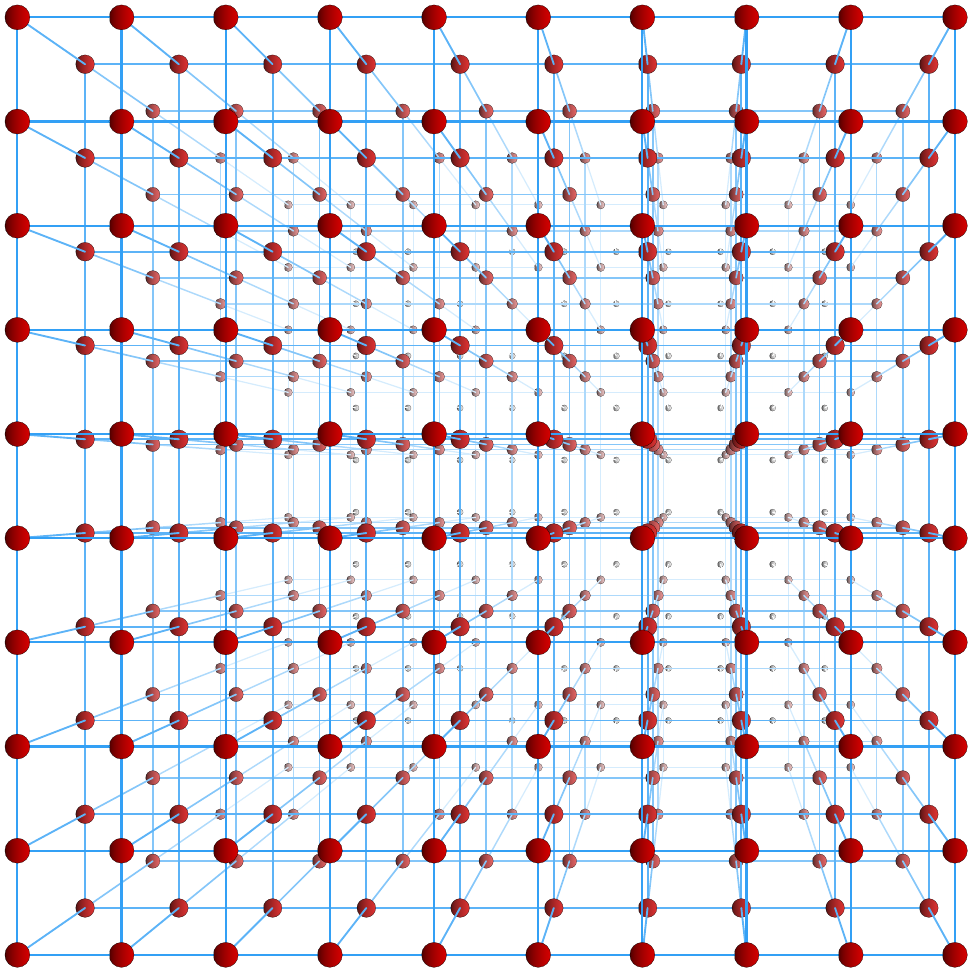}}
\end{center}
\caption{\label{fig:latt1}Left: space-time structure of the classical
  gauge field ${\cal A}^\mu$. Right: 3-dimensional cubic lattice.}
\end{figure}
In order to perform this calculation in practice, one should  be aware
of the following~:
\begin{itemize}
\item[{\bf i.}]~High energy collisions are nearly invariant under
  boosts in the longitudinal direction. This invariance has its
  simplest manifestation if one uses proper-time ($\tau\equiv\sqrt{2x^+
    x^-}$) and rapidity ($\eta\equiv\frac{1}{2}\log(x^+/x^-)$) as the
  coordinates inside the forward light-cone. When written in this
  system of coordinates, the classical Yang-Mills equations do not
  depend explicitly on rapidity, and thus become 1+2 dimensional equations.
\item[{\bf ii.}]~The sources $\rho_1$ and $\rho_2$ have support on the
  light-cone, where they are singular, i.e. proportional respectively
  to $\delta(x^-)$ and $\delta(x^+)$. These sources divide the space
  time in four distinct regions, as shown in the left figure
  \ref{fig:latt1}. The gauge field is identically zero in the region
  0, and it can be found analytically in the regions 1 and
  2~\cite{Kovch1}. In the region 3, the best one can do analytically
  is to obtain the value of the gauge fields and the conjugate
  electrical fields just above the forward light-cone, at a proper
  time $\tau=0^+$~\cite{KovneMW2}~:
\begin{eqnarray}
&
A_0^i=\alpha_1^i+\alpha_2^i\quad,\;
&
E_0^i=0\quad,\;
\alpha_n^i=\frac{i}{g}U_n^\dagger \partial^i U_n\quad(n=1,2)\; ,
\nonumber\\
&
A_{0\eta}=0\quad,\;
&
E_0^\eta=i\frac{g}{2}[\alpha_1^i,\alpha_2^i]\;,
\label{eq:init-LO-tau0}
\end{eqnarray}
where the Wilson lines $U_{1,2}(\x_\perp)$ read
\begin{equation}
U_1(\x_\perp)
=
{\rm P}\,e^{ig\int dx^- \frac{1}{\nabla_\perp^2}\rho_1(x^-,\x_\perp)}\; .
\label{eq:wilson-def}
\end{equation}
\end{itemize}

Given these remarks, we need only to solve numerically 1+2-dimensional
classical Yang-Mills equations inside the forward
light-cone~\cite{KrasnV3,KrasnV1,KrasnV2,KrasnNV2,KrasnNV1,Lappi1,Lappi3,KrasnNV3,KrasnNV4,LappiV1},
starting with the initial conditions (\ref{eq:init-LO-tau0}). Choosing
the time variable $\tau$ determines the Hamiltonian of the system, and
from there one can determine the Yang-Mills equations in Hamiltonian
form.  In order to handle them numerically, one must discretize space
on a cubic lattice (see the right figure \ref{fig:latt1}), while time
remains a continuously varying variable.
\begin{figure}[htbp]
\begin{center}
\hfill
\resizebox*{3cm}{!}{\includegraphics{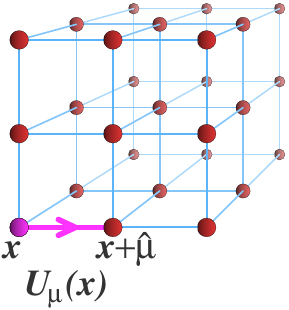}}
\hfill
\resizebox*{3.6cm}{!}{\includegraphics{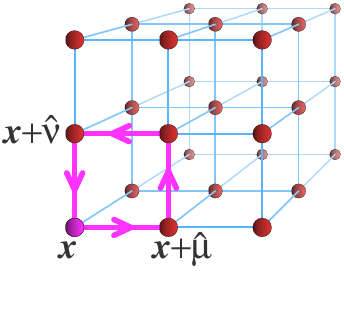}}
\hfill{\ }
\end{center}
\caption{\label{fig:latt2}Left: link variable. Right: elementary plaquette variable.}
\end{figure}
This is most easily done in the temporal
gauge $A^\tau=0$ (also called the Fock-Schwinger gauge in this
context).  After having adopted this gauge condition, the problem has
a residual gauge invariance, under any gauge transformation that
depends only on space. Naive discretizations based on the gauge
potentials $A^\mu$ are not very adequate because they lead to violations of
this residual gauge invariance. Instead, one should adopt Wilson's
formulation, in which the gauge potentials are traded in favor of link
variables (see the left figure \ref{fig:latt2}), i.e. Wilson lines
that span one elementary edge of the lattice
\begin{equation}
  {\colorc U_i(x)} \equiv {\rm P}\,\exp i\,{\colord g}\int_x^{x+\hat{\i}}ds\; A^i(s)\; .
\end{equation}
Under a residual gauge transformation, these links transform as
\begin{equation}
 {\colorc U_i(x)}\quad\to\quad {\colora\Omega(x)}\,{\colorc U_i(x)}\,{\colora \Omega^\dagger(x+\hat{\i})}\; .
\end{equation}
The electrical fields $E^i$ that appear in Hamilton's equations
transform covariantly,
\begin{equation}
 {\colorc E^i(x)}\quad\to\quad 
{\colora\Omega(x)}\,{\colorc E^i(x)}\,{\colora \Omega^\dagger(x)}\; ,
\end{equation}
and therefore they should live on the nodes of the lattice. In the
$A^\tau=0$ gauge, the Hamiltonian discretized in this fashion reads
\begin{eqnarray}
      {\cal H}&=&
\sum_{\vec\x;i}\frac{E^i(\x) E^i(\x)}{2}\nonumber\\
&&-\frac{6}{{\colord g^2}}\sum_{\vec\x;ij}1-\frac{1}{3}{\rm Re\ Tr}\,(\underbrace{U_i(x)U_j(x+\hat{\imath})U^\dagger_i(x+\hat{\jmath})U^\dagger_j(x)}_{\scriptsize\mbox{plaquette at the point $\vec\x$ in the $ij$ plane}})\; .
\end{eqnarray}
The only combinations of link variables that enter in this formula are
{\sl plaquettes} (i.e. the trace of the product of the four link variables that
form an elementary square on the cubic lattice), which are gauge
invariant. The Hamilton equations that can be derived from this
Hamiltonian form a large (but finite) set of ordinary differential
equations, that can be solved numerically by standard methods such as
the leapfrog algorithm.

\subsection{Structure of the classical color fields}
\begin{figure}[htbp]
\begin{center}
\resizebox*{7cm}{!}{\includegraphics{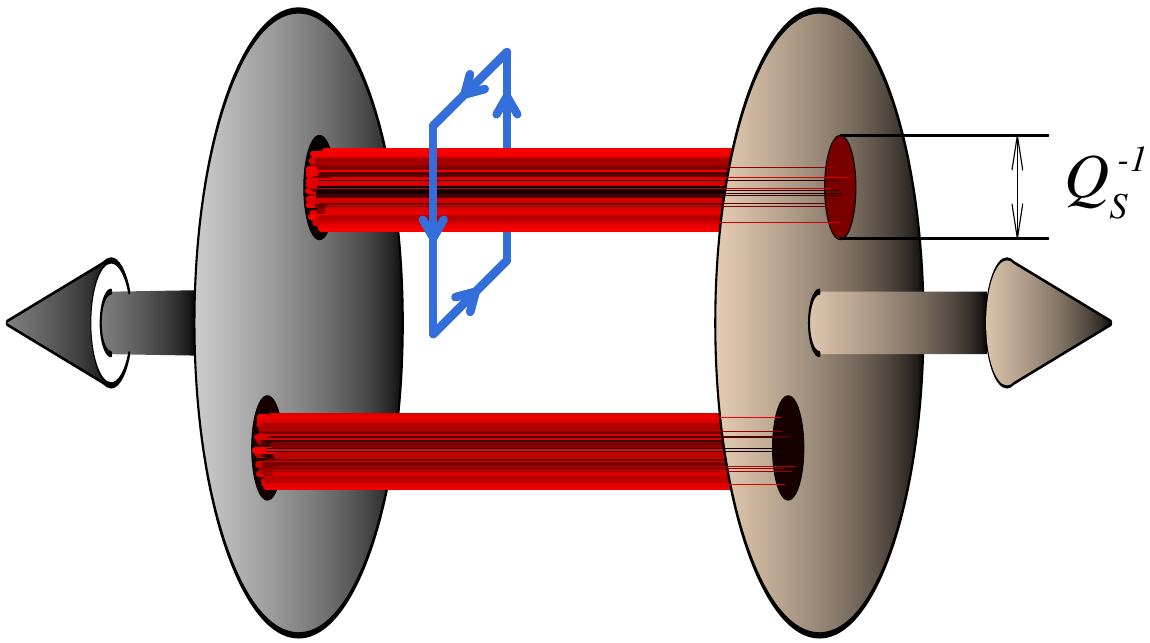}}
\end{center}
\caption{\label{fig:flux}Color flux tubes just after the collision.}
\end{figure}

At very short times after the collisions ($\tau\ll Q_s^{-1}$), the
classical chromo-electric and chromo-magnetic fields are parallel to
the collision axis~\cite{LappiM1}, as illustrated in the figure \ref{fig:flux}.
By studying how the expectation value of transverse Wilson
loops\cite{DumitNP1,DumitLN1},
\begin{equation}
W\equiv \left<{\rm P}\,\exp ig\int_{\gamma}dx^i {\cal A}^i\right>\; ,
\end{equation}
depends on the area enclosed by the loop, one can infer the typical
transverse size of these flux tubes. Indeed, one can roughly view the
argument of the exponential as the magnetic flux going through the
loop. It has been found that $W$ decreases roughly as
$\exp(-\#\times\mbox{Area})$ for areas larger than $Q_s^{-2}$, which indicates
that the fields are not correlated over transverse distances larger
than $Q_s^{-1}$. $Q_s^{-1}$ is therefore the typical radius of these
flux tubes.

\begin{figure}[htbp]
\begin{center}
\resizebox*{7cm}{!}{\includegraphics{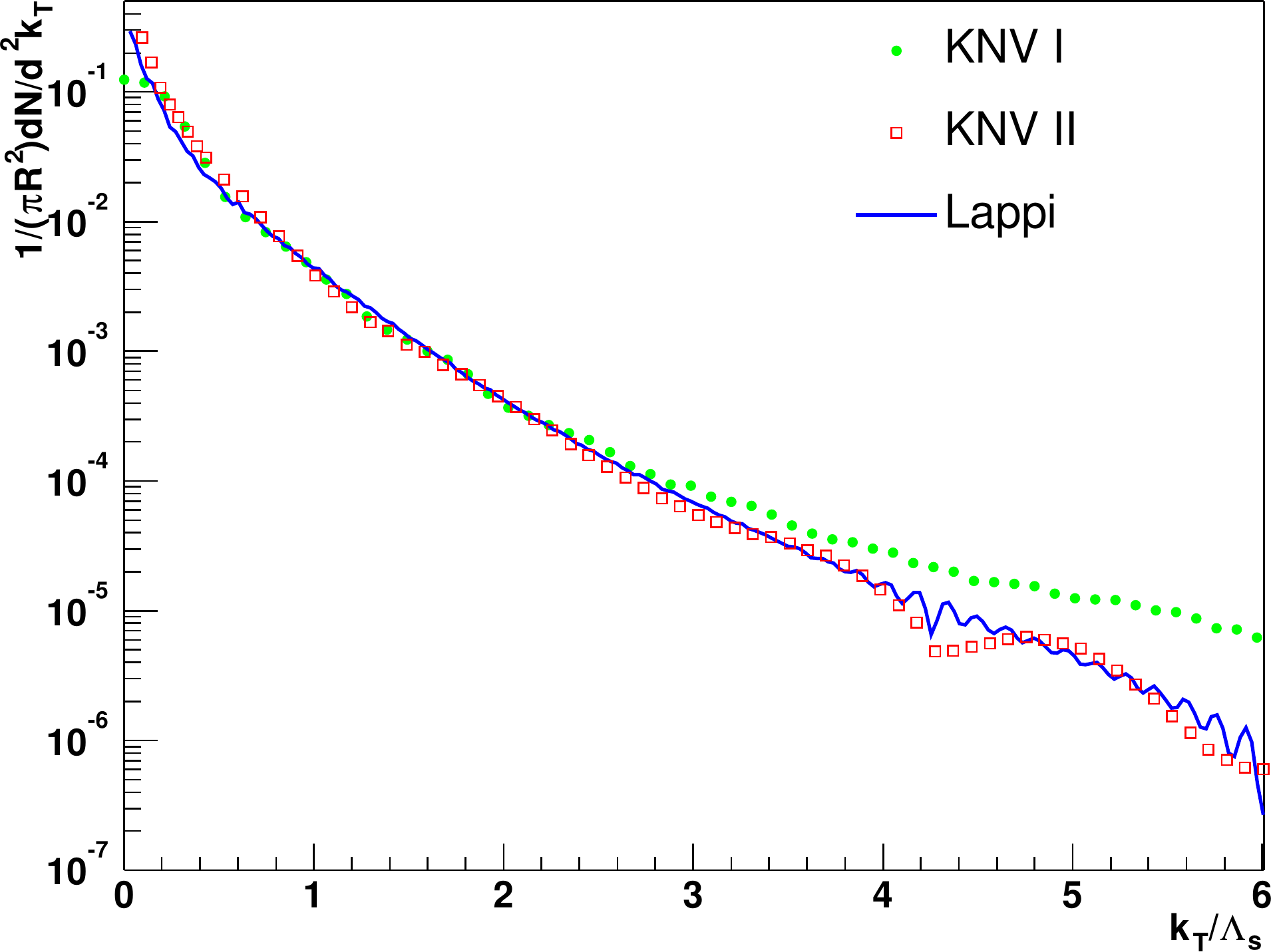}}
\end{center}
\caption{\label{fig:N1}Inclusive gluon spectrum.}
\end{figure}
From the classical gauge fields, one can compute the spectrum of
gluons produced in a heavy ion collision (see the figure
\ref{fig:N1}). At large transverse momentum, the spectrum decreases as
$k_\perp^{-4}$. Indeed, when $k_\perp \gg Q_s$, the saturation
criterion is not satisfied and one should recover the usual
perturbative results of the dilute regime.  In contrast, saturation
effects are quite large at small transverse momentum, $k_\perp\lesssim
Q_s$, where they produce a strong softening of the spectrum, while the
dilute result would still give a spectrum that grows as $k_\perp^{-4}$
at small $k_\perp$ (because there is no dimensionful scale other than
$k_\perp$ in the dilute calculation).  Let us close this section by a
remark concerning the energy dependence of the gluon multiplicity,
obtained by integrating the gluon spectrum over $k_\perp$. The result
is proportional to the transverse area $S_\perp$ multiplied by $Q_s^2$,
\begin{equation}
N_{\rm gluon}\sim \frac{S_\perp Q_s^2}{\alpha_s}\; .
\end{equation}
From this pocket formula, one sees that the energy dependence of the
gluon multiplicity is directly inherited from that of the saturation
momentum,
\begin{equation}
N_{\rm gluon}\sim x^{-0.3}\sim s^{0.15}\; .
\end{equation}
Note that there is no contradiction between the fact that the
multiplicity grows like a power of the collision energy and the
Froissart bound, that tells us that hadronic cross-sections cannot
grow faster than $\sigma\sim\log^2(s)$.  The difference between the
two is due to the fact that these two objects measure very different
things. The total cross-section measures the probability that the two
projectiles interact. As a probability, its growth is constrained by
unitarity. In contrast, the gluon multiplicity measures the ``amount
of stuff'' which is produced in a collision. It is not a probability,
and is not bound by the same constraints. More precisely, the
Froissart bound is related to the growth of the radius of the ``black
disk'' region with energy (this radius grows like the log of
energy). However, even after a certain region has become ``black''
(and thus its probability of interacting cannot grow anymore, because
it has reached the unitarity limit), the number of gluons produced in
this region will continue to grow like a power of energy (because the
number of gluons per unit area in the incoming projectiles continues
to increase like $Q_s^2$).

\section{Next to Leading Order}
\label{sec:NLO}
\subsection{Improved power counting}
From the power counting formula derived earlier, we expected for the
gluon spectrum an expansion of the form
\begin{equation}
      \frac{d{\colora N}}{d^3\vec\p}
      =
      \frac{1}{{\colorb g^2}}\;\Big[
	c_0+c_1\,{\colorb g^2}+c_2\,{\colorb g^4}+\cdots
	\Big]\; ,
\end{equation}
and we have seen in the previous section how to calculate the term
$c_0/g^2$. The following terms, $c_1, c_2 g^2,\cdots$, are given
respectively by the sum of 1-loop, 2-loop, etc... diagrams. When
calculating loops in the CGC framework, we have to recall that the
degrees of freedom have been divided into sources and fields,
separated by a cutoff $y_{\rm cut}$ in rapidity.  In the integration
over the loop momentum, we must use this cutoff in order to
prevent the loop momentum to go into the kinematical domain which is
described in terms of static sources. Failing to do this would lead to
a double counting of the contribution of the modes that lie in this region.

In general, loop diagrams will depend on this cutoff~\cite{AyalaJMV2}:
a graph with $n$ loops can contain up to $n$ powers of $y_{\rm
  cut}$. Therefore, the coefficients that appear in the $g^2$
expansion of the gluon spectrum can themselves be expanded in powers
of the cutoff as follows,
\begin{align}
  c_1
  &&=&&
  && &&
  c_{10}
  &&+&&
  {\colord c_{11}\;y_{\rm cut}}\nonumber
  \\
  c_2
  &&=&&
  c_{20}
  &&+&&
  c_{21}\;y_{\rm cut}
  &&+&&
\underbrace{{\colord c_{22}\;y_{\rm cut}^2}}
\\
  &&&&
  &&&&
  &&&&\mbox{\scriptsize Leading Log terms}\nonumber
\end{align}
The terms that have the maximal degree in $y_{\rm cut}$, i.e. a degree
equal to the number of loops, are called the {\sl leading log
  terms}\footnote{The terminology comes from the fact that $y_{\rm
    cut}$ is the logarithm of a cutoff on the longitudinal momentum.}.

The cutoff $y_{\rm cut}$ was introduced by hand when defining the CGC,
as a way of separating the two kinds of degrees of freedom, and it is
therefore not a physical parameter. Observables should not
depend upon it. As we shall see in this section, the leading log
cutoff dependence that arises from loop corrections to observables can
be absorbed into a redefinition of the probability distribution
$W[\rho]$. This redefinition turns $W[\rho]$ into a cutoff dependent
object, but its cutoff dependence is universal, which means that the
same distributions can be used for all inclusive observables.

Before we continue with a discussion of the leading log terms, let us
also mention the fact that the next-to-leading log corrections are now
known in some cases: in the BK equation (a mean field approximation of
the JIMWLK equation)~\cite{Balit3,BalitC1,KovchW1,GardiKRW1} and also
for the JIMWLK equation~\cite{Grabo1,BalitC10,KovneLM1,KovneLM3}. The running
coupling corrections have been used in some phenomenological
studies~\cite{AlbacAMSW1,AlbacK1,AlbacAMS1,AlbacAMS2,LappiM2} where
they appear to be quantitatively important.

\subsection{NLO result}
Let us give here a sketch of the proof of this factorization. The first
step is the derivation of an expression for the NLO correction to
inclusive observables. It turns out that there exists a formal
relationship between the LO and NLO contributions, valid for
any inclusive observable, that reads~\cite{GelisLV3,GelisLV4}:
\begin{equation}
{\colorb{\cal O}}_{_{\rm NLO}}
=
\Bigg[
 \frac{1}{2}\!\!
\int\limits_{_{\u,\v}}
\!\!
{\colorb{\bs\Gamma}_{2}(\u,\v)}\,
{\mathbbm T}_\u
{\mathbbm T}_\v
+\int\limits_{_{\u}}
{\colorb{\bs\alpha}(\u)}\,
{\mathbbm T}_\u
\Bigg]\;
{\colorb{\cal O}}_{_{\rm LO}}\; .
\label{eq:fact0}
\end{equation}
In this formula, the LO observable ${\cal O}_{_{\rm LO}}$ should be
viewed as a functional of the initial value of the classical field on
some space-like hypersurface (the integrations over the variables $\u$
and $\v$ are on this surface). The operator ${\mathbbm T}_\u$ is the
generator of shifts of this initial condition, in the sense that its
exponential translates the initial field ${\cal A}_{\rm init}$ in any
quantity that can be expressed in terms of the classical field
\begin{equation}
\exp\left[\int_\u {\bs\alpha}(\u){\mathbbm T}_\u\right]\;{\cal F}[{\cal A}_{\rm init}]={\cal F}[{\cal A}_{\rm init}+{\bs\alpha}]\; .
\end{equation}
The remarkable property of eq.~(\ref{eq:fact0}) is that the
coefficient functions ${\bs\Gamma}_2$ and ${\bs\alpha}$ are universal:
they do not depend on the observable under consideration. Note however
that although this formula is valid for all inclusive observables, it
is not true for exclusive observables.

\subsection{Classical phase-space formulation of quantum mechanics}
In the formula (\ref{eq:fact0}), the operator between the square
brackets acts only on the initial value of the classical fields, while
the time evolution from the initial time surface to the time where the
observable is evaluated remains classical. This is in fact a
completely general result in quantum mechanics: at the first order in
$\hbar$, the time evolution remains classical and $\hbar$ enters only
in the initial condition.  Let us make a digression to justify this
important point. This is best viewed if one rewrites the evolution
equation for the density operator,
\begin{equation}
    \frac{\partial\widehat{\rho}_\tau}{\partial\tau}=i\,{\colord \hbar} \,
    \big[\widehat{H},\widehat{\rho}_\tau\big]\; ,
\label{eq:VN}
\end{equation}
in terms of the Wigner transforms of $\widehat{\rho}_\tau$ and
$\widehat{H}$,
\begin{eqnarray}
    W_\tau(\x,\p)&\equiv& \int d\s \; e^{i\frac{\p\cdot \s}{\hbar}}\; \big<\x+\frac{\s}{2}\big|
    \widehat{\rho}_\tau\big|\x-\frac{\s}{2}\big>\nonumber\\
    {\cal H}(\x,\p)&\equiv& \int d\s \; e^{i\frac{\p\cdot \s}{\hbar}}\; \big<\x+\frac{\s}{2}\big|
    \widehat{H}\big|\x-\frac{\s}{2}\big>\; .
\end{eqnarray}
(The Wigner transform of $\widehat{H}$ is nothing but the classical
Hamiltonian of the system.) Note that in these Wigner transforms, the
variables $\x$ and $\p$ are classical phase-space variables, not
operators. It is straightforward to show that eq.~(\ref{eq:VN}) is
equivalent to
\begin{eqnarray}
    \frac{\partial W_\tau}{\partial\tau}
    &=&{\cal H}(\x,\p)\;
    \frac{2}{i\,{\colord \hbar}}\,\sin \left(\frac{i\,{\colord \hbar}}{2}
    \big(\stackrel{\leftarrow}{\partial}_\p\stackrel{\rightarrow}{\partial}_\x-\stackrel{\leftarrow}{\partial}_\x\stackrel{\rightarrow}{\partial}_\p\big)\right)\;W_\tau(\x,\p)\nonumber\\
  &=&\underbrace{\big\{{\cal H},W_\tau\big\}}_{\scriptsize\mbox{Poisson bracket}}
  +{\cal O}({\colord \hbar^2})\; .
\label{eq:Moyal}
\end{eqnarray}
The first line is known as the {\sl Moyal equation}. It is equivalent to
the von Neumann equation for $\widehat{\rho}_\tau$, except that it is
expressed entirely in terms of classical phase-space variables. This
equation makes the classical limit particularly transparent if one
expands in powers of $\hbar$ the operator that appears in its right
hand side. As one can see immediately, its zeroth order in $\hbar$ is
the usual Poisson bracket. This means that at the order $\hbar^0$, one
recovers classical Hamiltonian mechanics.  A remarkable feature of the
Moyal equation is that it has no term of order $\hbar^1$. This means
that at NLO, the time evolution of the system remains purely
classical. The only quantum effects at order $\hbar^1$ come via the
initial condition, through the fact that the support of the Wigner
distribution of a quantum state must have an extension of at least
$\hbar$. The first correction to the Moyal equation arises at the
order $\hbar^2$, i.e. at NNLO. At this order, one gets quantum
corrections both in the initial condition and in the time evolution
itself. This discussion also indicates that a formula such as
eq.~(\ref{eq:fact0}), where only the initial conditions are altered,
presumably does not exist beyond NLO.

\subsection{Cutoff dependence}
Eq.~(\ref{eq:fact0}) is very useful in order to extract the cutoff
dependence in inclusive observables at NLO, because this cutoff
dependence is already present in the operator that acts on ${\cal
  O}_{_{\rm LO}}$. If we keep only the terms that are linear in the
cutoff, we have~\cite{GelisLV3,GelisLV4,GelisLV5}
\begin{equation}
\frac{1}{2}\!\!
\int\limits_{_{\u,\v}}
\!\!
{\colorb{\bs\Gamma}_{2}(\u,\v)}\,
{\mathbbm T}_\u
{\mathbbm T}_\v
+\int\limits_{_{\u}}
{\colorb{\bs\alpha}(\u)}\,
{\mathbbm T}_\u
=
y_{\rm cut}^+\;{\colorb{\cal H}_1}
+
y_{\rm cut}^-\;{\colorb{\cal H}_2}\; .
\label{eq:fact1}
\end{equation}
In this equation, $y_{\rm cut}^+$ and $y_{\rm cut}^-$ are the cutoffs
corresponding to the right and left moving nucleus respectively, and
${\cal H}_{1,2}$ are operators known as the {\sl JIMWLK
  Hamiltonians}\footnote{If one expands this Hamiltonian at small
  $\rho$, one can recover the BFKL equation~ \cite{KuraeLF1,BalitL1}.}
of the two
nuclei~\cite{Balit1,JalilKMW1,JalilKLW1,JalilKLW2,JalilKLW3,JalilKLW4,IancuLM1,IancuLM2,FerreILM1},
\begin{equation}
{\cal H}
\equiv
{\frac{1}{2}\int_{\vec{\x}_\perp,\vec{\y}_\perp}\!
\frac{\delta}{\delta {\colorb \rho_a(\vec{\x}_\perp)}}
{\colorc \chi_{ab}(\vec{\x}_\perp,\vec{\y}_\perp)}
\frac{\delta}{\delta\colorb \rho_b(\vec{\y}_\perp)}}\; ,
\label{eq:HJIMWLK}
\end{equation}
where
\begin{eqnarray}
&&
{\colorc \chi_{ab}(\vec{\x}_\perp,\vec{\y}_\perp)}
\equiv
\frac{{\colorc\alpha_s}}{4\pi^3}
\int d^2\vec\z_\perp
\frac{(\vec\x_\perp-\vec\z_\perp)\cdot(\vec\y_\perp-\vec\z_\perp)}
{(\vec\x_\perp-\vec\z_\perp)^2(\vec\y_\perp-\vec\z_\perp)^2}
\nonumber\\
&&\qquad\qquad
\times
\left[\Big(1-{\colorb{\wt U}^\dagger(\vec\x_\perp){\wt U}(\vec\z_\perp)}\Big)
\Big(1-{\colorb{\wt U}^\dagger(\vec\z_\perp){\wt U}(\vec\y_\perp)}
\Big)\right]_{ab}\; .
\end{eqnarray}
In this equation, ${\wt U}$ is a Wilson line in the adjoint
representation, constructed from the gauge field $A^+$ such
that ${\bs\nabla}_\perp^2 A^+=-\rho$.

\subsection{Factorization}
\begin{figure}[htbp]
\begin{center}
\resizebox*{7cm}{!}{\includegraphics{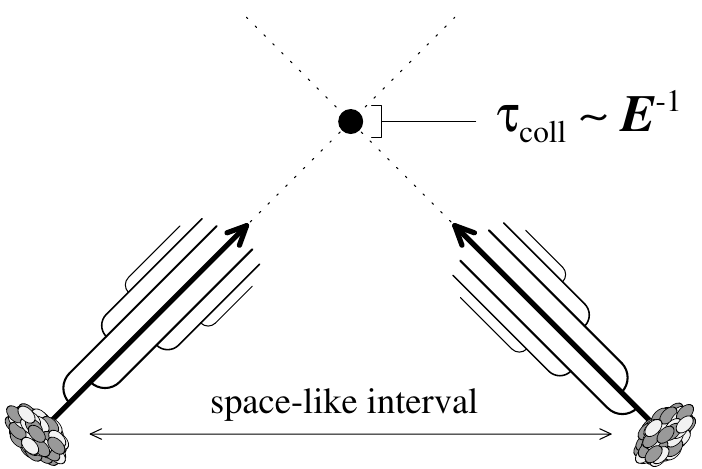}}
\end{center}
\caption{\label{fig:causal}Causality argument for factorization.}
\end{figure}

The formula (\ref{eq:fact1}) has two important consequences:
\begin{itemize}
\item[{\bf i.}]~This formula is the {\sl sum} of two terms
  corresponding to the two nuclei, but there is no cutoff dependent
  term mixing the two nuclei. This means that the cutoff dependent
  terms are intrinsic properties of the nuclei prior to their
  collision, and this is the reason why it is possible to eliminate
  them by redefining the distributions of the sources $\rho_{1,2}$ of
  the two projectiles.
\item[{\bf ii.}]~Since the operator in the square brackets in
  eq.~(\ref{eq:fact0}) is the same for all inclusive observables, the
  cutoff dependence is equally universal. This is the reason why it
  will be possible to define cutoff dependent distributions
  $W[\rho_{1,2}]$ such that they cancel the cutoff dependence of all
  observables.
\end{itemize}
The property {\bf i}, about the absence of mixing of the cutoff
dependence between the two nuclei, can be understood simply in terms
of causality. This is illustrated in the figure \ref{fig:causal}.
Indeed, the cutoff dependence arises from the phase-space integration
of the soft gluons emitted by bremsstrahlung. Since they are soft, the
formation time of these gluons is large: they cannot be emitted during
the very brief duration of the collision, so they have to be emitted
before the collision. Because the separation between the two nuclei is
space-like until the collision, causality forbids any cutoff dependent
term that would mix the two nuclei. The property {\bf ii} also follows
the same reasoning: gluon emissions that happen before the collision
should be the same for all observables measured after the collision.

If we compute observables for fixed configurations $\rho_1$ and
$\rho_2$ of the color charge densities in the two projectiles, there
is no way to get rid of the cutoff dependence. The only way to remove
it is to integrate over all the possible configurations of
$\rho_{1,2}$. The main ingredient in this manipulation is the fact
that the JIMWLK Hamiltonian ${\cal H}$ is a self-adjoint operator:
\begin{equation}
\int [D\rho]\;W\;\big({\colorb\cal H}\,{\cal O}\big)
=
\int [D\rho]\;\big({\colorb\cal H}\,W\big)\;{\cal O}\; .
\end{equation}
This property can be used to transfer the action of ${\cal H}$ from
the observable onto the distribution $W[\rho]$. From
eq.~(\ref{eq:fact1}), one can see that $\rho$-averaged quantities such
as
\begin{equation}
\frac{dN_1}{d^3\vec\p}
\;\empile{=}\over{\scriptsize\mbox{Leading Log}}\;
\int 
\big[D{\colora\rho_{_1}}\,D{\colorb\rho_{_2}}\big]
\;
{\colora W_1\big[\rho_{_1}\big]}\;
{\colorb W_2\big[\rho_{_2}\big]}
\;
\underbrace{\left.\frac{dN_1}{d^3\vec\p}\right|_{_{\rm LO}}}_{\rm fixed\ \rho_{1,2}}
\label{eq:N1fact}
\end{equation}
are independent of the cutoff, provided that the distributions
$W[\rho]$ themselves depend on the cutoff according to the JIMWLK
equation
\begin{equation}
\frac{\partial W}{\partial y} = {\cal H}\,W\; .
\end{equation}
From eqs.~(\ref{eq:fact0}) and (\ref{eq:fact1}), it is furthermore
obvious that the same factorization formula (with the same $W$'s)
applies to any inclusive observable.

In eq.~(\ref{eq:N1fact}), it is the evolution with rapidity of the
distributions $W$ that gives the gluon spectrum its rapidity
dependence. Indeed, the gluon spectrum for fixed $\rho_{1,2}$
that enters in the integrand is still independent of rapidity. From
the JIMWLK equation, one sees that the distributions $W$ evolve
significantly for changes of the rapidity of the order of $\Delta
y\sim \alpha_s^{-1}$. This factorization result thus predicts that the
gluon spectrum is rather flat in rapidity at weak coupling.

\subsection{Ridge correlations}
Eq.~(\ref{eq:N1fact}) can be generalized to the inclusive multigluon
spectrum. Recalling also eq.~(\ref{eq:Nn}), we obtain the following
factorization formula
  \begin{eqnarray}
    &&
    \frac{dN_n}{d^3\vec\p_{1}\cdots d^3\vec\p_{n}}
\;\empile{=}\over{\scriptsize\mbox{Leading Log}}\;
  \nonumber\\
  &&\qquad=
  \int 
  \big[D{\colora\rho_{_1}}\,D{\colorb\rho_{_2}}\big]
  \;
  {\colora W_1\big[\rho_{_1}\big]}\;
  {\colorb W_2\big[\rho_{_2}\big]}
  \;
  \left.\frac{dN_1}{d^3\vec\p_{1}}
  \cdots
  \frac{dN_1}{d^3\vec\p_{n}}\right|_{_{\rm LO}}\; .
\end{eqnarray}
This equation, valid at leading log accuracy, shows that at this order
the correlations between the produced gluons only originate from
correlations of the $\rho$'s of the incoming projectiles, since in the
integrand the $n$ gluons still appear completely factorized.  Since
the relevant rapidity interval for a significant JIMWLK evolution is
$\Delta y\sim \alpha_s^{-1}$, this is also the typical rapidity
distance over which the produced gluons will be correlated.

\begin{figure}[htbp]
\begin{center}
\resizebox*{7cm}{!}{\includegraphics{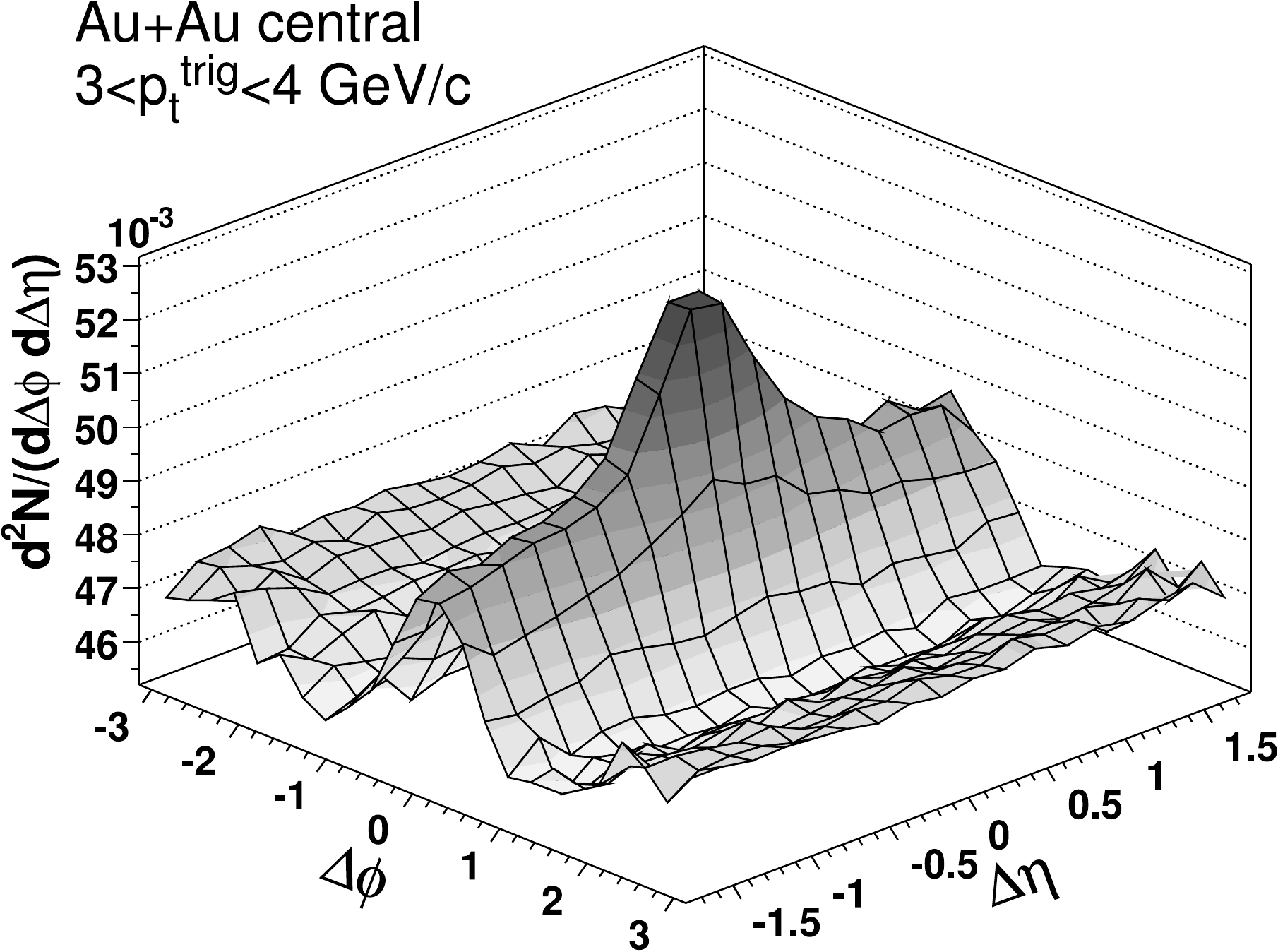}}
\end{center}
\caption{\label{fig:ridge-data}Two hadron correlation measured in heavy ion collisions~\cite{Abelea1}.}
\end{figure}

\begin{figure}[htbp]
\begin{center}
\resizebox*{10cm}{!}{\includegraphics{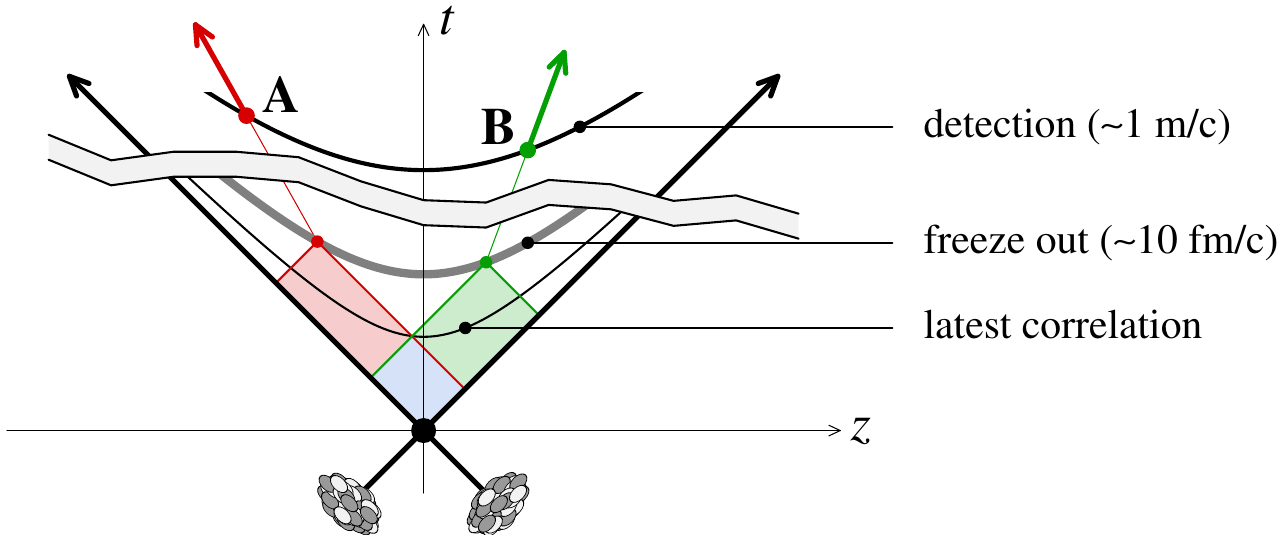}}
\end{center}
\caption{\label{fig:horizon}Origin of the rapidity correlations.}
\end{figure}

This is the basis of an interpretation of the peculiar shape of the
2-hadron correlations observed in heavy ion collisions. This
correlation function is represented in the figure
\ref{fig:ridge-data}, as a function of the relative azimuthal angle
and relative rapidity.
As one can see, the correlation is very narrow in azimuthal angle, and
very elongated in rapidity\footnote{The peak in the middle is a
  jet-like correlation, due to quasi-collinear splittings in the final
  state.}. Because of causality, the existence of a correlation
between particles that are widely separated in rapidity must originate
from phenomena that happened very shortly after the collision. This is
explained in the figure \ref{fig:horizon}.
Let us consider the time evolution of two particles A and B in
reverse, starting from their last interaction on the freeze-out
surface.  Obviously, by causality, they must come from the light-cones
represented respectively in red and green in the figure.  A
correlation is an event that had an influence on both the particle A
and the particle B. It must therefore have happened in the overlap
between these two light-cones, that we have represented in blue.  One
sees clearly that there is a maximal time at which this correlation could
possibly have been created. From the time of the freeze-out and the
rapidity separation of the two particles, it is easy to determine this
upper bound of the time,
\begin{equation}
\tau_{\rm correlation}\;\le\;\tau_{\rm freeze\ out}\;\;e^{-|\Delta y|/2}\; .
\end{equation}
This bound decreases very rapidly as one increases the rapidity
separation $\Delta y$. In heavy ion collisions, the order of magnitude
of the freeze-out time is $10$~fm/c. For instance, a correlation
between particles separated in rapidity by $\Delta y=6$ must be
produced before the time $0.5$~fm/c, which is well within the regime
where the CGC is still applicable. 

It is in fact easy to understand qualitatively the main features of
the observed correlation from the structure of the classical color
fields produced at early times in heavy ion
collisions~\cite{DumitGMV1,LappiSV1}.  As we have said before, these
fields are organized in flux tubes that have a typical transverse size
of $Q_s^{-1}$ and that remain coherent over rapidity intervals of
order $\alpha_s^{-1}$.
\begin{figure}[htbp]
\begin{center}
\resizebox*{6cm}{!}{\includegraphics{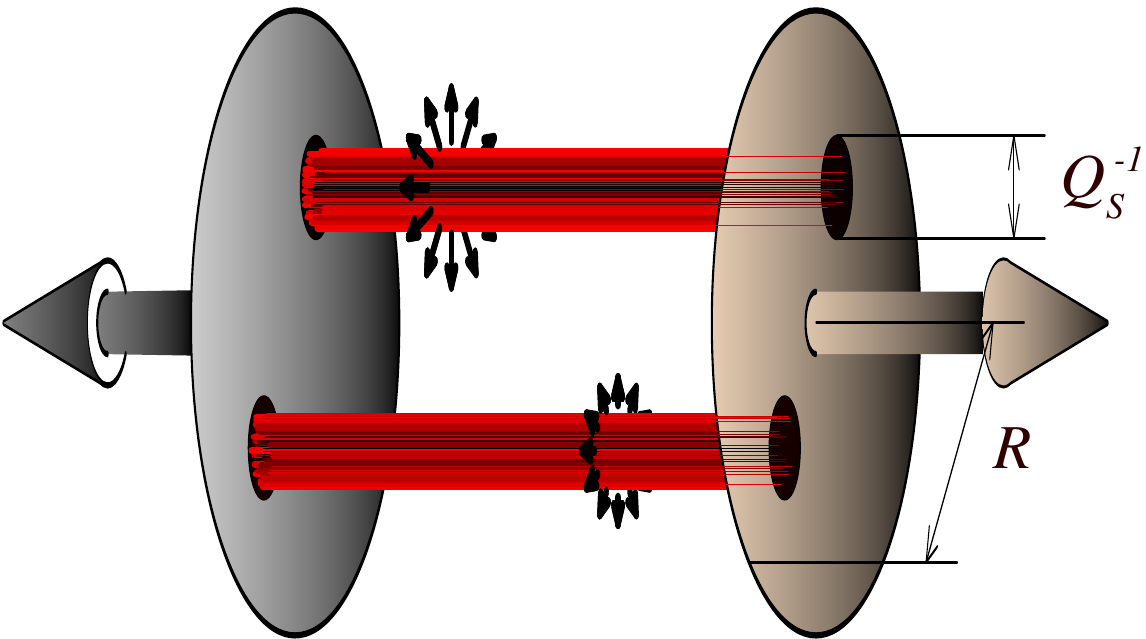}}\hfill
\resizebox*{6cm}{!}{\includegraphics{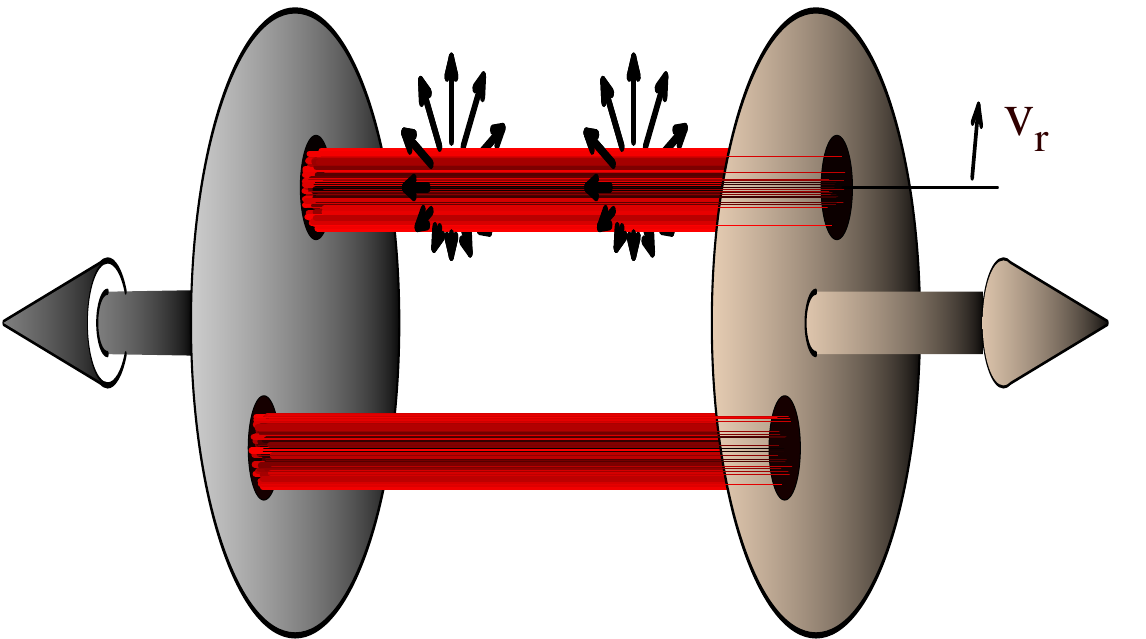}}
\end{center}
\caption{\label{fig:ridge}Rapidity correlations from color flux tubes.}
\end{figure}
Two gluons emitted from the same tube are correlated if they are
produced with a rapidity separation $\big|\Delta y\big| \lesssim
\alpha_s^{-1}$, but are not correlated if they come from two different
flux tubes\footnote{The fact that the chromo-electric and
  chromo-magnetic fields are purely longitudinal at early time does
  not seem to play any role in this argument. The important properties
  are their coherence length in rapidity and in the transverse
  plane.}. From the size of the flux tubes, we conclude that the
probability that two particles are correlated scales as $(RQ_s)^{-2}$
where $R$ is the transverse radius of the collision zone.

This explains the existence of a long range correlation in rapidity
between pairs of particles, but not why their correlation is peaked in
azimuthal angle. The 2-gluon correlation one gets from the CGC color
fields is nearly independent of the azimuthal angle, because on
average these gluons can be emitted in any transverse direction.
However, one should keep in mind that the above causality argument
applies only to the correlation in rapidity, not to the correlation in
azimuthal angle that can be produced much later.  These azimuthal
correlations are generated by the radial flow~\cite{Volos1,PruneGV1}
that develops subsequently and expels radially the matter produced in
the collision.  Simple relativistic kinematics indeed shows that if
one boosts a 2-particle spectrum independent of azimuth, it becomes
peaked around $\Delta\phi=0$ (the prominence of the peak increases
with the velocity of the boost)~\cite{DumitGMV1}.
\begin{figure}[htbp]
\begin{center}
\resizebox*{8cm}{!}{\includegraphics{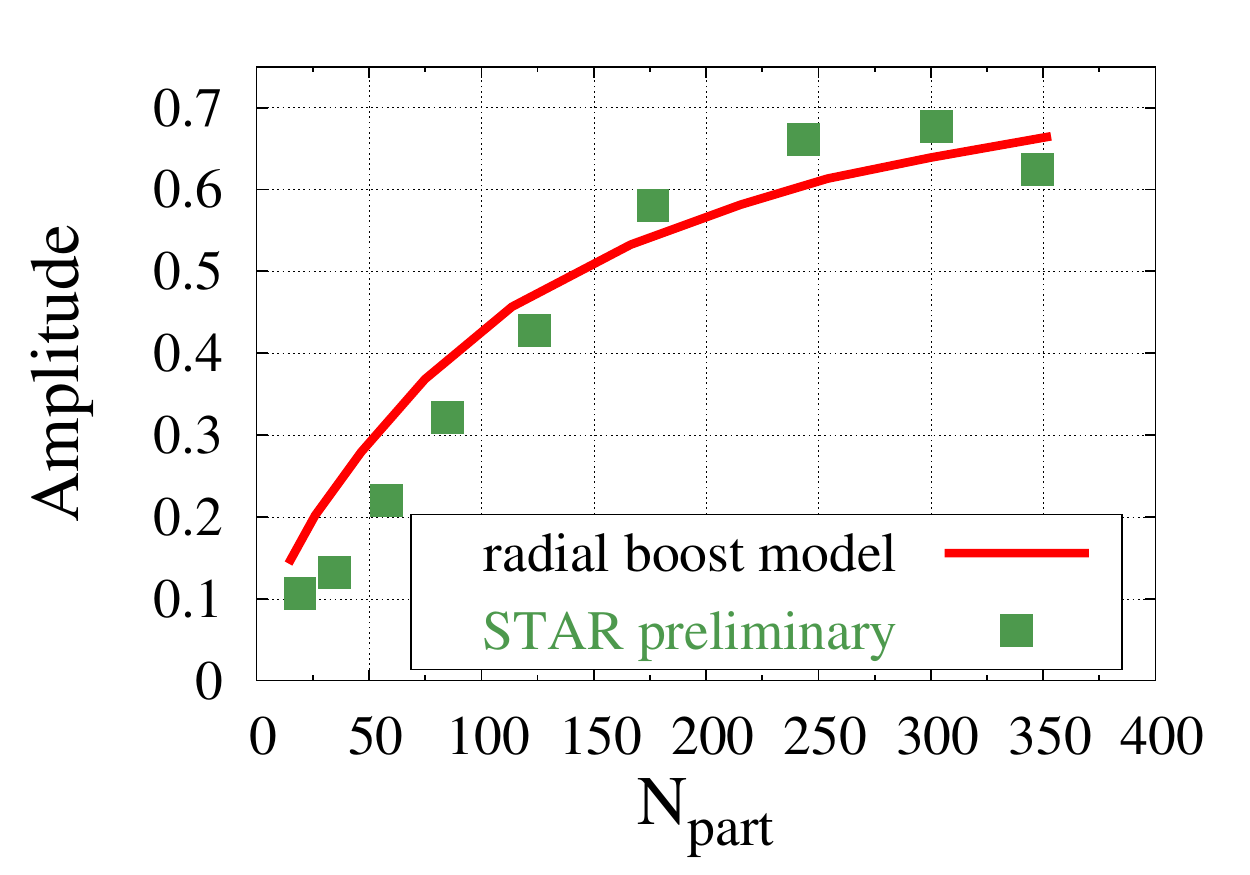}}
\end{center}
\caption{\label{fig:ridge1}Comparison of the strength of the
  azimuthal correlation in data and in a simple radial boost
  model \cite{GavinMM1} (see also \cite{Shury2}). }
\end{figure}
In the figure \ref{fig:ridge1}, we show a comparison of the strength
of the azimuthal correlation in data and in a very basic radial boost
model where a unique radial boost velocity is applied to a flat
spectrum (the radial velocity is estimated from the slope of $p_\perp$
spectra at small momentum).  The centrality dependence, which in this
model comes from the increase of the radial velocity with centrality,
is in fair qualitative agreement with the measurement.

\section{From the Color Glass Condensate to hydrodynamics}
\label{sec:matching}
\subsection{Requirements for hydrodynamics}
The CGC provides a self-contained QCD based framework for describing
heavy ion collisions from first principles. It also provides tools for
calculating inclusive observables at leading log accuracy,
i.e. leading order plus a resummation of all the leading log
contributions coming from higher loop diagrams. However, there is some
physics that plays an important role in heavy ion collisions but is
not easily captured in the CGC. The fact that the produced gluons and
quarks will eventually hadronize when their energy density falls below
the QCD critical energy density is obviously not present in the CGC
(at least, at any fixed loop order).

For this reason, CGC calculations can only describe the early stages
of the fireball expansion, and should be later on matched onto another
description such as relativistic
hydrodynamics~\cite{Adamsa3,Adcoxa1,Arsena2,Backa2,HuoviR1,Romat1,Teane1,RomatR1}. Since
the components of the energy momentum tensor can be computed in the
CGC framework, one could use it as initial data for the hydrodynamical
evolution.
\begin{figure}[htbp]
\begin{center}
\resizebox*{7cm}{!}{\includegraphics{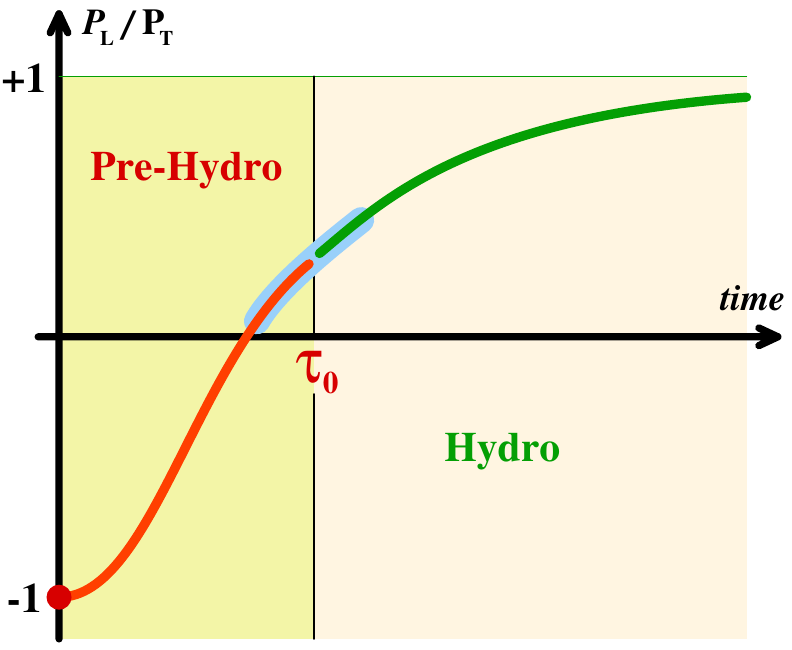}}
\end{center}
\caption{\label{fig:matching}Smooth matching between CGC and hydrodynamics. }
\end{figure}
Firstly, for such a matching to be possible, the CGC must bring the
system to a state that hydrodynamics can handle. This means that the
transverse and longitudinal pressure should not be too different (in
particular, the longitudinal pressure should not be negative), and
that the viscous effects (e.g. the ratio $\eta/s$ of the shear
viscosity to the entropy density) should be small.  

Moreover, when performing such a matching between two descriptions
that use different degrees of freedom, one should be careful to check
that the two descriptions are compatible in a certain time window. In
other words, there should be some range where the two models 
predict equivalent results. If this is the case, the precise time
$\tau_0$ at which the matching is realized is not important, and it
can be varied in this range without any incidence on the final
result\footnote{This is very similar to the factorization of the
  source distribution $W[\rho]$ in the calculation of inclusive
  observables. The independence with respect to the cutoff $y_{\rm
    cut}$ is possible because the static sources describe the same
  physics as the gauge fields in a certain range of longitudinal
  momentum.}.

\begin{figure}[htbp]
\begin{center}
\resizebox*{8cm}{!}{\includegraphics{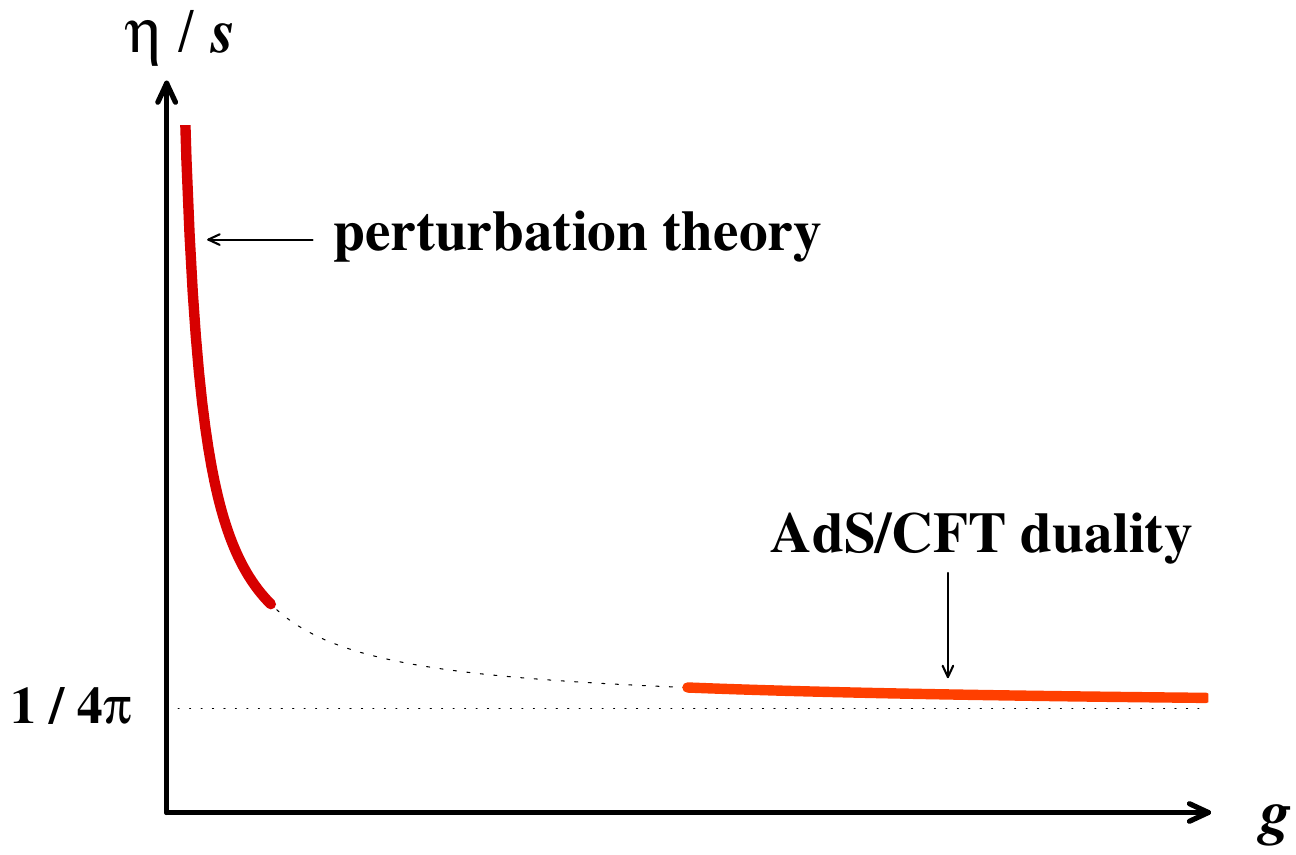}}
\end{center}
\caption{\label{fig:visco}Shear viscosity to entropy ratio in a gauge
  theory as a function of the coupling. }
\end{figure}

The typical behavior of the ratio $\eta/s$ in a gauge theory in
equilibrium is shown in the figure \ref{fig:visco}.
When the coupling is small, this ratio can be calculated in a weak
coupling expansion. For QCD, it is given by
\begin{equation}
\frac{\eta}{s}\approx\frac{5.1}{{\colorb g^4}\ln\left(\frac{2.4}{g}\right)}\; .
\end{equation}
This formula shows that $\eta/s$ is large at weak coupling. At large
coupling, this quantity cannot be calculated in QCD, but there is an
exact result for a supersymmetric cousin of QCD, the ${\cal N}=4$ SUSY
Yang-Mills theory\footnote{This gauge theory is conformal, and is dual
  to a string theory in an $\mbox{AdS}_5\times\mbox{S}_5$
  background. The large coupling limit of the gauge theory corresponds
  to the weak coupling limit of the string theory, in which gravity
  becomes classical and reduces to Einstein's equations.}:
$\eta/s=1/(4\pi)$~\cite{PolicSS1,PolicSS2}. From this plot, it seems
that only strongly coupled systems can have a small $\eta/s$ ratio.
There is however another possibility to evade this
conclusion. Firstly, one should recall the kinetic interpretation of
the ratio $\eta/s$,
\begin{equation}
\frac{\eta}{s}\sim \frac{\mbox{mean free path}}{\mbox{de Broglie wavelength}}\; .
\end{equation}
In a system where the degrees of freedom have a typical momentum $Q$,
the de Broglie wavelength is of order $Q^{-1}$, while the inverse mean
free path is given by 
\begin{equation}
(\mbox{{\colorb mean free path}})^{-1}
\sim \underbrace{\vphantom{\int_\k}g^4 Q^{-2}}_{\mbox{\scriptsize cross section}}\times \underbrace{\int_{\k} \; f_\k}_{\mbox{\scriptsize density}} \underbrace{\vphantom{\int_\k}{\colord (1+f_\k)}}_{{\mbox{\colord\scriptsize Bose}}\atop{\mbox{\colord\scriptsize enhancement}}}\; .
\end{equation}
The factor $1+f_\k$ under the integral is not needed when discussing
dilute plasmas, but is important if the occupation number is large.
In the CGC, just after the collision of two heavy ions, one has
$f_\k\sim g^{-2}$ up to $k\sim Q$. Therefore, in such a system one has
$\eta/s\sim g^0$, which is much smaller than the perturbative result
in equilibrium. It seems therefore plausible that the strong color
fields produced in heavy ion collisions may flow, not because they are
strongly coupled but because they are highly occupied.

\subsection{Expansion and free streaming}
\begin{figure}[htbp]
\begin{center}
\resizebox*{7cm}{!}{\includegraphics{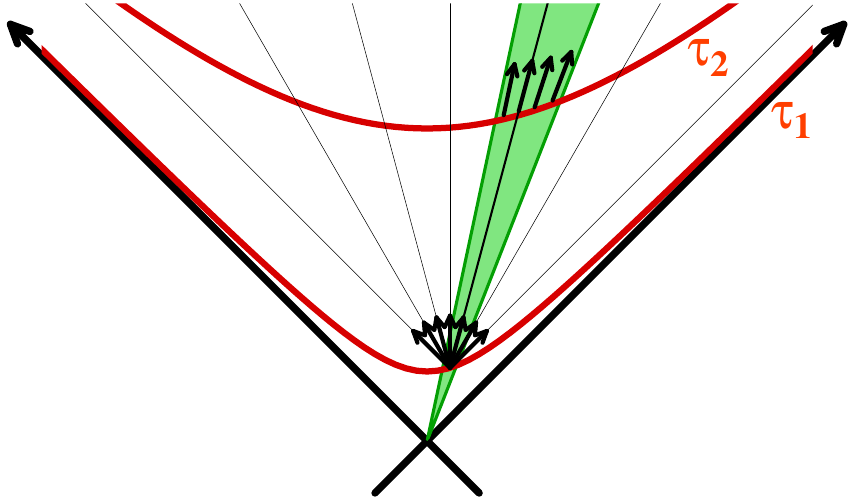}}
\end{center}
\caption{\label{fig:expansion}Role of the expansion in decreasing the
  longitudinal pressure.}
\end{figure}

The other main feature of heavy ion collisions is the very rapid
expansion of the system in the longitudinal direction, which causes a
redshifting of the longitudinal momenta. As illustrated in the figure
\ref{fig:expansion}, the pressure tensor tends to become anisotropic
because of this, unless the interactions are strong enough to overcome
the expansion.
The figure shows what the expansion does on non-interacting
particles. Starting from a nearly isotropic distribution of the
velocities at the time $\tau_1$, the expansion will ``filter out'' the
particles so that at the time $\tau_2$ only particles with the
momentum rapidity $y\approx \eta$ exist at the space-time rapidity
$\eta$. This means that, in the local comoving frame, the longitudinal
pressure is much smaller than the transverse pressure. A large pressure
anisotropy is a problem for hydrodynamics. Indeed, the difference
between the longitudinal and transverse pressures goes into the
viscous terms\footnote{There are now attempts to view hydrodynamics as
  an expansion around a non-isotropic background. In this formulation,
  this may be less of a
  problem~\cite{MartiS2,MartiRS1,FlorkMRS1,FlorkRS1,BazowHS1,Stric1,NopouRS1}.},
and large viscous corrections are a sign that hydrodynamics may
incorrectly reproduce the underlying dynamics.

\begin{figure}[htbp]
\begin{center}
\resizebox*{8cm}{!}{\includegraphics{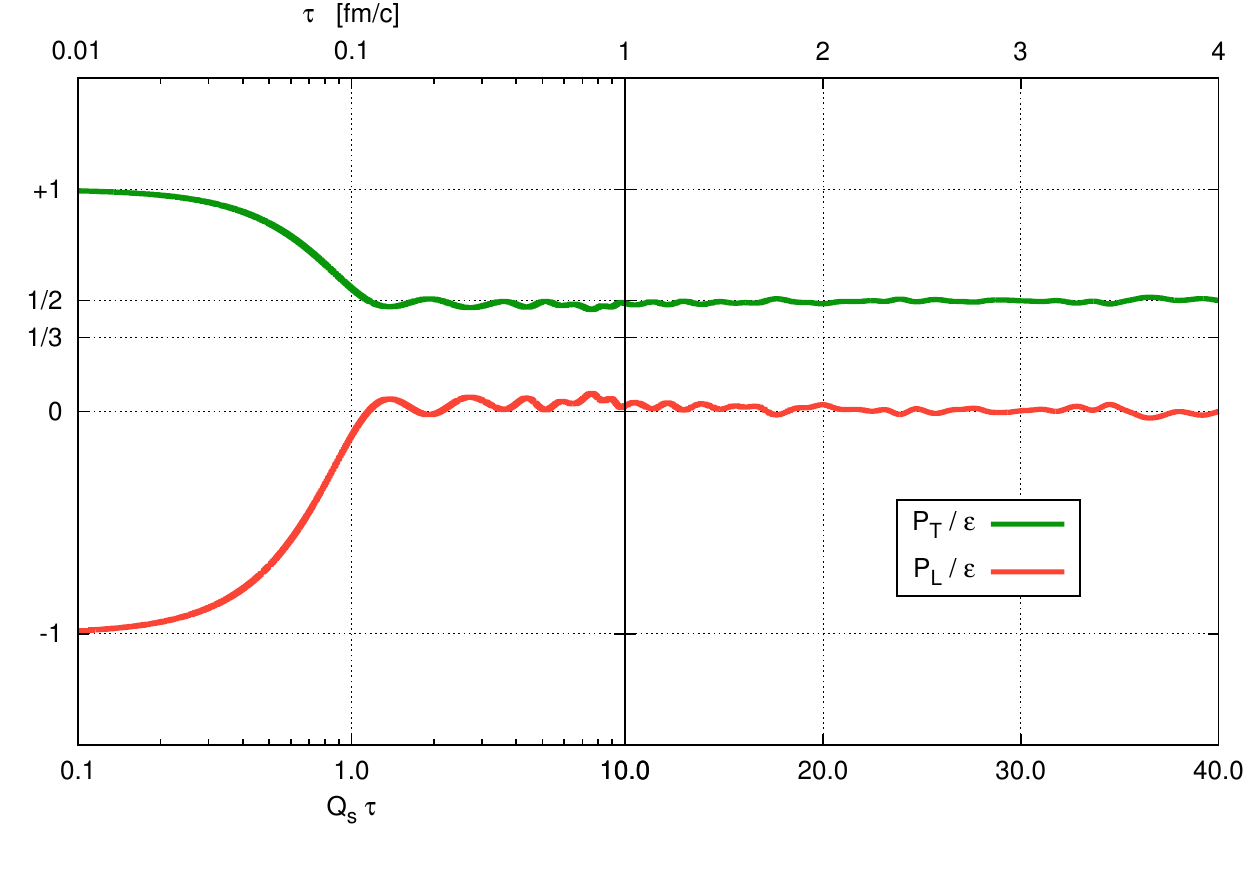}}
\end{center}
\caption{\label{fig:CGC-LO}Transverse and longitudinal pressure to
  energy density ratios, in the CGC at leading order.}
\end{figure}

Let us now describe the CGC prediction for the energy momentum tensor,
starting with the LO calculation\footnote{This discussion also applies
  to the LO result improved by the resummation of the leading log
  corrections. This does not change the conclusion of this paragraph
  since this resummation is totally absorbed into the rapidity
  evolution of the distributions $W[\rho]$.}. Immediately after the
collision, at $\tau=0^+$, it is known analytically that the
chromo-electric and chromo-magnetic fields are both parallel to the
collision axis. This peculiar structure of the color fields implies that the
energy momentum tensor is diagonal, $T^{\mu\nu}={\rm
  diag}(\epsilon,P_{_T},P_{_T},P_{_L})$, with\footnote{The negative
  longitudinal pressure in a longitudinal flux tube is the analogue of
  a string tension.} $P_{_T}=\epsilon$ and $P_{_L}=-\epsilon$. At
later times, the energy-momentum tensor must be determined numerically
by solving the classical Yang-Mills equations, and by computing
$T^{\mu\nu}$ from the classical gauge fields according to
eqs.~(\ref{eq:Tclass}). The results of this
calculation~\cite{LappiM1,FukusG1} are shown in the figure
\ref{fig:CGC-LO}.
After starting at $-1$, the ratio $P_{_L}/\epsilon$ increases and
becomes mostly positive at a time of order $Q_s\tau \sim 1$. However,
this calculation shows that the longitudinal pressure remains at all
times much smaller than the transverse one. Thus, the CGC at leading
order leads to a situation which is similar to free streaming, where
$\epsilon,P_{_T}\sim \tau^{-1}$ and $P_{_L}\ll \tau^{-1}$.
\begin{figure}[htbp]
\begin{center}
\resizebox*{7cm}{!}{\includegraphics{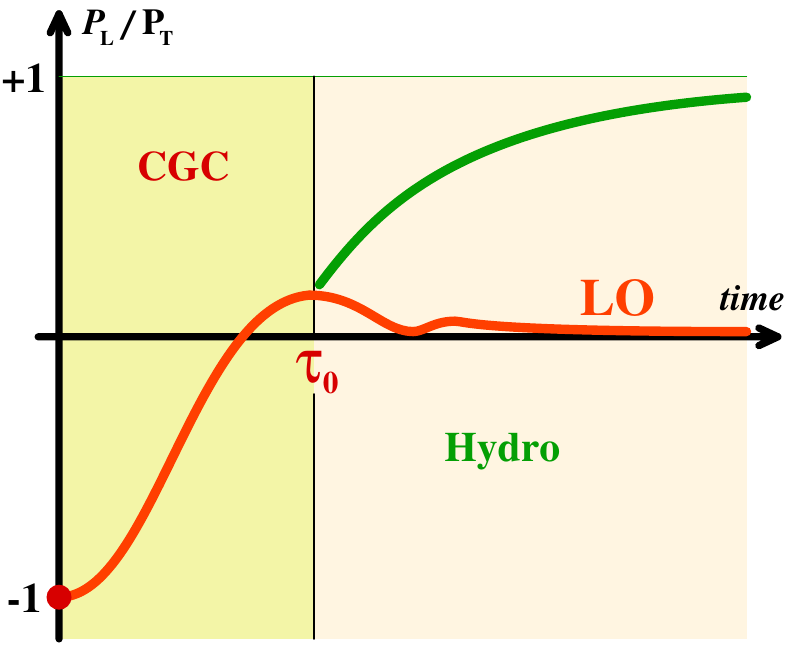}}
\end{center}
\caption{\label{fig:matching-LO}Matching between hydrodynamics and the CGC at LO.}
\end{figure}
This leads to an unsatisfactory matching between the CGC at LO and
hydrodynamics\footnote{In principle, this matching requires several
  steps: (1) compute $T^{\mu\nu}(x)$ from CGC, (2) find its time-like
  eigenvector such that $u_\mu T^{\mu\nu}(x)=\epsilon u^\nu$ (this
  defines the local rest frame, and the local energy density), (3)
  compute the pressure from some equation of state $P=f(\epsilon)$,
  (4) compute the viscous stress tensor as the difference between the
  full $T^{\mu\nu}$ and the ideal part (obtained from $\epsilon, P$
  and $u^\mu$). In many calculations using ``CGC initial conditions'',
  a simplified procedure is often employed, where one assumes that
  $u^\mu=(1,{\bs 0})$: (1) compute $\epsilon=T^{00}$, (2) define
  $P=f(\epsilon)$, (3) neglect the viscous stress tensor.}, as
illustrated in the figure \ref{fig:matching-LO}. Indeed, in
hydrodynamics the ratio $P_{_L}/P_{_T}$ increases to approach $1$
while it remains near zero in the CGC at LO.

\subsection{A simpler model study case: thermalization in a box}
Before presenting recent works on isotropization and
thermalization in heavy ion collisions, it is interesting to pause a
brief moment on the same question for a hypothetical system of gluons
enclosed in a fixed volume. In this case, the ultimate outcome is
completely clear from the start: the system will eventually
thermalize, and since its total energy is conserved one can predict
from the start what will be the equilibrium temperature. The only
issues are the timescale of the thermalization process, its possible
dependence on the nature of the initial condition, and the shape of
the gluon distribution at intermediate times before thermalization is
complete.

A particularly interesting class of initial conditions are the so
called ``overoccupied'' initial distributions, because they are
also realized in heavy ion collisions in the CGC framework.  The typical
CGC-like initial condition has modes occupied up to the saturation
momentum $Q_s$, with a large occupation number of order $g^{-2}$. It
is called overoccupied because it contains too many particles for its
energy density:
\begin{equation}
n\sim \frac{Q_s^3}{g^2}\quad,\quad\epsilon\sim\frac{Q_s^4}{g^2}\quad,\quad
n\epsilon^{-3/4}\sim g^{-1/2}\gg 1\; ,
\end{equation}
while the dimensionless combination $n\epsilon^{-3/4}$ should be of
order 1 in a system with the same energy density in equilibrium. With
such an initial condition, the mean free path in the system is
parametrically shorter than the thermalization time, which leaves
ample time for the system to forget its initial conditions long before
reaching thermal equilibrium. With this type of initial condition, the
hard scale of the system grows with time according to the following law\cite{KurkeM1,BlaizGLMV1,YorkKLM1}
\begin{equation}
\Lambda_{_{\rm H}}\sim Q_s\, (t/t_0)^{1/7}\; ,
\end{equation}
and the distribution scales as
\begin{equation}
f(t,p)\sim (t_0/t)^{4/7} {\rm f}(p/\Lambda_{_{\rm H}})\; .
\end{equation}
The thermalization time can be estimated as the time at which the hard
scale reaches the expected equilibrium temperature (i.e. the fourth
root of the initial energy density), $T_{\rm eq}\sim Q_s
g^{-1/2}$. This gives a thermalization time that has the following
parametric dependence on the coupling,
\begin{equation}
Q_st_{\rm eq}\sim g^{-7/2}\; .
\end{equation}
In a  recent work\cite{KurkeL1} using the effective kinetic
theory\footnote{This effective theory includes $2\to 2$ processes
  dressed by in-medium masses, and effective $1\to 2$ and $2\to 1$
  processes due to the quasi-collinear splitting of hard gluons by
  bremsstrahlung. For the latter, one must resum multiple scatterings,
  due to the Landau Pomeranchuk Migdal effect.}  of
ref.~\cite{ArnolMY5}, this estimate was made more quantitative and the
numerical prefactors computed to give
\begin{equation}
t_{\rm eq}\approx \frac{72}{1+0.12\log(\lambda^{-1})}\frac{1}{\lambda^2 T_{\rm eq}}\; ,
\end{equation}
where $\lambda\equiv g^2 N_c$.

This type of overoccupied initial condition has also led to
speculations about the possibility of forming a Bose-Einstein
condensate (BEC) in such systems, that would accommodate the excess of
particles~\cite{BlaizGLMV1}. Such a condensate can only be transient,
because the number of particles is not conserved in a relativistic
quantum field theory. Therefore, the true equilibrium state cannot
have a BEC.  Whether such a condensate can form as a transient
phenomenon depends on the magnitude of the initial overoccupation and
the rate of the number changing processes. In a scalar $\phi^4$
theory, where the rate of the inelastic processes is quite small
compared to the elastic one, such a condensate has been observed in a
number of numerical simulations\cite{EpelbG1,BergeS4}. The situation
is much less clear in QCD, because the inelastic rate is enhanced by
soft and collinear divergences. Unsurprisingly, kinetic theory
calculations (based on a small scattering angle approximation of the
matrix element) neglecting the inelastic processes do observe the
formation of a BEC \cite{BlaizLM2,ScardPPRG1,XuZZG1}. Extensions of
these QCD kinetic theory computations to include inelastic
processes\cite{HuangL1} suggest that a BEC may still appear after
including number changing processes, while other computations do not
see any evidence for it\cite{YorkKLM1,KurkeL1}.

Note that for initially underoccupied\footnote{In the bottom up
  scenario of ref.~\cite{BaierMSS2} (see also ref.~\cite{KurkeM2}), it
  has been argued that in heavy ion collisions the expansion may turn
  a CGC-like initial condition into an underoccupied distribution
  before full thermalization is reached.} initial conditions, the
evolution towards equilibrium is quite different. In particular, it is
a lot less universal because the mean free path associated with the
initial distribution is comparable to or longer than its value at
equilibrium. In this situation, the initial hard particles first
create a bath of soft particles by bremsstrahlung radiation, which
thermalizes quickly. Then the remaining hard particles lose their
energy by successive $1\to 2$ splittings induced by collisions on the
soft background.

\section{Weibel instability and resummation}
\label{sec:weibel}
\subsection{Unstable classical solutions}
\begin{figure}[htbp]
\begin{center}
\resizebox*{8cm}{!}{\rotatebox{-90}{\includegraphics{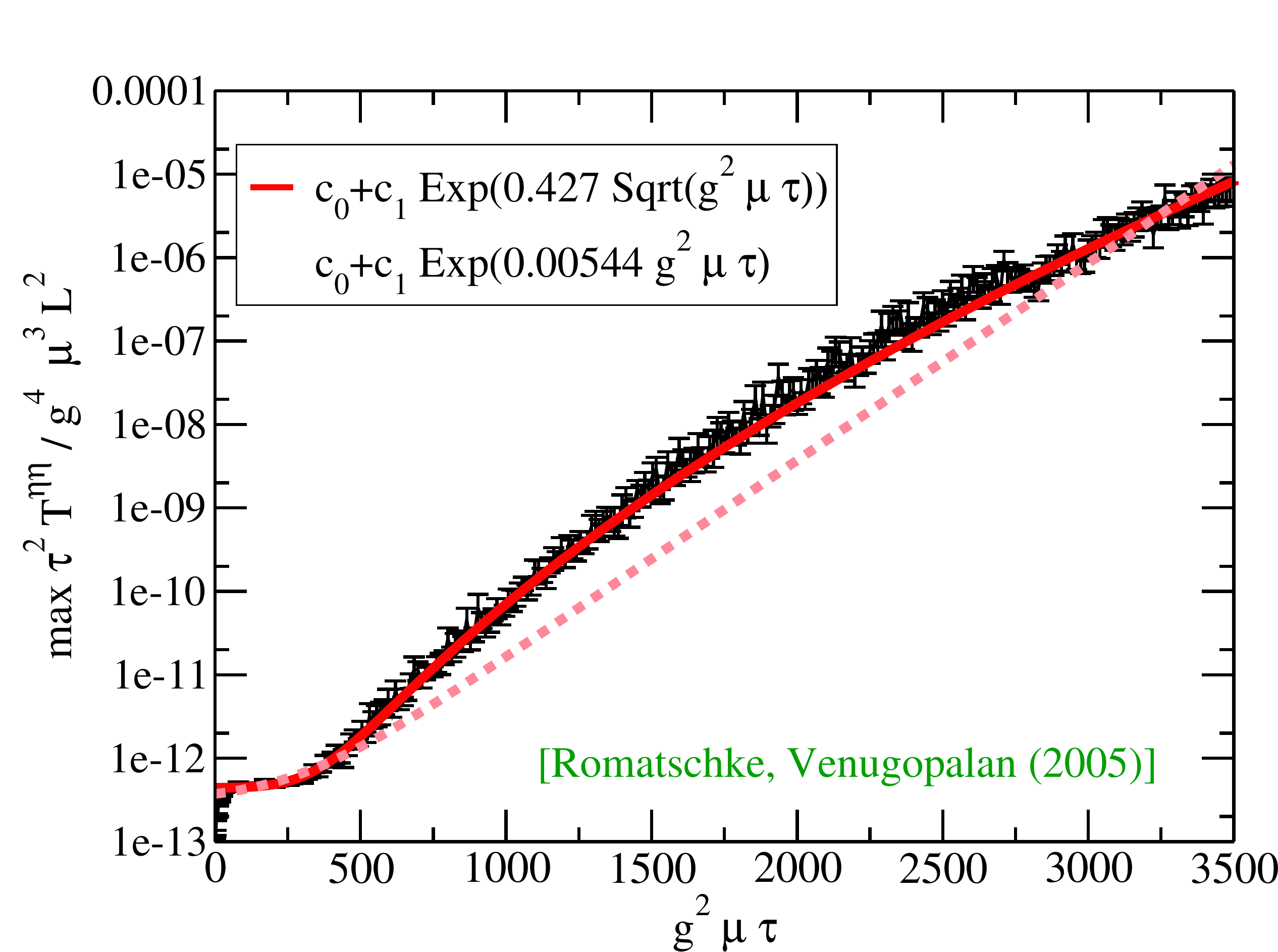}}}
\end{center}
\caption{\label{fig:insta}Growth of unstable modes in classical Yang-Mills dynamics.}
\end{figure}

\setcounter{footnote}{0} There is however a good reason to explore the
CGC beyond leading order: namely, the fact that the boost invariant
solutions of the classical Yang-Mills equations are
unstable~\cite{RomatV1,RomatV2,RomatV3,FukusG1,BiroGMT1,HeinzHLMM1,BolteMS1,FujiiI1,FujiiII1,KunihMOST1}. When
their initial condition is modified by a small but rapidity dependent
perturbation, the solution diverges from the unperturbed one. This is
illustrated in the figure \ref{fig:insta}, that shows a component of
the energy momentum tensor that should be very small at all times if
the perturbation was stable. Instead, it growths like an
exponential\footnote{The fact that $\sqrt{\tau}$ appears here instead
  of $\tau$ itself is due to the longitudinal expansion of the
  system. Because of the expansion, the equation that drives the
  growth of the perturbations is a Bessel equation instead of an
  ordinary wave equation.} $\exp\sqrt{\mu\tau}$ (the characteristic
growth rate $\mu$ is of the order of the saturation momentum $Q_s$).
These unstable modes in the classical Yang-Mills equations are closely
related to the Weibel instability that occurs in anisotropic plasmas
in QED and in
QCD~\cite{Mrowc3,Mrowc4,RebhaRS1,RebhaRS2,MrowcRS1,RomatS1,RomatS2,RebhaS1,RebhaSA1,ArnolLM1,ArnolLMY1,ArnolM1,ArnolM2,ArnolM3,ArnolMY4,DumitNS1,BodekR1,BergeGSS1,BergeSS2,KurkeM1,KurkeM2,AttemRS1}. More
details on how these instabilities of the classical solutions develop
can also be found in refs.~\cite{FujiiI1,FujiiII1,Fukus2,FukusG1}.

As we shall see in the next subsection, these instabilities are
disastrous for the power counting that we have exposed earlier, where
one keeps track only of the powers of $g^2$. Indeed, terms that have a
higher order in $g^2$ because they arise at a higher loop order may in
fact contain time dependent factors that increase exponentially with
time. These {\sl secular divergences} mean that fixed loop order
calculations are most likely unreliable (we will show an example of
this in the next subsection), and that the power counting should be
revisited and improved in order to capture the most important among
those terms. 

On the other hand, the existence of instabilities in the classical
solutions of the Yang-Mills equations may play a very important role
in the statistical equilibration of the system. If we consider this
system from the point of view of its classical phase-space, its
initial condition at $\tau=0^+$ occupies a very compact region in
phase-space. Indeed, at LO, the support of its Wigner distribution is
a single point, since the LO is purely classical. At NLO, this support
is broadened into a region of extension $\hbar$. If the dynamics was
completely stable (as in an integrable system for instance, where the
number of conservation laws equals the number of degrees of freedom),
the size of this support would remain roughly constant in time, and at
any subsequent time the system would still be far from statistical
equilibrium.  In contrast, an unstable dynamics will map this
initially compact support into the extended portion of the phase-space
allowed by the few conservation laws that remain valid. Measurements
performed after such an evolution should bring results that are more
in line with microcanonical equilibrium.

\subsection{Pathologies at fixed loop order}
These instabilities force us to reconsider the power counting that was
the basis for organizing the expansion in powers of $g^2$ of inclusive
observables. As we have seen before, the one-loop corrections --that
are formally of relative order $g^2$-- contain leading log terms
proportional to the cutoff $y_{\rm cut}$, that can be absorbed into a
redefinition of the distributions $W[\rho]$.  Because of the
instabilities, the 1-loop correction also contain some terms that grow
exponentially in time.  The best place to see this is via the following
expression for the 2-point function ${\bs\Gamma}_2(\u,\v)$ that enters
in the formula (\ref{eq:fact0}),
\begin{equation}
{\bs\Gamma}_2(\u,\v)
=
\int\frac{d^3\k}{(2\pi)^3 2\omega_\k}\;a_\k(\tau_0,\u)a_\k^*(\tau_0,\v)\; ,
\label{eq:gamma2}
\end{equation}
($\tau_0$ is the initial time surface on which eq.~(\ref{eq:fact0}) is
expressed) where the functions $a_\k(\tau,\x)$ are small perturbations
around the classical field encountered at leading order (for
Yang-Mills theory, these functions would also carry color, spin and
Lorentz indices not written explicitly in eq.~(\ref{eq:gamma2})). They
obey the equation
\begin{eqnarray}
&&\Big[{\cal D}_\rho {\cal D}^\rho \delta^\nu_\mu - {\cal D}_\mu {\cal D}^\nu
+ig\, {\cal F}_\mu{}^\nu\Big] {\colorb a_\k^\mu}=0\nonumber\\
&&{\colord\lim_{x^0\to-\infty}} {\colorb a_\k^\mu(x)} = \epsilon^\mu(\k)\;e^{ik\cdot x}
\; ,
\label{eq:pert}
\end{eqnarray}
in which the covariant derivatives contain the LO classical field. The
plane wave initial condition used in eq.~(\ref{eq:pert}) for these
small perturbations can be understood from the explicit derivation of
the formula (\ref{eq:fact0}).

The existence of instabilities means that, for some range of $\k$, the
mode functions $a_\k$ grow exponentially with time. The naive power
counting derived so far implicitly assumed that the function
${\bs\Gamma}_2$ is of order $g^0$, but we never looked at its time
dependence. From eq.~(\ref{eq:gamma2}), it is now obvious that it will
become very large if some of the $a_\k$'s are unstable.  Eventually,
these exponential factors in ${\bs\Gamma}_2$ will compensate the $g^2$
that comes from the loop, and these terms will be as large as the LO
terms.
\begin{figure}[htbp]
\begin{center}
\resizebox*{6.3cm}{!}{\rotatebox{-90}{\includegraphics{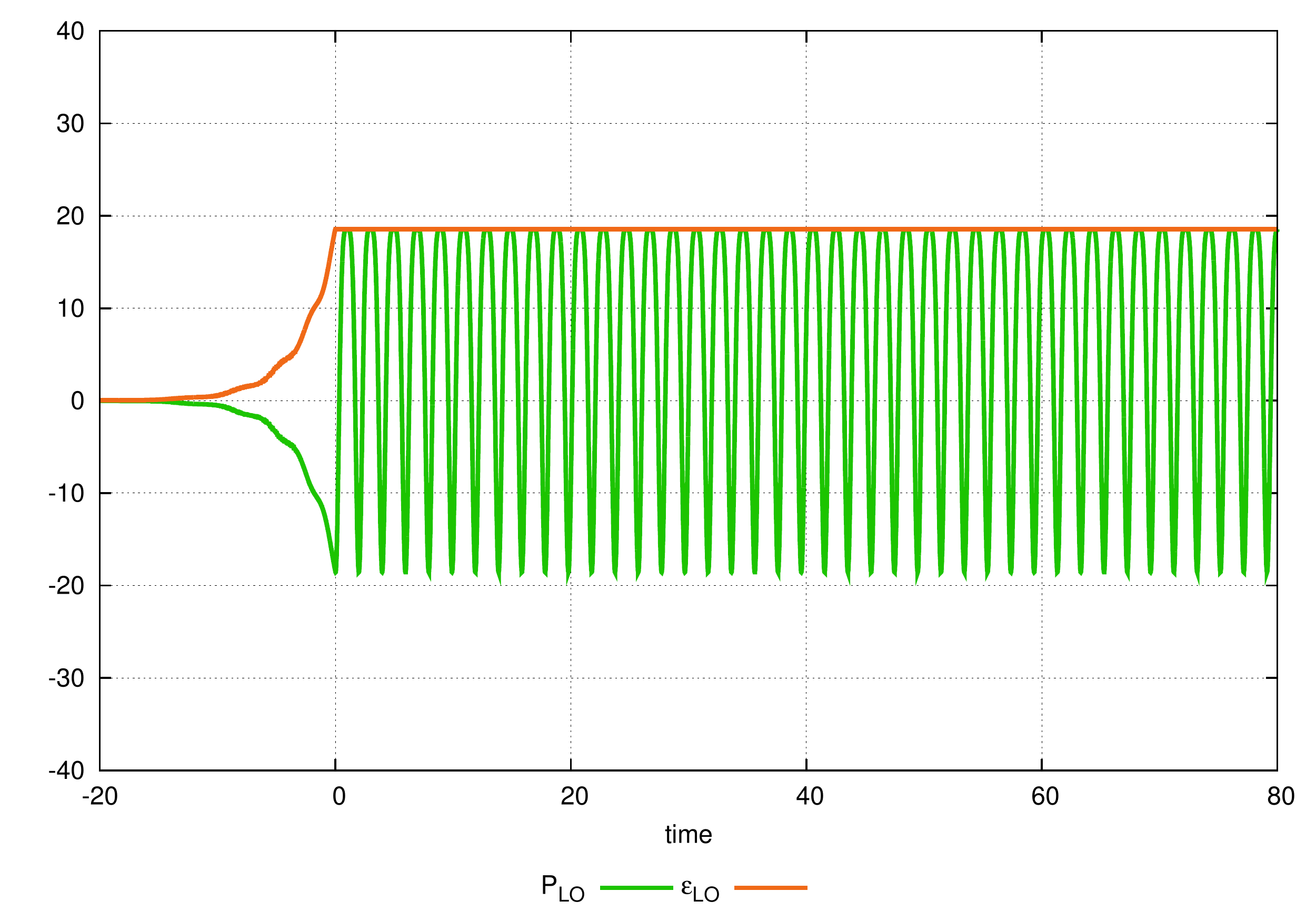}}}\hfill
\resizebox*{6.3cm}{!}{\rotatebox{-90}{\includegraphics{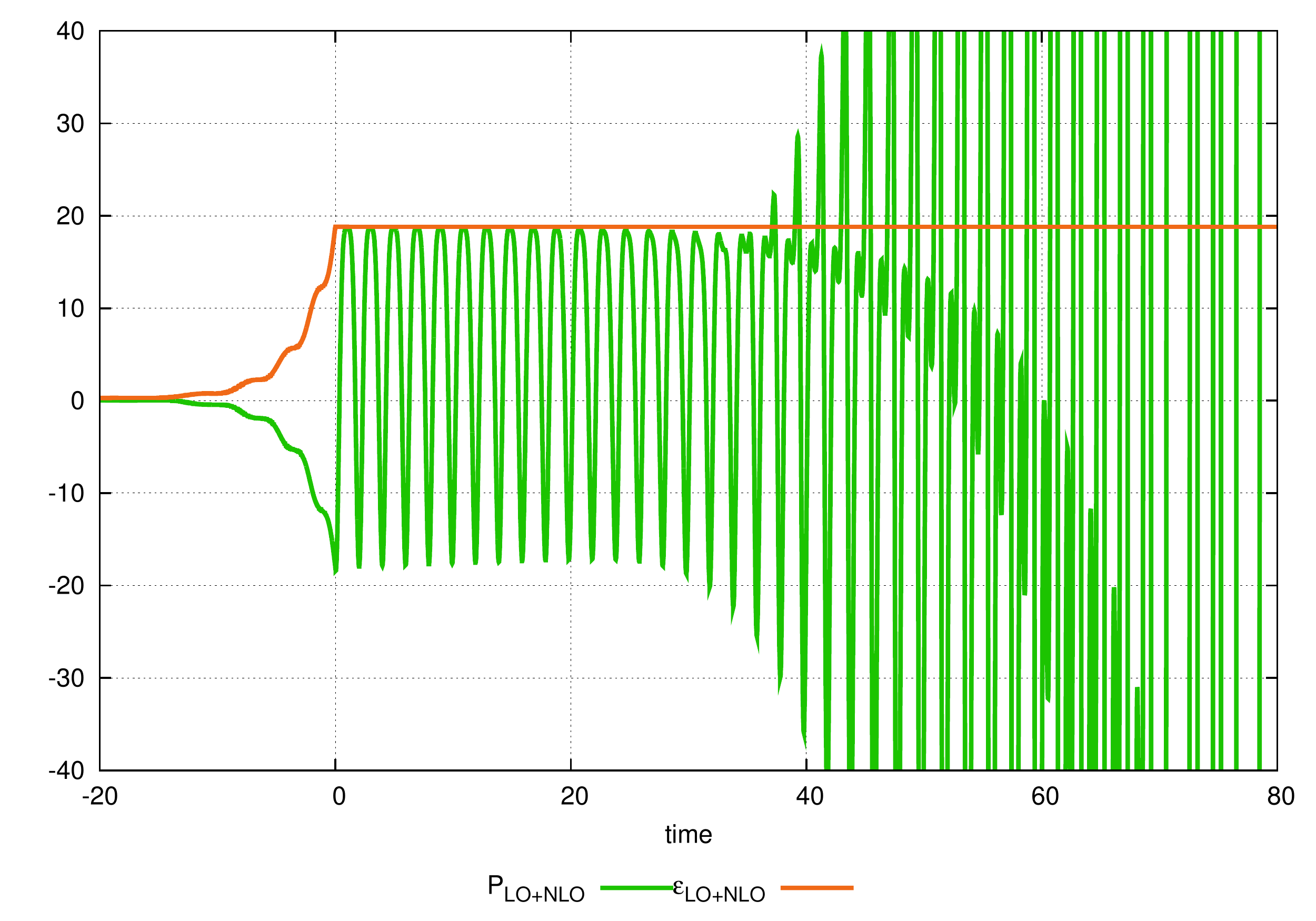}}}
\end{center}
\caption{\label{fig:phi4}Effect of parametric instabilities on the
  perturbative expansion of the energy-momentum tensor in a scalar
  $\phi^4$ theory.}
\end{figure}
This statement is illustrated in the figure \ref{fig:phi4}, where we
compare the energy density and pressure at LO and LO+NLO in a $\phi^4$
scalar field theory~\cite{DusliEGV1}, for which the solutions of the
classical field equations of motion are also unstable\footnote{The
  instability in the $\phi^4$ theory is of a totally different nature,
  since it is caused by parametric resonance~\cite{GreenKLS1}. However, the detailed
  mechanism of the instability is not important in this
  discussion.}. This computation shows clearly that the fixed order
LO+NLO result cannot be trusted after some time, because it becomes
much larger than the LO result (note that this happens only for the
pressure, because the energy density is protected from this
exponential growth by energy conservation).

Although a similar NLO calculation has not been done in the CGC
framework, one can guess what would happen. The Weibel instabilities
would produce an unbound growth of the longitudinal pressure, and this
time the ratio $P_{_L}/P_{_T}$ would be driven to arbitrarily large
values exceeding unity.
\begin{figure}[htbp]
\begin{center}
\resizebox*{7cm}{!}{\includegraphics{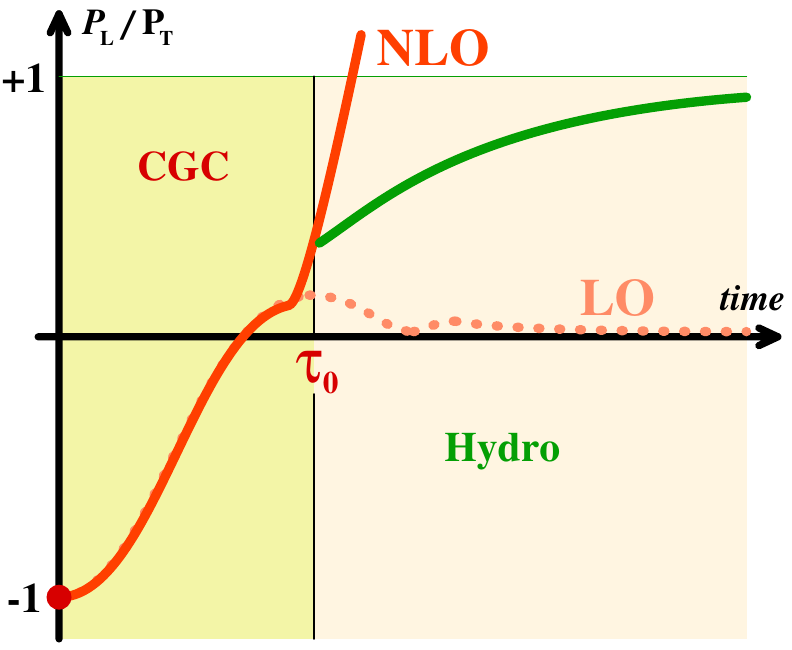}}
\end{center}
\caption{\label{fig:matching-NLO}Matching between hydrodynamics and the CGC at NLO.}
\end{figure}
Attempting to match such a NLO CGC initial condition to hydrodynamics
would not be better than at LO, as illustrated in the figure
\ref{fig:matching-NLO}.

\subsection{Resummation of the leading terms}
\begin{figure}[htbp]
\begin{center}
\resizebox*{5cm}{!}{\includegraphics{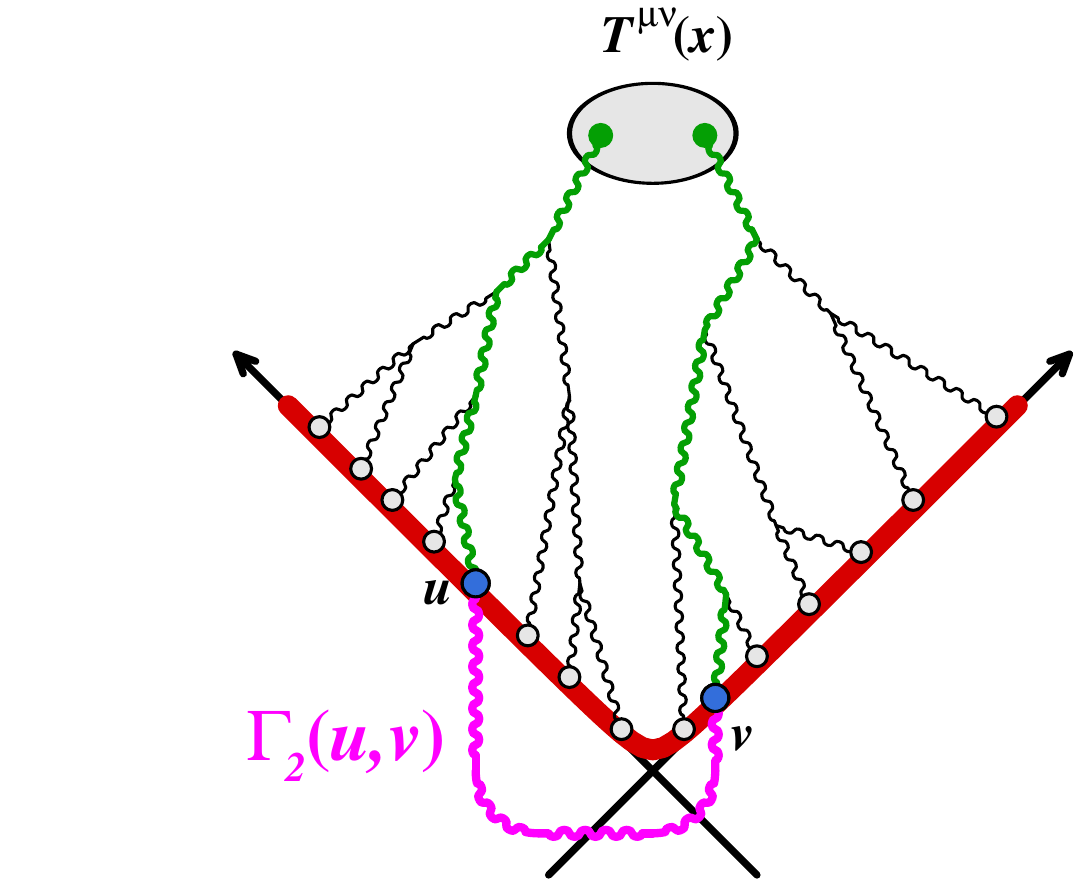}}\hfill
\resizebox*{5cm}{!}{\includegraphics{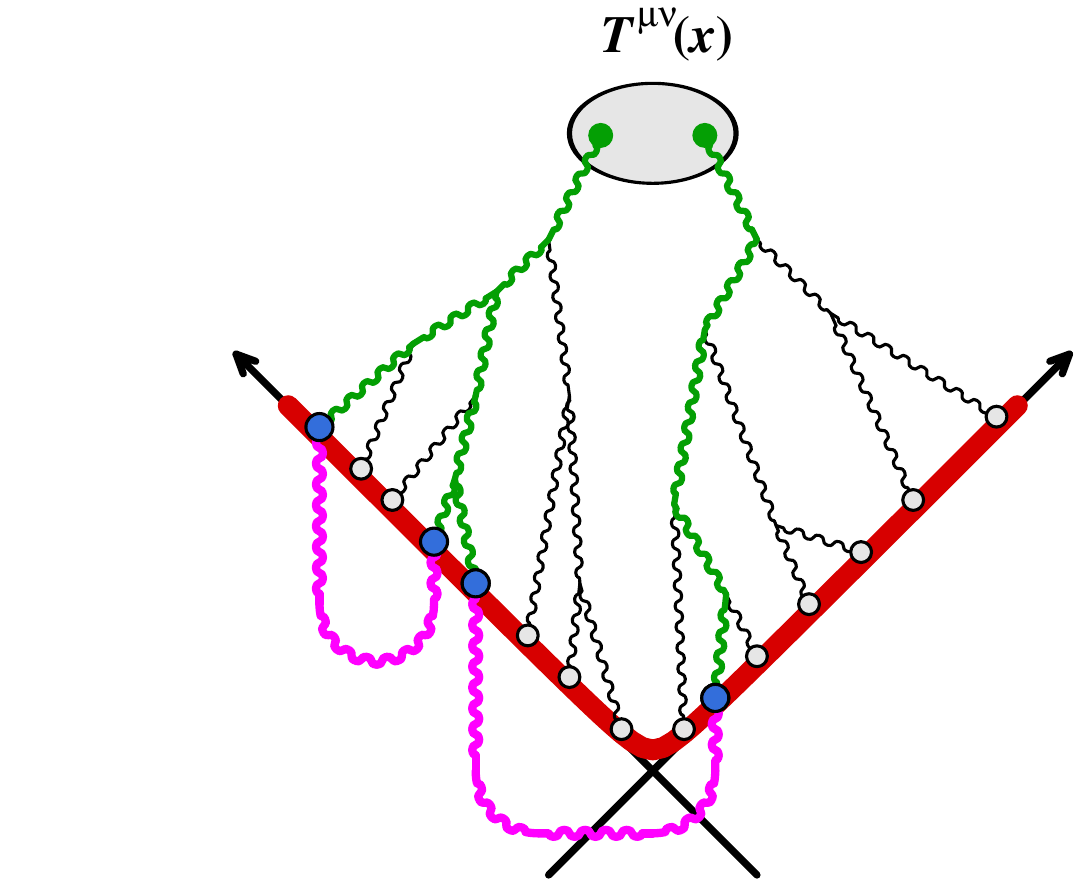}}\hfill
\resizebox*{5cm}{!}{\includegraphics{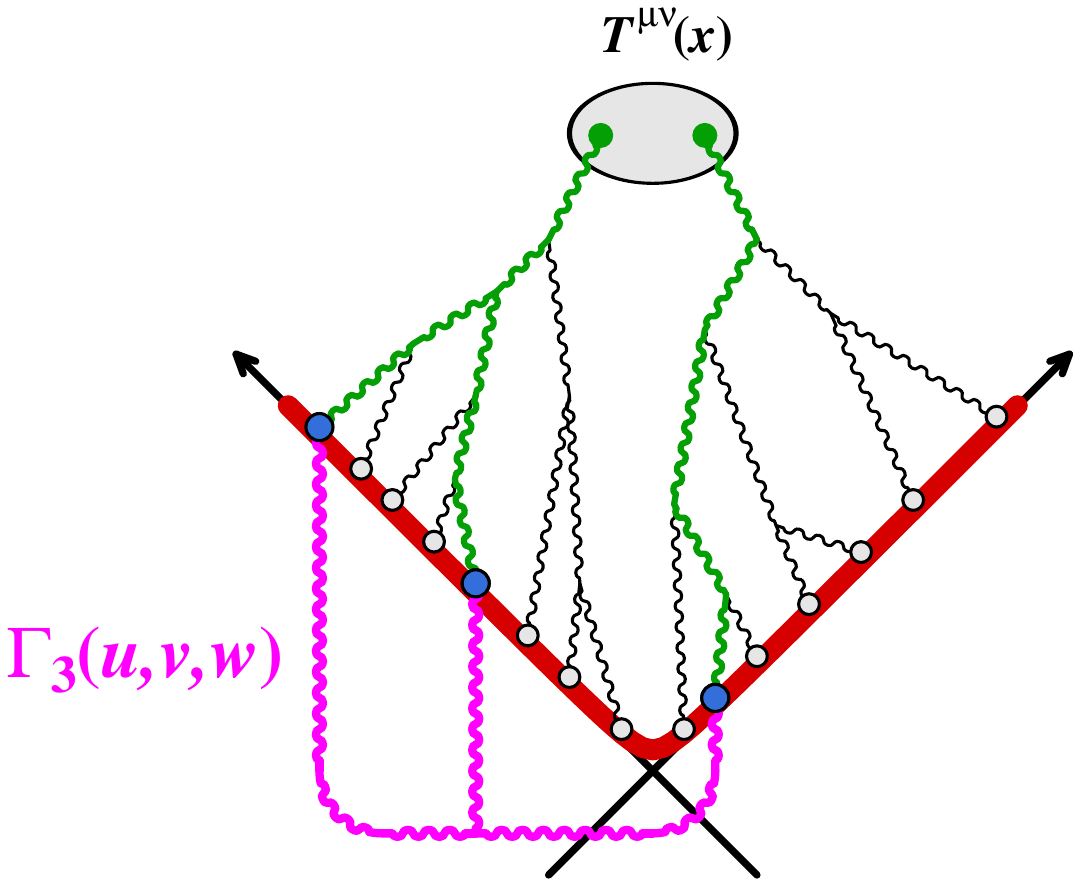}}
\end{center}
\caption{\label{fig:resum}Improved power counting taking into account
  the growth of the unstable modes.}
\end{figure}

The same problem would in fact occur at any fixed loop order. The only way to
improve the situation is to examine each loop order and to keep the
most important terms at each order. For this, we need first to modify
the power counting rules that we have established earlier, in order to
keep track of the unstable modes. Let us first examine the graph that
contributes at 1-loop, represented in the top-left corner of the
figure \ref{fig:resum}, in conjunction with the formula (\ref{eq:fact0}).
In this formula, each of the derivatives with
respect to the initial classical field creates a perturbation to this
classical field, that we have indicated by green propagators in the
figure \ref{fig:resum} (in the top-left graph, only the term
proportional to ${\bs\Gamma}_2$, that has second derivatives with respect
to the initial fields, has been represented -- the term with only one
derivative has a slower growth with time). The ``standard'' power
counting\footnote{For the term in ${\bs\alpha}{\mathbbm T}$ in
  eq.~(\ref{eq:fact0}), the standard counting has a factor $g$ from
  ${\bs\alpha}$ and a factor $g$ from the operator ${\mathbbm
    T}$. However, we will not need to consider this term further since
  it has a subleading growth in time.} would assign a factor 1 to
${\bs\Gamma}_2$ and a factor $g$ to each of the derivatives ${\mathbbm
  T}_\u$ (represented by a blue dot in the graphs). Thus the NLO
correction to the energy-momentum tensor is expected to be of order
$g^0$, while the LO is of order $g^{-2}$.  From this diagrammatic
representation, it also easy to count the number of perturbations of
the classical field. Each of them will develop into a factor of order
$\exp(\sqrt{\mu\tau})$ ($\mu$ is of order $Q_s$). Thus, the expansion
of $T^{\mu\nu}$ is more accurately written\footnote{This formula
  indicates the worst possible behavior. It is possible that some
  components of the energy-momentum tensor will not be affected by the
  instability, as was the case in the scalar field theory considered
  in the figure \ref{fig:phi4}.}  as~:
\begin{equation}
T^{\mu\nu}= c_0\,g^{-2}+c_1\,g^0\,e^{2\sqrt{\mu\tau}}+\cdots\; ,
\end{equation}
where the coefficients $c_0$, $c_1$ do not grow exponentially with
time. From this pocket formula, one can deduce at which time the naive
loop expansion breaks down. This is the time when the one-loop result
becomes as large as the leading order, i.e.
\begin{equation}
\tau_{\rm max}\sim \mu^{-1}\log^2(1/g^2)\; .
\end{equation}
Up to a logarithmic factor, this time is of the order of the inverse
saturation momentum.

At two loops, the naive power counting tells us that we should get
terms of order $g^2$ (i.e. $g^4$ relative to the leading
order). However, not all the terms have the same growth in time. In
the figure \ref{fig:resum}, we have represented two types of 2-loop
contributions, in order to illustrate these differences. In the
top-right graph, the two loops are the seed of four perturbations to
the classical field, while in the bottom graph, only three of these
perturbations are created. The latter term will therefore have a
subleading behavior in time. Moreover, one sees that the
distinguishing feature of the top-right graph is that it can be
generated by acting twice with the quadratic part of the operator that
appears in eq.~(\ref{eq:fact0}). This reasoning extends to all
orders. At the $n$-th loop order, the maximal number of field
perturbations that can be seeded on the light-cone is $2n$, and the
corresponding graphs are generated by acting $n$ times with this
quadratic operator. The sum of all these leading terms can
be obtained by
\begin{equation}
T^{\mu\nu}_{\rm resummed}
\equiv
\exp\Bigg[\frac{1}{2}
\int\limits_{\u,\v}{\bs\Gamma}_2(\u,\v)
{\mathbbm T}_\u
{\mathbbm T}_\v
\Bigg]\;\;T^{\mu\nu}_{_{\rm LO}}\; .
\label{eq:resum}
\end{equation}
Note that the Taylor coefficients of the exponential correspond
precisely to the symmetry factors of graphs such as the top-right
diagram of the figure \ref{fig:resum} (the $1/2!$ of the second Taylor
coefficient gives the symmetry factor that corresponds to the freedom
of swapping the two ${\bs\Gamma}_2$'s that are hanging below the light
cone). 

Moreover, if the 2-point function ${\colorb\bs{\Gamma}_2}$ used in
eq.~(\ref{eq:resum}) is precisely the one that enters in the NLO
result (\ref{eq:fact0}), then one has
\begin{equation}
  T^{\mu\nu}_{\rm resummed}=T^{\mu\nu}_{_{\rm LO}}+T^{\mu\nu}_{_{\rm
      NLO}}+\cdots
\end{equation}
In other words, this resummation contains the exact LO and NLO
contributions, and a subset of all the higher loop contributions. It
is important to keep in mind that, starting at the 2-loop order, it is
an approximation which is not equivalent to the complete underlying
theory. This will have important consequences that we will discuss later.

\section{Classical statistical approximation (CSA)}
\label{sec:resum}
\subsection{Reformulation as a Gaussian average}
At this point, the resummation performed via the eq.~(\ref{eq:resum})
is quite formal. Three questions must be addressed: (1) can this formula
in terms of functional derivatives be evaluated in a practical way?
(2) does eq.~(\ref{eq:resum}) lead to results whose time dependence is
bounded?  (3) when doing this, are we introducing other pathologies?

In order to answer the first question, one should recall the following
identity:
\begin{equation}
    e^{\frac{{\colord\alpha}}{2}\partial_x^2}\;{\colorc f(x)}
  =
  \int\limits_{-\infty}^{+\infty}{\colora dz}\;
  \frac{e^{-{\colora z^2}/2{\colord\alpha}}}{\sqrt{2\pi\alpha}}\;
  {\colorc f(x+{\colora z})}\; .
\end{equation}
This formula can be established e.g. by applying a Fourier transform
to both sides. Although we cast it here in a space of functions of a
single variable, this formula can be generalized to operators that are
Gaussian in derivatives over a functional space. It enables us to
rewrite eq.~(\ref{eq:resum}) as\footnote{Such a Gaussian averaging
  procedure has also been reached in other
  approaches~\cite{PolarS1,Son1,KhlebT1,MichaT1,FukusGM1}.}
\begin{eqnarray}
{\colorb T^{\mu\nu}_{_{\rm resummed}}}
=
\int[D{\colord a}]\,
\exp\Bigg[-\frac{1}{2}
\int\limits_{_{\u,\v}}
{\colord a(\u)}{\colorb{\bs\Gamma}^{-1}_{2}(\u,\v)}{\colord a(\v)}\Bigg]
\;
T^{\mu\nu}_{_{\rm LO}}[{\cal A}_{\rm init}+{\colord a}]\; .
\label{eq:resum1}
\end{eqnarray}
This resummation procedure, where one averages classical trajectories
over an ensemble of initial conditions, is known as the {\sl Classical
Statistical Approximation} (CSA).

From this equation, one can easily see that the problem of the
unbounded growth of the fluctuations has been cured. Indeed, this
resummation has promoted the linearized perturbations\footnote{One
  would recover the pathological behavior of the NLO result by
  linearizing the equation of motion for the classical field of
  initial condition ${\cal A}_{\rm init}+a$.} that appear in the NLO
contribution into an integral part of the non-linear classical field
(the initial condition of the classical field is modified by the
perturbation, but its evolution remains fully non-linear). In any
theory where the potential prevents the fields from running away to
infinity, this guarantees that the resummed quantity will not diverge
in time.

\subsection{Practical implementation}
In the form of eq~(\ref{eq:resum1}), the procedure for evaluating the
resummed energy-momentum tensor\footnote{Although the discussion here
  is centered on the energy-momentum tensor, the same resummation can
  be applied to any inclusive quantity.} is quite straightforward:
\begin{itemize}
\item[{\bf 1.}] Determine the 2-point function
  ${\colorb{\bs\Gamma}_{2}(\u,\v)}$ that defines the Gaussian
  fluctuations, for the initial time $Q_s\tau_0$ of interest. This is
  an initial value problem, whose outcome is uniquely determined by
  the state of the system at $x^0=-\infty$, and depends on the history
  of the system from $x^0=-\infty$ to $\tau=\tau_0$. This problem is
  solvable analytically as long as the fluctuations remain weak,
  $a^\mu\ll Q_s/g$. If they grow larger, the fluctuations start to
  interact non-linearly, and their spectrum becomes non-Gaussian. To
  avoid this, the initial time should be chosen such that
  $Q_s\tau_0\ll 1$.
\item[{\bf 2.}] Solve the classical Yang-Mills equations from $\tau_0$
  to $\tau_f$. In high energy collisions, the problem as a whole is
  boost invariant, but individual field configurations are now
  rapidity dependent because of the fluctuating part superimposed to
  their initial condition. Therefore, unlike in the CGC at LO, the
  classical Yang-Mills equations must now be solved in 3+1 dimensions.
\item[{\bf 3.}] Do a Monte-Carlo sampling of the fluctuating initial
  conditions.
\end{itemize}

\begin{figure}[htbp]
\begin{center}
\resizebox*{4.5cm}{!}{\includegraphics{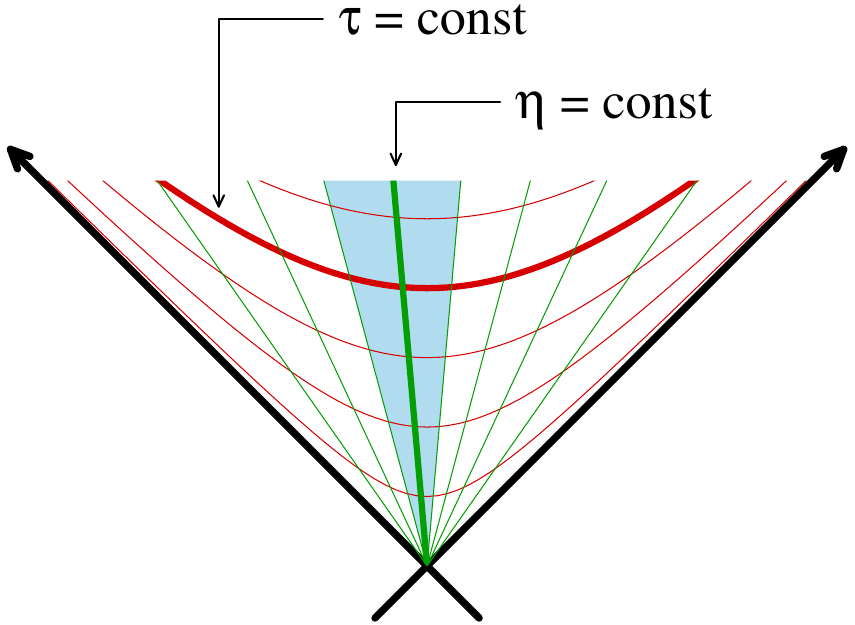}}\hfill
\resizebox*{8cm}{!}{\includegraphics{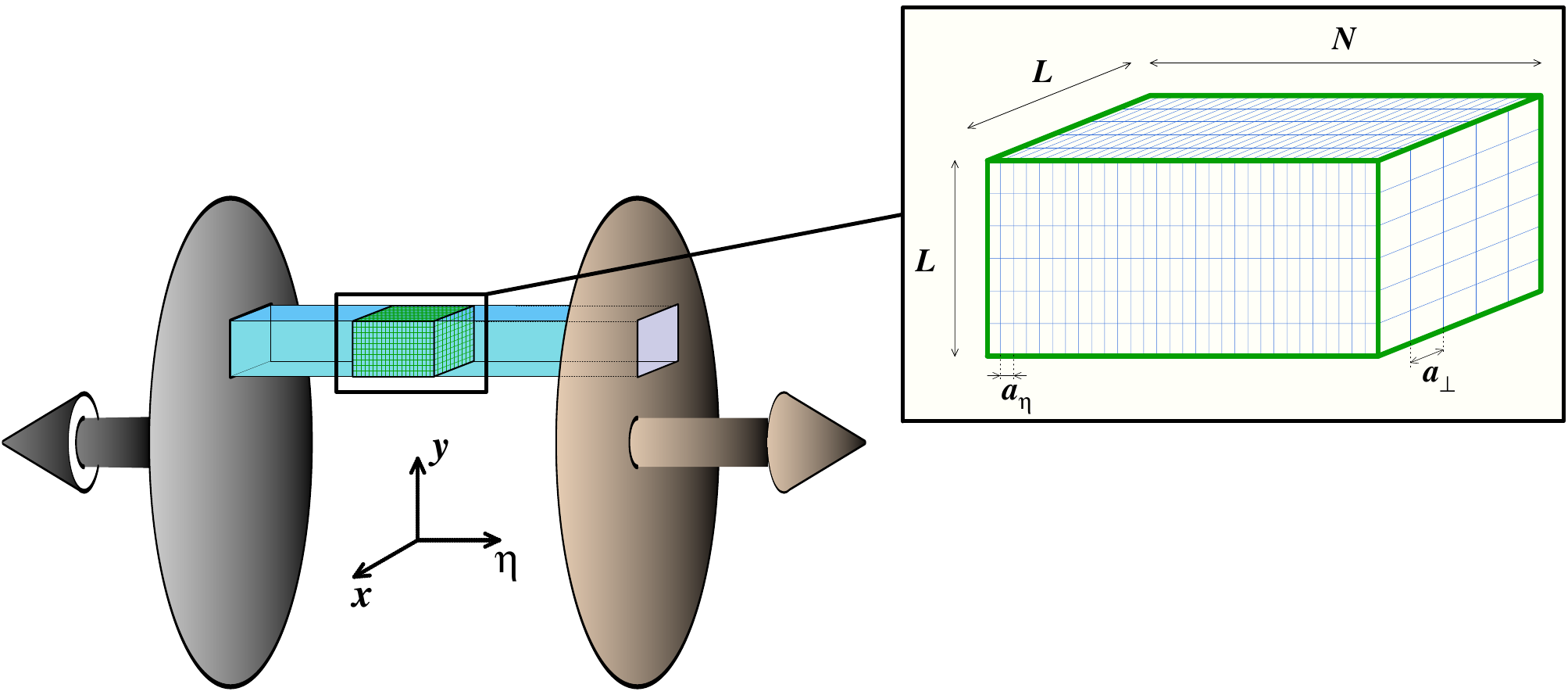}}
\end{center}
\caption{\label{fig:simul}Lattice setup for numerical implementations of the classical statistical method.}
\end{figure}

The setup for doing this is the same as the one described when
discussing the CGC at LO, except for the fact that one must keep
rapidity explicitly. One must discretize the classical Yang-Mills
equations in the system of coordinates $\tau,\eta,\x_\perp$. The
lattice used in these computations usually represents a sub-volume of
the interaction region that expands in the longitudinal direction, as
illustrated in the figure \ref{fig:simul}.
This implies that the lattice spacing in the $z$ coordinate is time
dependent. In order to be able to resolve the physically relevant
scales at the final time $\tau_f$ of the simulation, it is usually
necessary to have a larger number of lattice spacings in the
longitudinal direction.

\subsection{CSA in quantum mechanics}
The Classical Statistical Approximation has an analogue in ordinary
quantum mechanics, which is helpful to understand what approximation
is being made when one uses it. The starting point is the
transformation that goes from the von Neumann equation (\ref{eq:VN})
for the density operator to the Moyal equation (\ref{eq:Moyal}) for
its Wigner transform. The Moyal equation is still an exact
representation of the quantum evolution of the system, although it is
expressed in terms of objects that depend on classical phase-space
variables. However, we have seen that if one expands the operator in
the right hand side of eq.~(\ref{eq:Moyal}) to lowest order in
$\hbar$, one recovers the classical Poisson bracket. In this
approximation, the evolution of the system is thus purely
classical. Note however that the initial condition can remain fully
quantum in this approximation. For instance, if the system is
initially in a pure quantum state $\big|\psi\big>$, its initial
Wigner distribution would be
\begin{equation}
W_0(\x,\p)\equiv \int d^3\r\; e^{i\p\cdot\r}\;
\big<\x+\frac{\r}{2}\big|\psi\big>
\big<\psi\big|\x-\frac{\r}{2}\big>
=
\int d^3\r\; e^{i\p\cdot\r}\;\psi^*(\x+\frac{\r}{2})\psi(\x-\frac{\r}{2})\; .
\end{equation}
If the initial state $\big|\psi\big>$ is known, then no approximation
is needed in the CSA for the initial Wigner distribution. In this
analogy, the Gaussian initial Wigner distribution that we have
obtained in the previous subsection would correspond to starting from
a coherent state, i.e. a state whose Wigner distribution is a Gaussian
of width $\hbar$ centered on some classical configuration.

\subsection{CSA from the path integral}
It is also instructive to see how the CSA can be derived from the path
integral formulation of quantum field theory. Since we aim at
calculating the expectation value of observables in a system
initialized in a known state, we must use the Schwinger-Keldysh
formalism
\begin{equation}
\left<{\cal O}\right>
=
\int\big[D\varphi_\pm^{(i)}(\x)\big]\;\rho_0(\varphi_+^{(i)},\varphi_-^{(i)})
\int\big[D\phi_\pm(x)\big]\;
e^{iS[\phi_+]-iS[\phi_-]}\;{\cal O}(\phi)\; ,
\label{eq:path-SK}
\end{equation}
where the field is doubled into a $+$ component (that represents the
evolution in the amplitude) and a $-$ component (that describes the
conjugate amplitude). In this formulation, the initial state of the
system is represented by a density operator $\widehat{\rho}_0$, whose
matrix elements $\rho_0(\varphi_+^{(i)},\varphi_-^{(i)})$ control the
distribution of the initial values of the fields $\phi_\pm$. The
second integral is restricted to fields whose boundary value at the
initial time is
\begin{equation}
\phi_\pm(t_0,\x)=\varphi_\pm^{(i)}(\x)\; .
\end{equation}
Inclusive observables do not put any constraint on the final state,
and therefore the fields have no specific boundary condition at
$t=+\infty$ in the above path integral, except $\phi_+=\phi_-$ when the measurement is done.

From eq.~(\ref{eq:path-SK}), one should define new fields as the
sum and difference of $\phi_+$ and $\phi_-$,
\begin{equation}
\phi_2\equiv\frac{\phi_++\phi_-}{2}\quad,\qquad \phi_1 \equiv \phi_+-\phi_-\; .
\end{equation}
The difference $S[\phi_+]-S[\phi_-]$ is obviously odd in the field
$\phi_1$, and it is trivial to verify that the term linear in $\phi_1$
comes as a prefactor of the classical equation of motion for
$\phi_2$. In addition, in an interacting theory, there are terms that
are cubic in $\phi_1$,
\begin{equation}
S[\phi_+]-S[\phi_-]
={\colorb\phi_1}\cdot{\colord\frac{\delta S[\phi_2]}{\delta\phi_2}}
+
{\colora\mbox{terms cubic in }{\colorb\phi_1}}\; .
\label{eq:action12}
\end{equation}

We are seeking an approximation that applies in the regime of strong
fields, e.g. when the fields are excited by a large external source like
in the CGC framework, which implies that $\phi_2$ is large. Since
$\phi_+$ and $\phi_-$ are the fields in the amplitude and conjugate
amplitude respectively, their difference is a quantum effect whose
magnitude is suppressed by $\hbar$. Therefore, in such a situation, we
have $\phi_1\ll\phi_2$, and it is natural to neglect the cubic term in
$\phi_1$ in the action. After doing this, the field $\phi_1$ becomes a
Lagrange multiplier for the classical equation of motion for
$\phi_2$. The evolution of $\phi_2$ is now deterministic, and the only
remaining fluctuations are those inherited from the average over the
initial density matrix $\rho_0(\varphi_+^{(i)},\varphi_-^{(i)})$. In
the Hamiltonian formulation of the classical equation of motion for
$\phi_2$, the average over $\varphi_+^{(i)},\varphi_-^{(i)}$ becomes
an average over $\varphi_2^{(i)}$ and its conjugate momentum
$\Pi_2^{(i)}$, with a distribution obtained as the Wigner transform of
$\rho_0$~:
\begin{equation}
W_0[\varphi_2^{(i)},\Pi_2^{(i)}]
\equiv
\int\big[D\chi\big]\;
e^{i\int \chi\cdot\Pi_2}\;
\rho_0(\varphi_2^{(i)}+\frac{\chi}{2},\varphi_2^{(i)}-\frac{\chi}{2})\; .
\end{equation}

\subsection{CSA in perturbation theory}
\label{sec:pert-CSA}
The path integral derivation of the classical statistical
approximation also clarifies what this approximation amounts to in
perturbation theory. This knowledge will be useful later when we
discuss the non-renormalizability of this approximation. Let us
discuss this in the simple framework of a scalar field theory with a
$\phi^4$ interaction term.

The first step is to express the transformation from the $\phi_\pm$
fields to the $\phi_{1,2}$ fields as a ``rotation'', in order to
obtain the diagrammatic rules in this new basis~\cite{AurenB1,EijckKW1,Gelis3,AartsS3,EpelbGW1}. From the propagators
in the $\pm$ basis,
\begin{align}
    &G^0_{++}(p)=\frac{i}{p^2-m^2+i\epsilon}\;,&&G^0_{--}(p)=\frac{-i}{p^2-m^2-i\epsilon}\nonumber\\
    &G^0_{+-}(p)=2\pi\theta(-p^0)\delta(p^2-m^2)\;,&&G^0_{-+}(p)=2\pi\theta(p^0)\delta(p^2-m^2)\; ,
    \label{eq:SKprops}
\end{align}
one can define a set of new propagators by a linear transformation on the two indices
\begin{eqnarray}
{\mathbbm G}_{\alpha\beta}^0\equiv 
\sum_{\epsilon,\epsilon^\prime=\pm}
\Omega_{\alpha\epsilon}\Omega_{\beta\epsilon^\prime}
{G}_{\epsilon\epsilon^\prime}^0\; ,
\label{eq:SK-rotation}
\end{eqnarray}
with the following transformation matrix 
\begin{equation}
\Omega_{\alpha\epsilon}
\equiv
\begin{pmatrix}
1 & -1 \\
1/2 & 1/2
\label{eq:rot}
\end{pmatrix}\; .
\end{equation}
The free rotated propagators are
\begin{equation}
{\mathbbm G}_{\alpha\beta}^0
=
\begin{pmatrix}
0 & G_{_A}^0\\
G_{_R}^0& G_{_S}^0\\
\end{pmatrix}\; ,
\label{eq:12prop}
\end{equation}
where we define
\begin{equation}
G_{_R}^0 = G_{++}^0-G_{+-}^0\;,\;
G_{_A}^0 = G_{++}^0-G_{-+}^0\;,\;
G_{_S}^0 = \frac{1}{2}(G_{+-}^0+G_{-+}^0)\; .
\end{equation}
(The subscripts R, A and S mean respectively for {\sl retarded}, {\sl
  advanced} and {\sl symmetric}.) Note that the symmetric propagator
depends on the initial state of the system, while the retarded and
advanced ones are independent of this initial data (they only reflect
how modes propagate in the theory under consideration). Under the
rotation of eq.~(\ref{eq:rot}), the vertices are transformed into
\begin{eqnarray}
&&\Gamma_{1111}=\Gamma_{1122}=\Gamma_{2222}=0\nonumber\\
&&\Gamma_{1222}=-ig^2\; ,\quad \Gamma_{1112}=-ig^2/4\; .
\label{eq:12vert}
\end{eqnarray}
(The ones not listed explicitly here should be obtained by
permutations of the indices.) The CSA simply amounts to setting to
zero the vertex $\Gamma_{1112}$ (and its permutations) wherever it
appears in the diagrammatic expansion, while all the other Feynman
rules are unmodified.

\section{Applications of  the CSA to heavy ion collisions}
\label{sec:appl}
The classical statistical approximation has recently been applied to
heavy ion collisions in two sets of works, that mostly differ by the
nature of the initial conditions that were used.

\subsection{CGC Initial conditions}
\begin{figure}[htbp]
\begin{center}
\resizebox*{4.5cm}{!}{\includegraphics{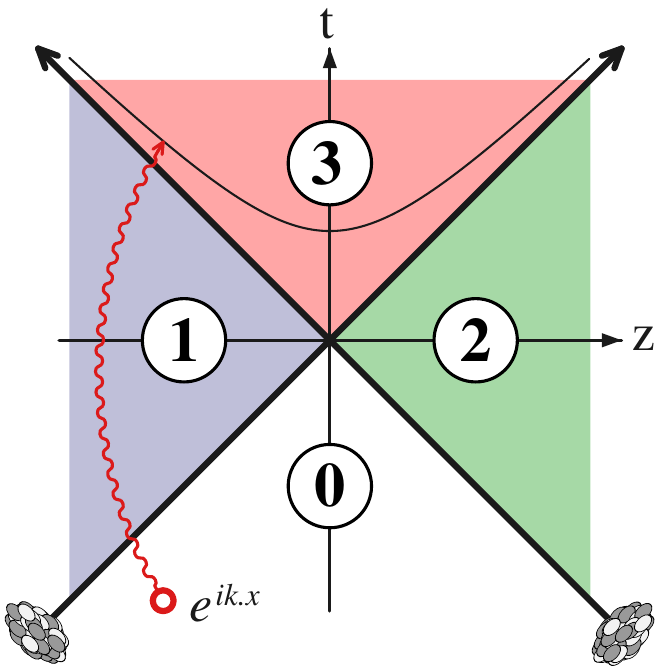}}\hfill
\resizebox*{6cm}{!}{\includegraphics{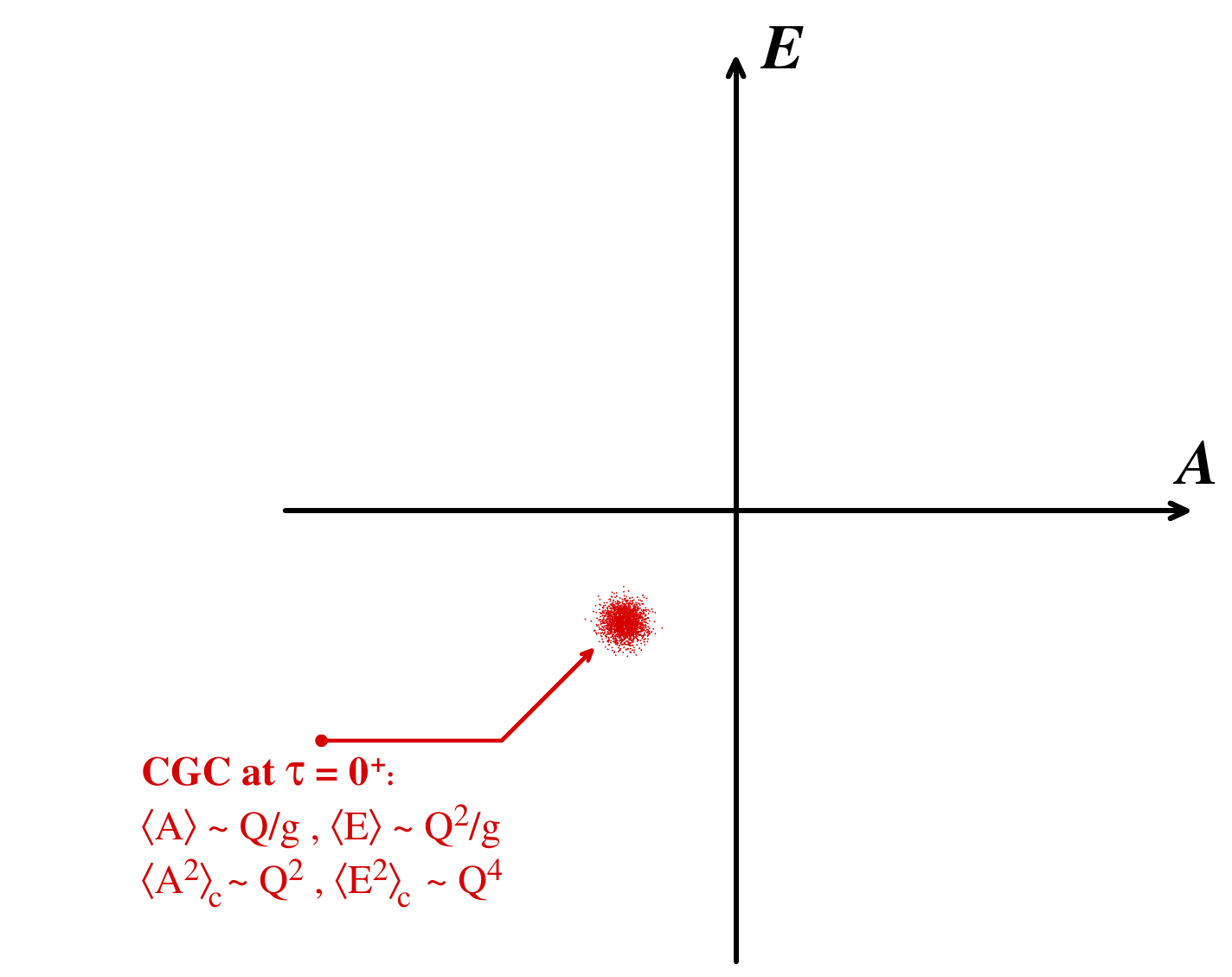}}
\end{center}
\caption{\label{fig:fluct}CGC spectrum of initial fluctuations at $Q_s\tau_0\ll 1$.}
\end{figure}

In a strict application of the CSA to the description of heavy ion
collisions in the Color Glass Condensate framework, the Gaussian
ensemble of initial conditions arises from the exponentiation of the
1-loop result. In other words, the CGC description of heavy ion
collisions is an initial value problem, and the state of the system at
$\tau=0^+$ is therefore prescribed uniquely from the fact that the
system was in the vacuum state at $x^0=-\infty$. The initial Gaussian
ensemble of fields is characterized by the following mean values and
variance (written here in a very sketchy way, without color and Lorentz
indices):
\begin{eqnarray}
&&\big<{\cal A}^\mu\big>={\cal A}_{_{\rm LO}}^\mu\nonumber\\
&&{\bs\Gamma}_2(\u,\v)
 = \int\frac{d^2\k}{(2\pi)^3 2\omega_\k}\;
{\colorb a_\k(\tau_0,\u) a_\k^*(\tau_0,\v)}\nonumber\\
&&\Big[{\cal D}_\rho {\cal D}^\rho \delta^\nu_\mu - {\cal D}_\mu {\cal D}^\nu
+ig\, {\cal F}_\mu{}^\nu\Big] {\colorb a_\k^\mu}=0\nonumber\\
&&{\colord\lim_{x^0\to-\infty}} {\colorb a_\k(x)} = e^{ik\cdot x}\; .
\label{eq:fluct3}
\end{eqnarray}
The mean value is already known from the LO
calculation~\cite{KovneMW2}, given in
eqs.~(\ref{eq:init-LO-tau0}). From these equations, one sees that the
determination of the variance ${\bs\Gamma}_2$ amounts to solving the
linearized Yang-Mills equations for all the mode functions ${\colorb
  a_\k^\mu}$ (this formula comes from the derivation of the NLO
contribution)~\cite{EpelbG2}. As long as we stay in a
regime where these perturbations are not yet enhanced by
instabilities, the variance is small compared to the mean value
squared, and the Gaussian distribution of the initial fields is a
narrow distribution centered on the LO color fields (see the right
figure \ref{fig:fluct}). The mode functions describe how plane waves
with a well defined momentum, color and polarization at $x^0=-\infty$
are distorted while they propagate over the LO color background field
${\cal A}^\mu_{_{\rm LO}}$.
The linearized Yang-Mills equations must be solved for each mode
function, up to the time $\tau_0$ (in the forward light-cone) at which
the numerical simulation will start (left figure
\ref{fig:fluct}). Explicit formulas for these mode functions were
derived in ref.~\cite{EpelbG2}, in the Fock-Schwinger gauge $A^\tau=0$
used in these computations.  For given quantum numbers~: ${\colora
  \nu}$ (the Fourier conjugate of the rapidity $\eta$),
${\k_\perp,\lambda,c}$, the gauge potential and the associated
  electrical fields read
\begin{align}
&a^{\colorc i}= \beta^{+{\colorc i}}+\beta^{-{\colorc i}}
&&a^{\colorc\eta}= {\cal D}^{\colorc i}
\Big(\frac{\beta^{+{\colorc i}}}{2+i{\colora\nu}}-\frac{\beta^{-{\colorc i}}}{2-i{\colora\nu}}\Big)
\nonumber\\
&
\vphantom{\Bigg(}
e^{\colorc i}= -i{\colora\nu}\Big(\beta^{+{\colorc i}}-\beta^{-{\colorc i}}\Big)
&&e^{\colorc\eta}=-{\cal D}^{\colorc i}\Big(\beta^{+{\colorc i}}-\beta^{-{\colorc i}}\Big)
\; ,
\end{align}where

\begin{equation}
\beta^{+{\colorc i}}\equiv
e^{\frac{\pi{\colora\nu}}{2}}\Gamma(-i{\colora\nu}) e^{i{\colora\nu}\eta}\,
{\colorb{U}_1^\dagger(\x_\perp)} 
\int\limits_{\p_\perp} e^{i\p_\perp\cdot\x_\perp}\,
{\colorb\widetilde{U}_1(\p_\perp+{\colora\k_\perp})}
\left(\frac{p_\perp^2\tau}{2{\colora k_\perp}}\right)^{i{\colora\nu}}
\!
\Big(\delta^{{\colorc i}{\colord j}}-2\frac{p_\perp^{{\colorc i}} p_\perp^{{\colord j}}}{p_\perp^2}\Big){\colord \epsilon}^{{\colord j}}_{\colora\lambda}
\end{equation}
and
\begin{equation}
\beta^{-{\colorc i}}\equiv
e^{-\frac{\pi{\colora\nu}}{2}}\Gamma(i{\colora\nu}) e^{i{\colora\nu}\eta}\,
{\colorb{U}_2^\dagger(\x_\perp)} 
\int\limits_{\p_\perp} e^{i\p_\perp\cdot\x_\perp}\,
{\colorb\widetilde{U}_2(\p_\perp+{\colora\k_\perp})}
\left(\frac{p_\perp^2\tau}{2{\colora k_\perp}}\right)^{-i{\colora\nu}}
\!
\Big(\delta^{{\colorc i}{\colord j}}-2\frac{p_\perp^{{\colorc i}} p_\perp^{{\colord j}}}{p_\perp^2}\Big){\colord \epsilon}^{{\colord j}}_{\colora\lambda}\; .
\end{equation}
${U}_{1,2}$ are the Wilson lines that describe the color source
content of the two colliding nuclei (they are defined in terms of
$\rho_{1,2}$ in eq.~(\ref{eq:wilson-def})).

These mode functions have been used in ref.~\cite{EpelbG3} in order to
compute the time evolution of the components of the energy-momentum
tensor from CGC initial conditions. The main difficulty and source of
uncertainty in these calculation is the subtraction of the ultraviolet
divergences.  Firstly, since the energy-momentum tensor is a dimension
four operator, it picks up a pure vacuum contribution that behaves as
the fourth power of the inverse lattice spacing. This contribution can
be removed by subtracting the result of a second computation done with
the same lattice parameters, but without the colliding nuclei. A large
statistics in the Monte-Carlo evaluation of the average over the
initial fields is required for this subtraction. A second kind of
ultraviolet contribution, that depends on the background field, was
also observed to affect the energy density and the
pressure\footnote{This contribution is possibly due to the
  non-renormalizability of the CSA -- see the subsection
  \ref{sec:non-ren}.}, but not the transverse pressure. Since at this
level of approximation, the energy momentum tensor is traceless:
\begin{equation}
\epsilon=P_{_L}+2P_{_T}\; ,
\end{equation}
and obeys Bjorken's law due to energy-momentum conservation,
\begin{equation}
\partial_\tau(\tau\epsilon)+P_{_L}=0\; ,
\end{equation}
the only possible form of an ultraviolet sensitive contribution that
does not affect $P_{_T}$ is a term $const\times\tau^{-2}$ that affects
equally $\epsilon$ and $P_{_L}$. This is indeed what was
observed. Lacking a more precise understanding of this term, the
constant was fitted in order to subtract this term from $\epsilon$ and
$P_{_L}$. After this subtraction, the early time behavior of the
ratios $P_{_L}/\epsilon$ and $P_{_T}/\epsilon$ closely follows the LO
up to $Q_s\tau\sim 1$, which is indeed expected since this is before
the unstable modes may have affected the time evolution.
\begin{figure}[htbp]
\begin{center}
\resizebox*{9cm}{!}{\includegraphics{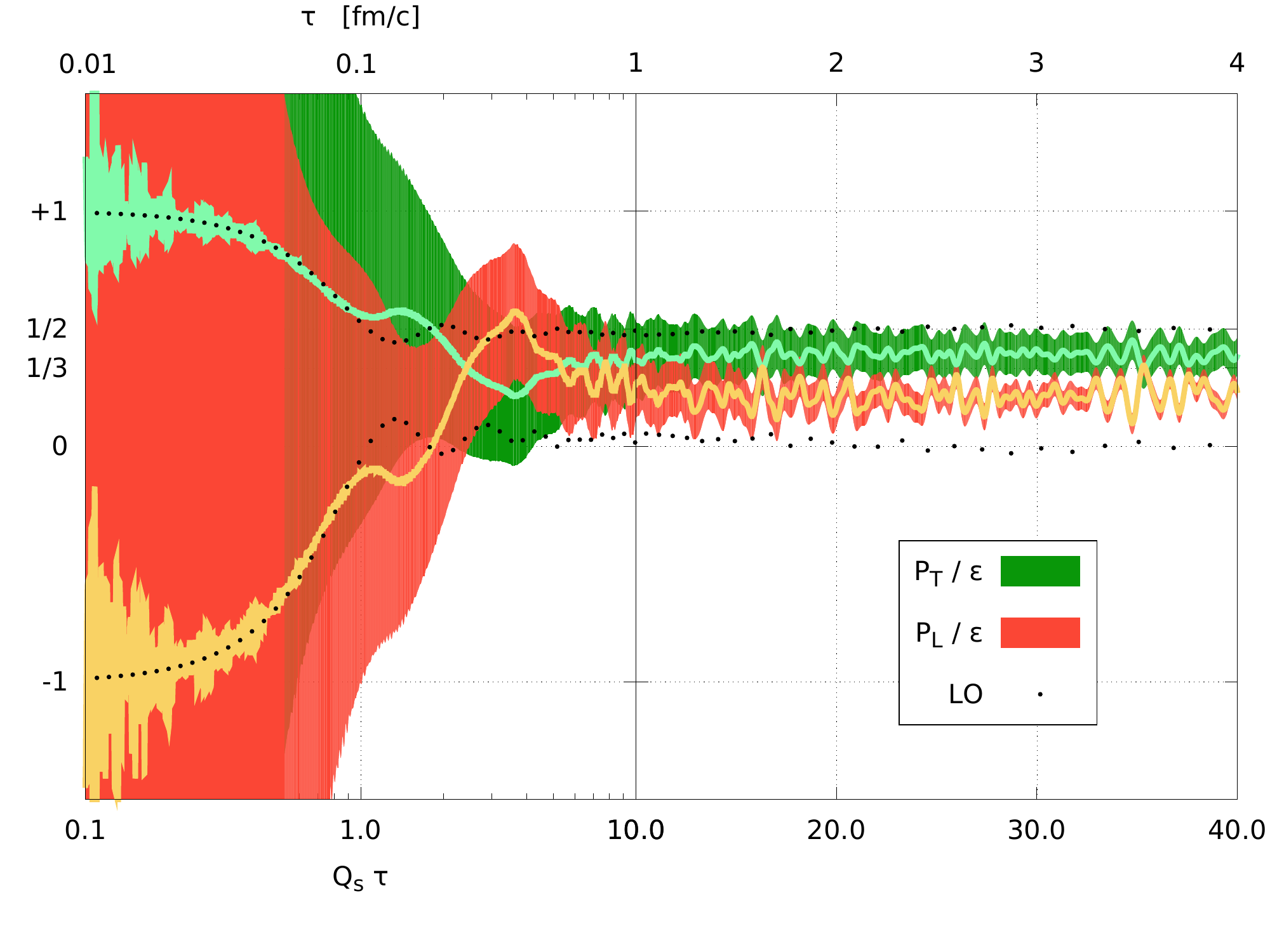}}
\end{center}
\caption{\label{fig:CSA-vac}Behavior of $P_{_L}/\epsilon$ and
  $P_{_T}/\epsilon$ in the CSA with CGC initial conditions. The upper
  time scale, in $fm/c$, is based on a saturation momentum
  $Q_s=2$~GeV.}
\end{figure}
For a moderately large coupling $g=0.5$, it was found that the
longitudinal pressure increases significantly compared to its value in
the LO calculation (see the figure \ref{fig:CSA-vac}), and now becomes
a sizable fraction of the transverse pressure.

\subsection{Particle-like initial conditions}
Alternatively, one may depart somewhat from the CGC framework and use
the same approximation scheme with a different Gaussian ensemble of
initial conditions. The simplest model of initial conditions one may
consider is an ensemble of fields that describe a distribution of free
particles~\cite{BergeBSV1,BergeBSV2,BergeBSV3,BergeBSV4}. In this
case, the center of the Gaussian is ${\cal A}^\mu\equiv 0$, and
the variance is constructed from the in-vacuum mode functions (derived
in \cite{DusliGV1} for gluons in the Fock-Schwinger gauge) as follows~:
\begin{equation}
\big<{\cal A}^\mu\big>=0\; ,\qquad {\bs\Gamma}_2(\u,\v) =
  \int\limits_{{\rm modes\ }\k} {\colorc f_0(\k)}\; {\colorb a_\k(\u)
    a_\k^*(\v)}\;,\qquad {\colorb a_\k(x)} \equiv e^{ik\cdot x}\; ,
  \label{eq:inco}
\end{equation}
where $f_0(\k)$ is the initial momentum distribution of these
particles. It is plausible that the CGC initial state at $\tau_0=0^+$
may evolve after some time into an incoherent distribution of gluons
such as the one described by eqs.~(\ref{eq:inco}), but this certainly
deserves a thorough study.
\begin{figure}[htbp]
\begin{center}
\resizebox*{6cm}{!}{\includegraphics{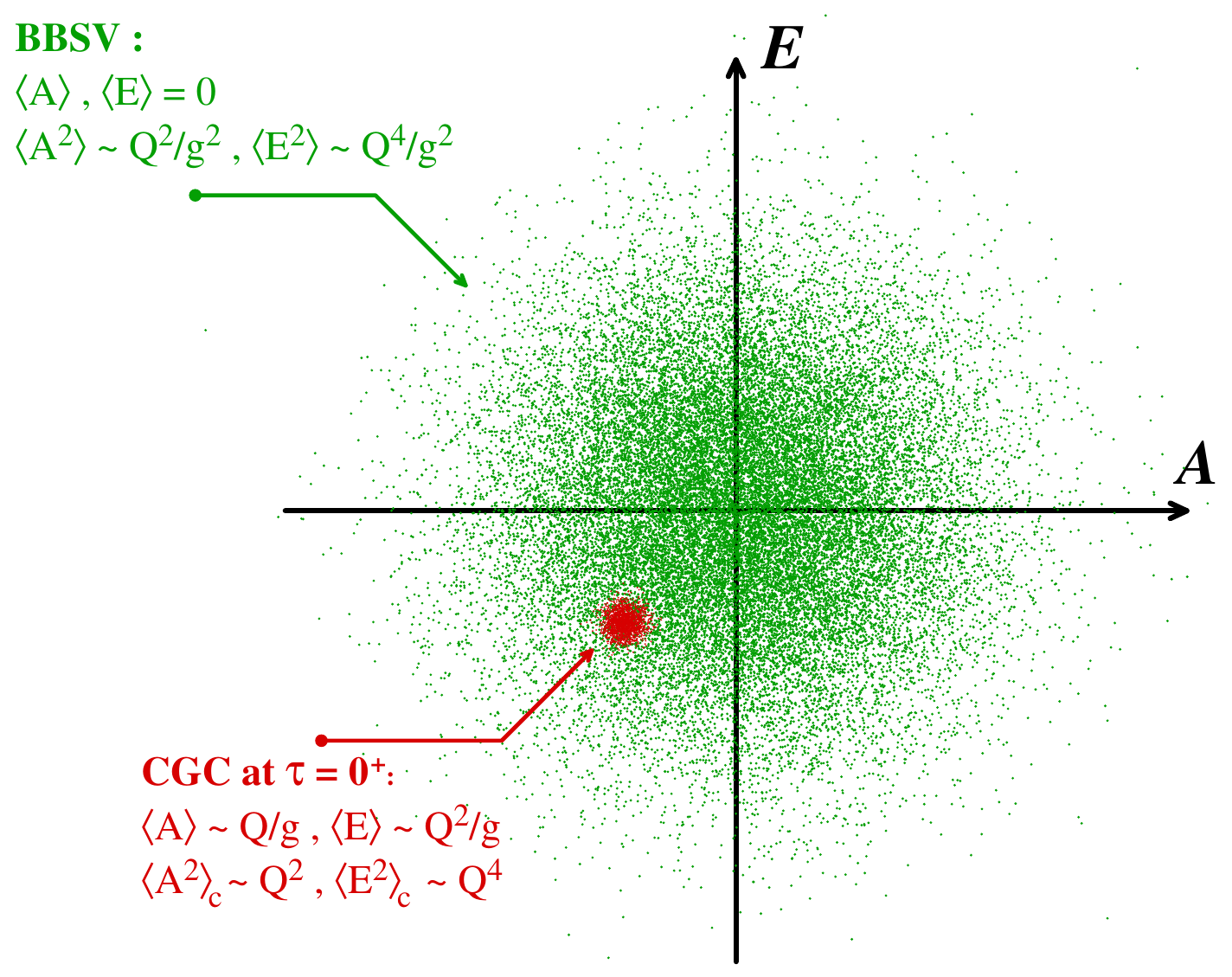}}
\end{center}
\caption{\label{fig:ico}Illustration of the difference between the
  initial conditions (\ref{eq:fluct3}) and (\ref{eq:inco}) for the
  Wigner distribution of the initial fields.}
\end{figure}

The difference and complementarity between the initial conditions in
eqs.~(\ref{eq:fluct3}) and (\ref{eq:inco}) becomes more transparent if
one recalls the symmetric 2-point Green's function in the
Schwinger-Keldysh formalism in the presence of a bath of particles,
\begin{equation}
G_{_S}(k)=2\pi\big(\frac{1}{2}+f_0(\k)\big)\delta(k^2)\; .
\end{equation}
In this formula, the term in $f_0(\k)$ represents an initial particle
distribution, while the term in $1/2$ corresponds to pure vacuum
quantum fluctuations. It thus becomes clear that the CGC initial
conditions at $\tau=0^+$ can be viewed as vacuum fluctuations that are
somewhat altered by the presence of the LO color background field. It
is the minimal modification that quantum mechanics can bring to a
classical state, promoting it to a (pure) coherent state, while the
particle-like initial conditions defined by eqs.~(\ref{eq:inco}) are a
mixed state from the point of view of quantum mechanics.

This type of particle-like initial condition was used in
refs.~\cite{BergeBSV1,BergeBSV2,BergeBSV3,BergeBSV4} in order to
study the evolution of a longitudinally system, in Yang-Mills theory
and also in the $\phi^4$ scalar field theory\cite{BergeBSV4}. 
\begin{figure}[htbp]
\begin{center}
\resizebox*{9cm}{!}{\includegraphics{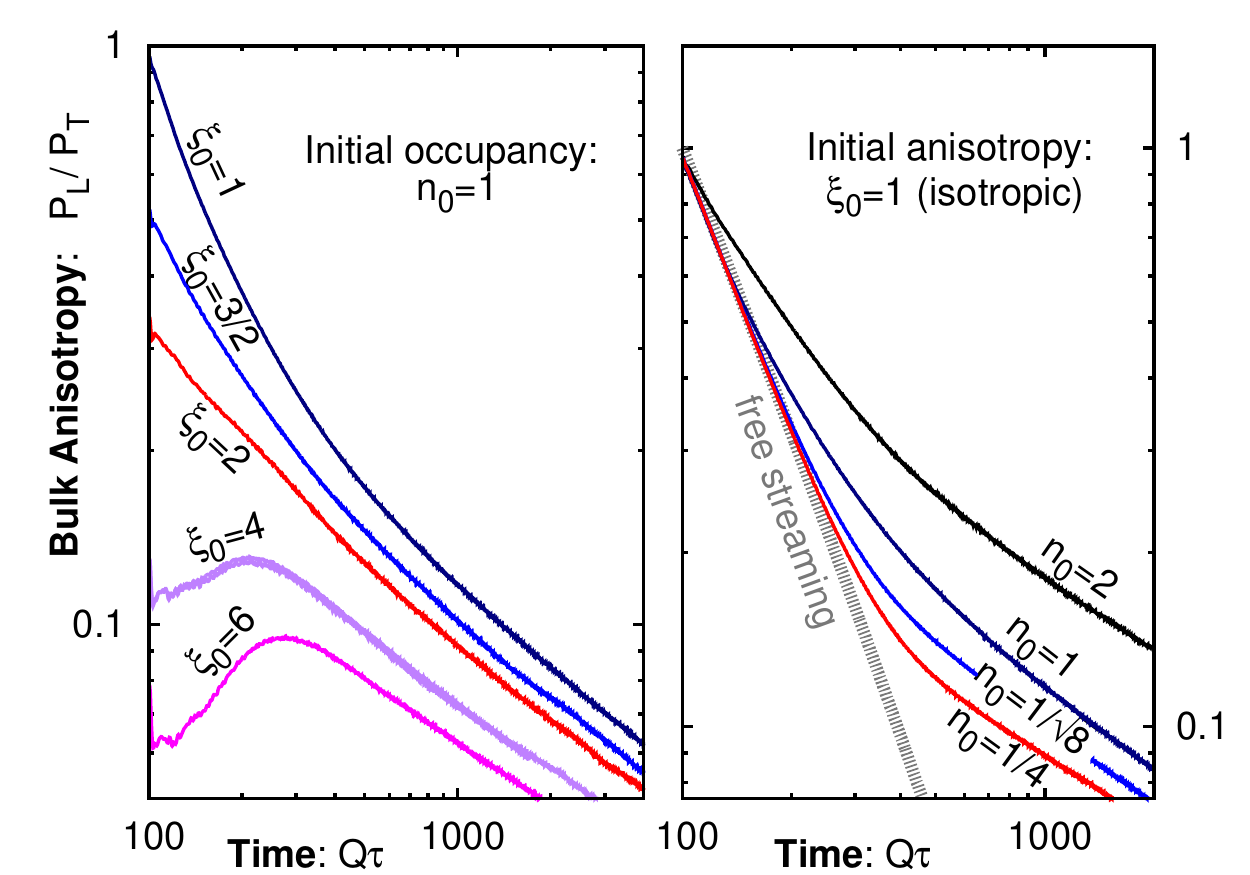}}
\end{center}
\caption{\label{fig:CSA-part}Behavior of $P_{_L}/P_{_T}$ in the CSA
  with particle-like initial conditions in Yang-Mills theory, for
  various anisotropies ($\xi_0$) and densities ($n_0$) of the initial
  particle distribution. From Ref.~\cite{BergeBSV1}.}
\end{figure}
Note that, if one chooses an initial distribution proportional to the
inverse coupling $g^{-2}$, then one can completely scale out the
coupling from the calculation by an appropriate rescaling of the
fields. In these works, the shape of the initial
distribution was controlled by two parameters: $n_0$ (that controls the
overall normalization) and $\xi_0$ (that controls the
anisotropy). Some results regarding the behavior of the ratio
$P_{_L}/P_{_T}$ are shown in the figure \ref{fig:CSA-part}.  Despite
substantial variations of these parameters, the system was always
observed to reach a scaling regime in which on has the following
approximate behaviors,
\begin{equation}
f(\p)\sim \tau^{-2/3}\quad,\quad p_\perp \sim \tau^0\quad,\quad
p_z\sim \tau^{-1/3}\quad,\quad \frac{P_{_L}}{P_{_T}}\sim \tau^{-2/3}\; .
\label{eq:CSA-part-scaling}
\end{equation}
Note that these scaling laws are not the free streaming ones (where
$p_z\sim \tau^{-1}$, $f(\p)\sim\tau^0$ and ${P_{_L}}/{P_{_T}}\sim
\tau^{-2}$) which means that this system interacts significantly but
not strongly enough in order to overcome the expansion, at least in
the classical approximation. The scaling behavior of the longitudinal
momentum $p_z\sim \tau^{-1/3}$ can be understood
semi-analytically\cite{BaierMSS2} if the elastic scattering rate is
dominated by small angle scatterings. Quite surprisingly, the same
scaling behavior of $p_z$ was also observed\cite{BergeBSV4} in this
approximation in the scalar $\phi^4$ theory, despite the fact that the
leading order scattering cross-section in this theory is dominated by
large angle scatterings\footnote{See the section \ref{sec:large-angle}
  for a discussion of the possible interplay between the classical
  approximation and large angle scatterings in anisotropic systems.}.

In this approximation, this scaling behavior of
eq.~(\ref{eq:CSA-part-scaling}) will continue forever, and the
pressure tensor becomes more and more anisotropic over time. However,
since this is a classical approximation, it is known to break at least
when the occupation number $f(\p)$ becomes of order one.
\begin{figure}[htbp]
\begin{center}
\resizebox*{8cm}{!}{\includegraphics{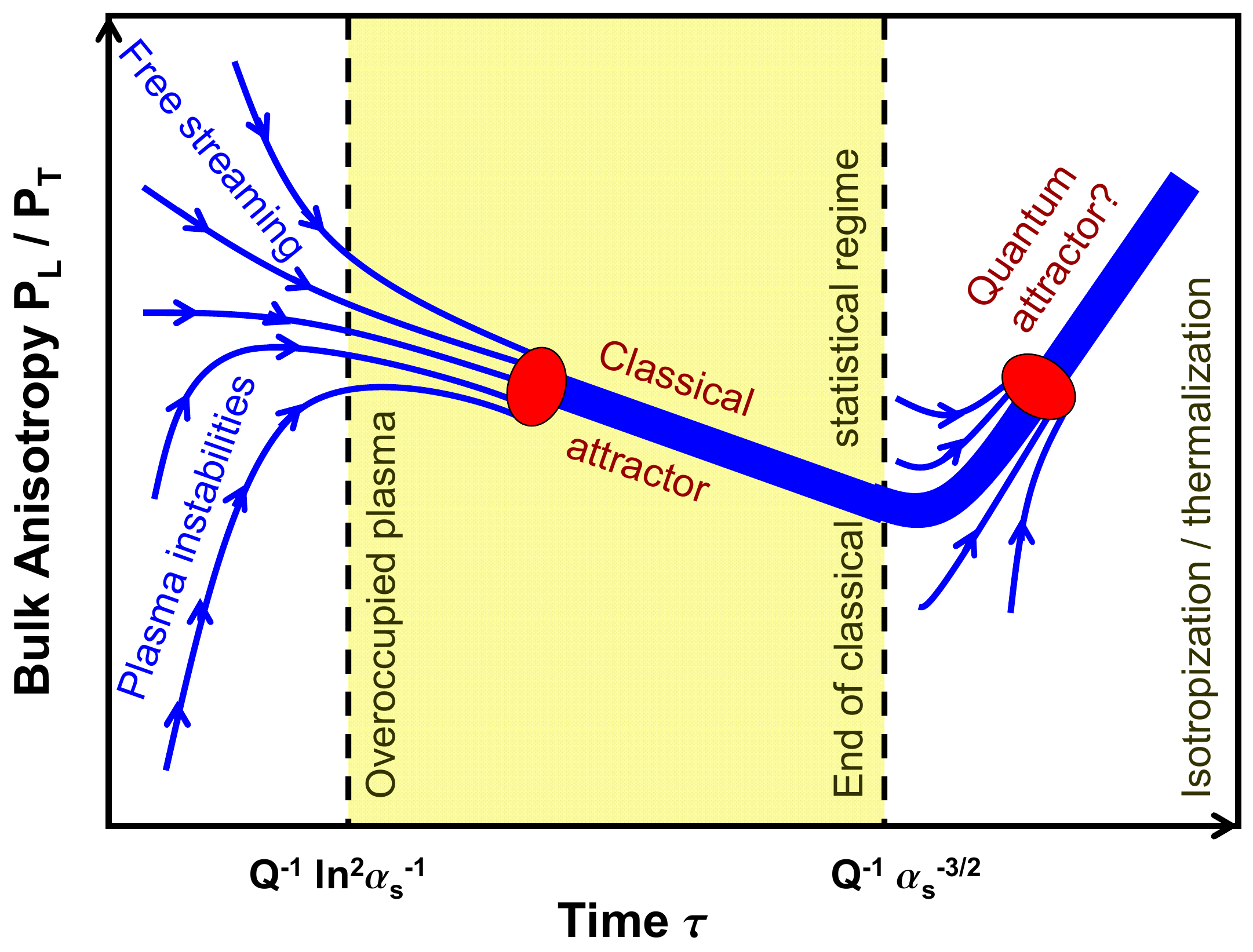}}
\end{center}
\caption{\label{fig:CSA-part1}Thermalization/isotropization scenario
  that emerges from CSA studies with particle-like initial
  conditions. From Ref.~\cite{BergeBSV1}.}
\end{figure}
This is expected to happen when $Q_s\tau \sim g^{-3}$, which
corresponds to the end of the first stage of the bottom-up scenario of
ref.~\cite{BaierMSS2}. If this scenario --illustrated in the figure
\ref{fig:CSA-part1}-- is confirmed, the isotropization of the system
would only begin after this time, while the occupation number in the
system is no longer large. The quantum corrections would eventually
bring the energy-momentum tensor to a quasi-isotropic shape compatible
with nearly ideal hydrodynamics. However, what would be more difficult
to understand in this scenario is why the shear viscosity to entropy
ratio is small. Indeed, in a weak coupling system with occupation
numbers of order 1 or below, this ratio is expected to be
parametrically large $\eta/s\sim g^{-4}$.

\section{Limitations of the classical statistical approximation}
\label{sec:limit}
The main appeal of the CSA is the fact that it is quite
straightforward to implement, (unlike other schemes like the
two-particle irreducible approach, that has never been applied to
Yang-Mills theory so far), while at the same time staying very close
to the dynamics of the gauge fields so that it is a natural extension
of the CGC framework. However, it is not without problems and
limitations, that we discuss in this section.

\subsection{Ultraviolet divergences and non-renormalizability}
\label{sec:non-ren}
A very important difference between the initial conditions
(\ref{eq:fluct3}) and (\ref{eq:inco}) is the momentum dependence of
the spectrum of field fluctuations. In eqs.~(\ref{eq:inco}) the large
momentum behavior of the spectrum is controlled by the initial
particle distribution $f_0(\k)$, and therefore it does not extend to
infinity if one chooses a $f_0(\k)$ that has a compact support. In
contrast, the vacuum fluctuations that are the source of the field
fluctuations in the CGC initial conditions have a spectrum which is
flat up to $k=\infty$.

This difference in the spectra at large momentum leads to very
different dependences on the ultraviolet cutoff (i.e. the inverse
lattice spacing) in numerical implementations of the CSA using these
two types of initial conditions. This issue can be phrased as follows:
starting from a renormalizable quantum field theory (e.g. Yang-Mills
theory, or a $\phi^4$ scalar field theory), does the CSA preserve its
renormalizability?  It turns out that the answer to this question
depends on the spectrum of the initial field fluctuations. This can be
studied perturbatively\cite{EpelbGW1} by using the retarded-advanced
basis, and by using the fact that the CSA amounts to dropping the
vertex that has 3 indices of type $1$.

From studies performed in the context of quantum field theory at
finite temperature~\cite{BodekMS1,ArnolSY1,AartsS1,AartsS3,AartsNW1},
it has been known for a long time that a particle-like spectrum that
falls at least as fast as $k^{-1}$ leads to a super-renormalizable
approximation. In this case, it is sufficient to perform a finite number
of subtractions in order to make predictions that do not depend on the
ultraviolet cutoff.  This applies directly to initial conditions of
type (\ref{eq:inco}), provided that the initial distribution $f_0(\k)$
falls quickly enough with momentum.

The situation is quite different with the vacuum-like CGC initial
conditions, because they have a flat spectrum of fluctuations.  By
using the perturbative CSA described in the subsection
\ref{sec:pert-CSA}, it has been shown in Ref.~\cite{EpelbGW1} that the
CSA is a non renormalizable approximation of the underlying quantum
field theory when this type of initial condition is used. For
instance, for the self-energy $\Sigma_{12}$ at two loops (in a
$\phi^4$ scalar field theory), the CSA gives an ultraviolet divergent
result \setbox1\hbox to
2.5cm{\resizebox*{2.5cm}{!}{\includegraphics{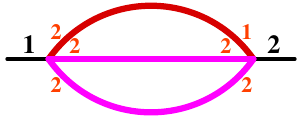}}}
\begin{equation}
\mbox{Im}\,\raise -4.5mm\box1
=
-\frac{g^4}{1024\pi^3}
\;\left({\colorb \Lambda_{_{\rm UV}}^2}-\frac{2}{3}p^2\right)\; .
\end{equation}
Since this divergence occurs in the imaginary part of a correlator, it
cannot be removed by a counterterm added to the action (otherwise that
would make the action non Hermitian). The consequence of this is that
one cannot take the continuum limit in computations based on the CSA
with vacuum-like initial conditions. This 2-loop divergence in a
self-energy has a counterpart in the classical approximation of the
Boltzmann equation, as we shall see in the next subsection. 

Non-renormalizable graphs can even be found in some 1-loop 4-point
functions, such as the following $\Gamma_{1122}$ function~:
\setbox1\hbox to 2.8cm{\resizebox*{2.8cm}{!}{\includegraphics{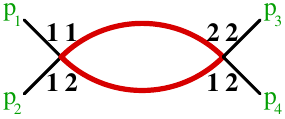}}}
\setbox2\hbox to 1.1cm{\resizebox*{1.1cm}{!}{\includegraphics{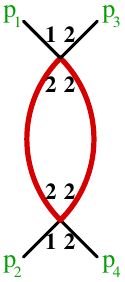}}}
\setbox3\hbox to 1.1cm{\resizebox*{1.1cm}{!}{\includegraphics{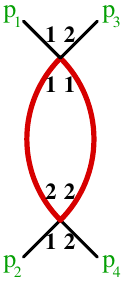}}}
\setbox4\hbox to 1.1cm{\resizebox*{1.1cm}{!}{\includegraphics{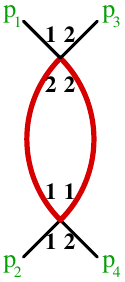}}}
\setbox5\hbox to 1.1cm{\resizebox*{1.1cm}{!}{\includegraphics{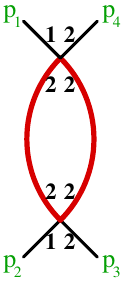}}}
\setbox6\hbox to 1.1cm{\resizebox*{1.1cm}{!}{\includegraphics{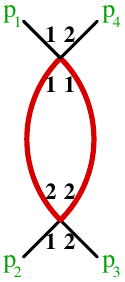}}}
\setbox7\hbox to 1.1cm{\resizebox*{1.1cm}{!}{\includegraphics{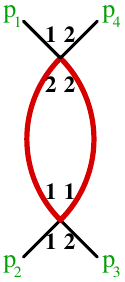}}}
\begin{eqnarray}
-i\Gamma_{1122}^{\rm 1\ loop}
&=
\underbrace{\raise -5mm\box1}_{\mbox{S channel}}
&+
\underbrace{
\raise -11mm\box2
+
\raise -11mm\box3
+
\raise -11mm\box4}_{\mbox{T channel}}
\nonumber\\
&&
+
\underbrace{
\raise -11mm\box5
+
\raise -11mm\box6
+
\raise -11mm\box7}_{\mbox{U channel}}\; .
\label{eq:1122-1loop}
\end{eqnarray}
Here, we have shown all the vertex assignments that would appear in the
retarded-advanced basis. Calculating and adding up all these
contributions would lead to an ultraviolet finite result, as is
expected in a renormalizable theory since the bare Lagrangian has no
$\phi_1^2\phi_2^2$ operator. Note however that some of these graphs
contain a $1112$ vertex, and would therefore be discarded in the CSA. 
In this approximation, we only have

\setbox2\hbox to 1.1cm{\resizebox*{1.1cm}{!}{\includegraphics{G1122-T1}}}
\setbox5\hbox to 1.1cm{\resizebox*{1.1cm}{!}{\includegraphics{G1122-U1}}}
\begin{equation}
-i\big[\Gamma_{1122}\big]_{{\rm CSA}}^{\rm 1\ loop}
=
\raise -11mm\box2
+
\raise -11mm\box5\; ,
\label{eq:1122-CSA}
\end{equation}
the result of which is given by
\begin{equation}
-i\big[\Gamma_{1122}\big]_{{\rm CSA}}^{\rm 1\ loop}
=-\frac{g^4}{64\pi}\left[
{\rm sign}(t)+{\rm sign}(u)
+2\,\Lambda_{_{\rm UV}}
\left(
\frac{\theta(-t)}{|\p_1+\p_3|}
+
\frac{\theta(-u)}{|\p_1+\p_4|}
\right)
\right]\; ,
\label{eq:1122-1loop-1}
\end{equation}
with
\begin{equation}
t\equiv (p_1+p_3)^2\quad,\quad u\equiv (p_1+p_4)^2\; ,
\end{equation}
the standard Mandelstam variables and where $\Lambda_{_{\rm UV}}$ is
the ultraviolet cutoff on 3-momentum. Despite the zero superficial
degree of divergence of these graphs, they contain a linear
divergence. Moreover, the coefficient of the divergent terms is
non-polynomial in the momenta, implying that it is a non-local
ultraviolet divergence.  Such non-renormalizable contributions will
appear at 2-loops and beyond in the expectation value of inclusive
quantities like the energy-momentum tensor. Because of them, CSA
calculations performed with initial conditions that contain vacuum
fluctuations should be performed with an ultraviolet cutoff (i.e. the
inverse lattice spacing) which is not too large compared to the
physical scales.

\subsection{CSA in kinetic theory} 
Assessing the cutoff dependence within the CSA itself is very costly,
because it requires repeating the same calculation several times with
smaller and smaller lattice spacings~\cite{BergeBSV3}.  A much cheaper
way of understanding the interplay between the classical approximation
and the dependence on the ultraviolet cutoff is to consider the same
approximation at the level of kinetic theory\footnote{Here, we employ
  kinetic theory as a way to assess some formal aspects of the
  underlying quantum field theory, like its dependence on the
  ultraviolet cutoff. A number of works\cite{BlaizLM2,HuangL1,RuggiSPG1,PugliPG2,PugliPG1,ScardPPRG1,UphofFSWX1,FochlUXG1,GreifBXG1,XuZZG1} have also used kinetic theory
  as a tool for studying thermalization in models of heavy ion
  collisions.}. Let us start with the Boltzmann equation, with the
collision term expressed in terms of self-energies in the
Schwinger-Keldysh formalism~:
\begin{equation}
[\partial_t+\v_\p\cdot{\bs\nabla}]\,f(p)=\underbrace{\frac{i}{2\omega_\p}
\left[f(\p)\Sigma_{-+}(P)-(1+f(\p))\Sigma_{+-}(P)\right]}_{{\cal C}_\p[f]}\; .
\label{eq:Boltz-plain}
\end{equation}
In order to perform in kinetic theory the same approximation as in the
CSA, one should first rewrite the collision term in the
retarded-advanced basis~:
\begin{equation}
C_\p[f]=
\frac{i}{2\omega_\p}
\left[
{\bs\Sigma}_{11}(P)
+\left(f(\p)+\frac{1}{2}\right)({\bs\Sigma}_{21}(P)-{\bs\Sigma}_{12}(P))
\right]\; .
\label{eq:coll12}
\end{equation}
(${\bs\Sigma}_{11}$ is imaginary, as well as
${\bs\Sigma}_{21}(P)-{\bs\Sigma}_{12}(P)$). At this point, we just
need to calculate the self-energies that appear in the right hand side
(at 2-loops if we want $2\to 2$ collisions) by neglecting the $1112$
vertex. The two versions of the CSA, with or without vacuum
fluctuations, simply correspond to keeping the $1/2$ in the factor
$1/2+f(\p)$ that appears in the propagator $G_{22}$ or
not\cite{MuellS1,Jeon3,MathiMT1}. It is easy to check that the
classical approximation with no vacuum fluctuations amounts to keeping
only the terms that are cubic in the particle distribution, while the
classical approximation with vacuum fluctuations leads to the
following collision term
\begin{eqnarray}
&&\frac{g^4}{4\omega_\p}
\int_{\k}\int_{\p'}\int_{\k'}
(2\pi)^4\delta(P+K-P'-K')\;
\nonumber\\
&&\qquad\times
\big[(f(\p')+\frac{1}{2})(f(\k')+\frac{1}{2})(1+f(\p)+f(\k))
\nonumber\\
&&\qquad-
(f(\p)+\frac{1}{2})(f(\k)+\frac{1}{2})(1+f(\p')+f(\k'))
\big]\; ,
\label{eq:Cclass1}
\end{eqnarray}
that has the same cubic and quadratic terms as the exact collision term,
but also some spurious terms that are linear in the particle
distribution.  

The fixed points of these ``classical'' Boltzmann equations are
\begin{align}
\mbox{without vacuum fluctuations~:~}&&f(\p)=\frac{T}{\omega_\p-\mu}
\nonumber\\
\mbox{with vacuum fluctuations~:~}&&f(\p)=\frac{T}{\omega_\p-\mu}-\frac{1}{2}
\; ,
\end{align}
($\omega_\p\equiv\sqrt{p^2+m^2}$) i.e. respectively the first and the
first two terms in the expansion of a Bose-Einstein distribution
around low energy.

The parameters $T$ and $\mu$ that appear in the asymptotic
distributions can be determined from conservation laws. In the
Boltzmann equation with only elastic scatterings, both the energy and
the particle number are conserved. For the CSA with vacuum
fluctuations, these conservation laws lead to
\begin{eqnarray}
n&=&n_c+\int\frac{d^3\p}{(2\pi)^3}\;\left(
\frac{T}{\omega_\p-\mu}-\frac{1}{2}
\right)\nonumber\\
\epsilon&=&n_c m+\int\frac{d^3\p}{(2\pi)^3}\;\omega_\p\;\left(
\frac{T}{\omega_\p-\mu}-\frac{1}{2}
\right)\; .
\label{eq:conserv}
\end{eqnarray}
In this exercise, we allow the formation of a Bose-Einstein condensate
with a particle density $n_c$, in case one wishes to consider highly
populated initial conditions. Note that in any case, there are only
two unknowns in these equations: $T$ and $\mu$ if there is no
condensation, or $T$ and $n_c$ if there is condensation (in which case
$\mu=m$). From these two equations, we can determine the two unknown
parameters. However, since the integrals are ultraviolet divergent, it
is necessary to introduce a cutoff $\Lambda$ on $|\p|$, which turns
$T$, $\mu$ and $n_c$ into $\Lambda$-dependent quantities~\cite{EpelbGTW1}.
\begin{figure}[htbp]
\begin{center}
\resizebox*{8cm}{!}{\includegraphics{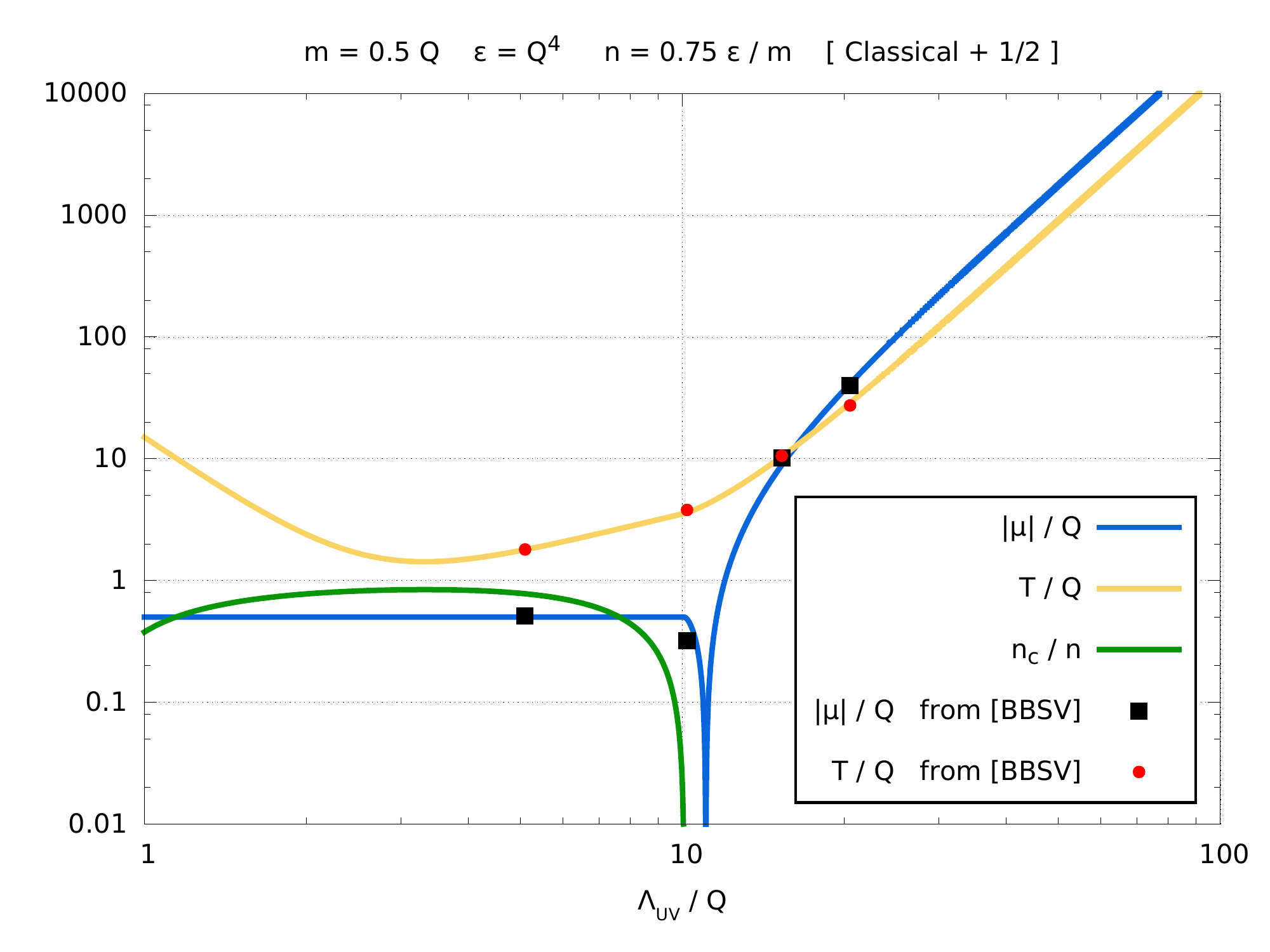}}
\end{center}
\caption{\label{fig:Tmu}Evolution of $T,\mu$ and $n_c$ as a function
  of the ultraviolet cutoff. The points reproduce the values listed
  in the figure 10 of Ref.~\cite{BergeBSV3}, obtained with a classical
  statistical field simulation.}
\end{figure}
The solution of eqs.~(\ref{eq:conserv}) is shown in the figure
\ref{fig:Tmu} ($Q$ is a physical momentum scale that characterizes the
initial momentum distribution of the particles), where we have also
superimposed results from ref.~\cite{BergeBSV3} obtained by lattice
classical statistical simulations. These curves show a very strong
dependence on the ultraviolet cutoff when it becomes much larger than
the physical scales, as expected given the fact that the CSA with
vacuum fluctuations is not renormalizable. On the other hand, there is
a region where the cutoff is a few times the physical scale, and where
the parameters that characterize the asymptotic distribution are
rather insensitive to the cutoff. Therefore, simulations using the CSA
with vacuum fluctuations as initial conditions should preferably be
performed with an ultraviolet cutoff chosen in that range, in order to
minimize the sensitivity of the results on the cutoff.

\subsection{Quantum corrections in anisotropic systems}
\label{sec:large-angle} 
Note that this non-renormalizability problem arises only when one uses
the CSA in conjunction with a spectrum of initial conditions that
represents vacuum fluctuations, as in eqs.~(\ref{eq:fluct3}). When
using particle-like initial conditions, such as those described in
eqs.~(\ref{eq:inco}), the CSA is ultraviolet finite provided that the
initial particle distribution falls faster than $1/k$. Because of
this, if one forgets that this type of initial condition is not
derived directly from the CGC, this implementation of the CSA may seem
better since one does not need to worry about the dependence on the
ultraviolet cutoff.

\begin{figure}[htbp]
\begin{center}
\resizebox*{4.5cm}{!}{\includegraphics{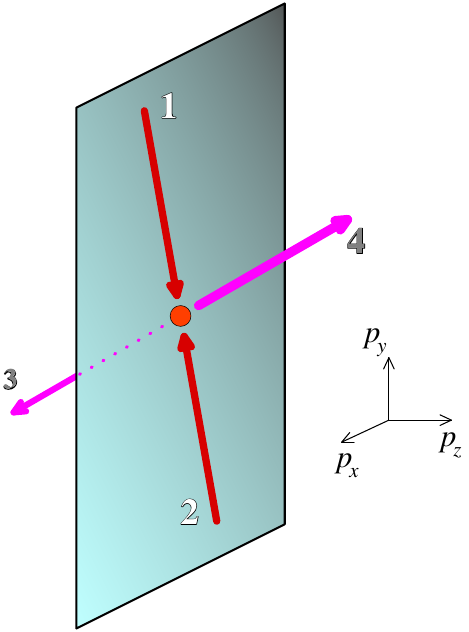}}
\end{center}
\caption{\label{fig:iso}$2\to 2$ scattering contributing to isotropization.}
\end{figure}

However, this choice of initial conditions in the CSA leads to a
different kind of problem when used in studies of isotropization in
heavy ion collisions, for the following reason. For the purpose of
this argument, let us again reason in terms of the
Boltzmann equation. For $2\to 2$ collisions, it reads
\begin{eqnarray}
\partial_t f_3
&\sim&
g^4\int_{124}
\cdots
\big[f_1f_2({f_3+f_4})-{f_3f_4}(f_1+f_2)\big]
\nonumber\\
&&\quad+g^4\int_{124}\cdots\big[f_1f_2-{f_3f_4}\big]\; .
\end{eqnarray}
(Here we are tracking the distribution of particles of momentum $p_3$
-- see the figure \ref{fig:iso}.)
 In this equation, we have separated the purely classical
terms (in $f^3$) from the subleading $f^2$ terms.  When only
particle-like fluctuations are included, the CSA applied to the
Boltzmann equation keeps only the $f^3$ terms, and neglects all the
other terms. In contrast, the CSA where vacuum fluctuations are
included has the $f^3$ and $f^2$ terms, and also some unphysical terms
that are linear in $f$ (see Eq.~(\ref{eq:Cclass1})).

Consider now a situation where the particle distribution is strongly
anisotropic, with a support in $p_z$ which is squeezed compared to the
support in $\p_\perp$. Starting from a generic initial condition of this
form, the unapproximated Boltzmann equation will in general lead
to isotropization because two purely transverse particles can be
scattered outside of the transverse plane\footnote{This is obvious in
  the $\phi^4$ scalar field theory, where the leading term in the two
  body cross-section is point-like and where scatterings at large
  angle are dominant.  This is also the case in the CGC, thanks to
  screening. Indeed, since the occupation number is of order $g^{-2}$,
  the Debye mass is of order $Q_s$. Since this is comparable to the
  typical momentum of the gluons, large angle scatterings should be important at early times.}. But does this still happen in
approximations of the Boltzmann equation?  In the figure
\ref{fig:iso}, one has $p_3^z=-p_4^z\not=0$, and given our assumption
about the support of the particle distribution, this means that
$f_3=f_4\approx 0$. In the collision term of the Boltzmann equation, a
number of terms therefore vanish~:
\begin{eqnarray}
\partial_t f_3
&\sim&
g^4\int_{124}
\cdots
\big[f_1f_2(\underbrace{f_3+f_4}_{0})-\underbrace{f_3f_4}_{0}(f_1+f_2)\big]
\nonumber\\
&&\quad+g^4\int_{124}\cdots\big[f_1f_2-\underbrace{f_3f_4}_{0}\big]\; .
\end{eqnarray}
In particular, all the cubic terms are zero. The problem is that these
are the only terms that are kept in the CSA with no vacuum
fluctuations.  The only non-zero term is the term in $f_1 f_2$, which
would be present in the CSA only if vacuum initial fluctuations are
present. From this discussion, the particle-like initial conditions in
the CSA, despite their appeal since they lead to UV finite results,
may be inappropriate because they lead to missing the most important
contribution to isotropization. In other words, in an anisotropic
system, the classical approximation could break down long before the
occupation number becomes of order 1. 

\begin{figure}[htbp]
\begin{center}
\resizebox*{9cm}{!}{\includegraphics{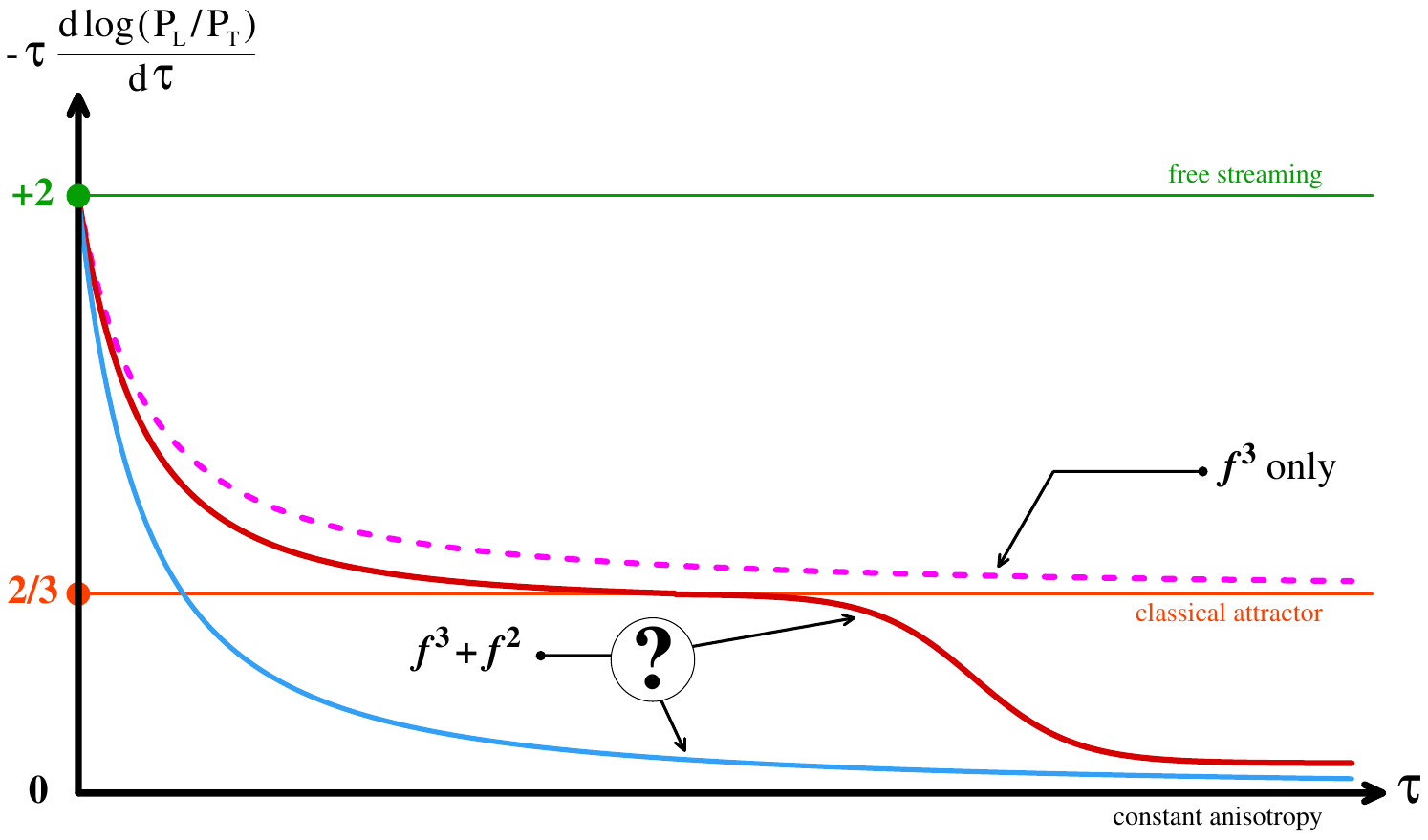}}
\end{center}
\caption{\label{fig:iso1}Behavior of the logarithmic derivative of the
  ratio $P_{_L}/P_{_T}$. Dotted curve: classical approximation where
  one keeps only the $f^3$ terms. Red curve: full collision term, if
  there exists a ``classical attractor''. Blue curve: full collision
  term, if the $f^2$ terms prevent the classical attractor.}
\end{figure}

This may also have an incidence on the behavior at intermediate
times. With only the $f^3$ terms, the ratio $P_{_L}/P_{_T}$ decreases
like $\tau^{-2/3}$ at late times, and therefore the system never
isotropizes in this approximation. On the other hand, the system is
expected to isotropize eventually with the complete Boltzmann equation
($f^3$ and $f^2$ terms), with a ratio $P_{_L}/P_{_T}\sim \tau^0$ at
late times. If the $f^2$ terms are truly negligible over some extended
period of time, then the full Boltzmann equation should lead to the
red curve in the figure \ref{fig:iso1},
in which the system spends some time stuck into a ``classical
attractor'' before eventually leaving it in order to isotropize. In
contrast, if the fact that the $f^2$ terms are always dominant in the
tail plays an important role, then one may instead get the blue curve.
The Boltzmann equation offers an interesting playground in order to
test these possibilities, since it can be solved
with and without the $f^2$ terms at comparable computational costs.

\section{Summary}
A lot of progress has been made in the past 10 years in QCD-based studies
of the early stages of heavy ion collisions. The most promising
framework for these studies is the Color Glass Condensate, which allows
one to systematically include the non-linear effects that prevail when
the gluon occupation numbers are large.

The description of the relevant degrees of freedom in the wavefunction
of the incoming nuclei and how they evolve as one varies the energy of
the collision is entering a very mature stage, since one now knows
this evolution at next-to-leading log accuracy. A lot more work will be 
needed before these NLL evolution equations can be implemented in
phenomenological studies of heavy ion collisions, but this is a very
important step towards more quantitative results.

Regarding the collision itself, i.e. how two objects described by
means of the CGC interact while they collide, one has now a much clearer
understanding of how observables can be expressed at leading order in
terms of a classical solution of the Yang-Mills equations (and how the
inclusive nature of an observable translates into retarded boundary
conditions for this classical solution), and how to calculate their
next-to-leading order corrections in terms of linearized perturbations
around this classical solution.

The recent years have also witnessed a renewed interest in the
question of the isotropization and thermalization of the system
produced in heavy ion collisions. This issue has indeed been made more
pressing by the many phenomenological successes of hydrodynamical
models in describing the expansion of the fireball as a nearly perfect
fluid, something which is at odds with the LO CGC results. From recent
works, there appears to be two tools of choice for these studies:
kinetic theory (in the small scattering angle approximation, or the
more sophisticated effective kinetic theory of ref.~\cite{ArnolMY5})
and the classical statistical approximation, which is more directly
related to the CGC framework. Although this problem seems to have now
reached a satisfactory level of understanding --both qualitatively and
quantitatively-- in the simpler case of a system confined in a fixed
volume, the situation is at the moment somewhat unclear in the more
realistic situation of a longitudinally expanding system, for which
two different results have been obtained in the classical statistical
approximation with different initial conditions. These initial
conditions mostly differ in whether some vacuum quantum fluctuations
are included or not, and a prime question to clarify in the near
future will be to understand to what extent quantum fluctuations play
a role in isotropization and thermalization.

\section*{Acknowledgements}
This work is supported by the Agence Nationale de la Recherche project 11-BS04-015-01.

%\bibliography{biblio}
%\bibliographystyle{unsrt}

\end{document}